\documentclass[a4paper,fleqn,usenatbib]{mnras}
\usepackage{Times}
\usepackage[T1]{fontenc}
\usepackage{ae,aecompl}

\usepackage{graphicx}	
\usepackage{amsmath}	
\usepackage{amssymb}	

\title[SC4K: Ly$\alpha$ emitters at $z\sim2-6$]{Slicing COSMOS with SC4K: the evolution of typical Ly$\alpha$ emitters and the Ly$\alpha$ escape fraction from $\bf z\sim2$ to $\bf z\sim6$}

\author[Sobral, Santos, Matthee et al.]{David Sobral$^{1,2}$\thanks{E-mail: d.sobral@lancaster.ac.uk}, S\'ergio Santos$^{1}$, Jorryt Matthee$^{2}$, Ana Paulino-Afonso$^{1,3,4}$, \newauthor Bruno Ribeiro$^{5}$, Jo\~{a}o Calhau$^{1}$, Ali~A. Khostovan$^{6}$ \\ 
$^{1}$ Department of Physics, Lancaster University, Lancaster, LA1 4YB, UK \\
$^{2}$ Leiden Observatory, Leiden University, P.O.\ Box 9513, NL-2300 RA Leiden, The Netherlands \\
$^{3}$ Instituto de Astrof\'{\i}sica e Ci\^{e}ncias do Espa\c{c}o, Universidade de Lisboa, OAL, Tapada da Ajuda, PT1349-018 Lisboa, Portugal \\
$^{4}$ Departamento de F\'{i}sica, Faculdade de Ci\^{e}ncias, Universidade de Lisboa, Edif\'{i}cio C8, Campo Grande, PT1749-016 Lisboa, Portugal\\
$^{5}$ Centro de Computa\c{c}\~{a}o Gr\'afica, CVIG, Campus de Azur\'em, PT4800-058 Guimar\~{a}es, Portugal\\
$^{6}$ Department of Physics and Astronomy, University of California, 900 University Ave., Riverside, CA 92521, USA}

\date{Accepted 2018 February 9. Received 2018 February 7; in original form 2017 December 12}

\pubyear{2017}

\begin{document}
\label{firstpage}
\pagerange{\pageref{firstpage}--\pageref{lastpage}}
\maketitle

\begin{abstract}
We present and explore deep narrow- and medium-band data obtained with the Subaru and the Isaac Newton telescopes in the $\sim2$\,deg$^2$ COSMOS field. We use these data as an extremely wide, low-resolution ($R\sim20-80$) IFU survey to slice through the COSMOS field and obtain a large sample of $\sim4000$ Ly$\alpha$ emitters (LAEs) from $z\sim2$ to $z\sim6$ in 16 redshift slices (SC4K). We present new Ly$\alpha$ luminosity functions (LFs) covering a co-moving volume of $\sim10^8$\,Mpc$^{3}$. SC4K extensively complements ultra-deep surveys, jointly covering over 4\,dex in Ly$\alpha$ luminosity and revealing a global ($2.5<z<6$) synergy LF with $\alpha=-1.93^{+0.12}_{-0.12}$, $\rm \log_{10}\Phi^*_{\rm Ly\alpha}={-3.45^{+0.22}_{-0.29}}$\,Mpc$^{-3}$ and $\rm \log_{10}L^*_{\rm Ly\alpha}={42.93^{+0.15}_{-0.11}}$\,erg\,s$^{-1}$. The Schechter component of the Ly$\alpha$ LF reveals a factor $\sim5$ rise in $\rm L^*_{\rm Ly\alpha}$ and a $\sim7\times$ decline in $\Phi^*_{\rm Ly\alpha}$ from $z\sim2$ to $z\sim6$. The data reveal an extra power-law (or Schechter) component above L$_{\rm Ly\alpha}\approx10^{43.3}$\,erg\,s$^{-1}$ at $z\sim2.2-3.5$ and we show that it is partially driven by X-ray and radio AGN, as their Ly$\alpha$ LF resembles the excess. The power-law component vanishes and/or is below our detection limits above $z>3.5$, likely linked with the evolution of the AGN population. The Ly$\alpha$ luminosity density rises by a factor $\sim2$ from $z\sim2$ to $z\sim3$ but is then found to be roughly constant ($1.1^{+0.2}_{-0.2}\times10^{40}$\,erg\,s$^{-1}$\,Mpc$^{-3}$) to $z\sim6$, despite the $\sim0.7$\,dex drop in UV luminosity density. The Ly$\alpha$/UV luminosity density ratio rises from $4\pm1$\% to $30\pm6$\% from $z\sim2.2$ to $z\sim6$. Our results imply a rise of a factor of $\approx2$ in the global ionisation efficiency ($\xi_{\rm ion}$) and a factor $\approx4\pm1$ in the Ly$\alpha$ escape fraction from $z\sim2$ to $z\sim6$, hinting for evolution in both the typical burstiness/stellar populations and even more so in the typical ISM conditions allowing Ly$\alpha$ photons to escape.
\end{abstract}


\section{Introduction}\label{sec:introduction}

Understanding how galaxies form and evolve across cosmic time is a complex challenge which requires identifying and studying the inter-dependencies of key physical mechanisms over a range of environments \citep[see e.g.][]{Schaye2015,Crain2015,Henriques2015,Muldrew2018}, informed by a variety of observations \citep[e.g.][]{Muzzin2013}. It is now well established that the star formation rate density (SFRD) of the Universe evolves with redshift, peaking at $z\sim2-3$ \citep[e.g.][]{Lilly1996,Karim2011,Sobral2013,Madau2014} and declining at even higher redshift \citep[e.g.][]{Bouwens2015,Khostovan2015}, but several questions related to the physics of such evolution remain unanswered.

In order to unveil the evolution of physical properties of galaxies and active galactic nuclei (AGN) across time one requires self-consistent selection methods which can be applied across redshift. The Lyman Break selection \citep[e.g.][]{Koo1980,Steidel1993,Giavalisco1996} has been successfully used to produce large samples of galaxies up to $z\sim10$ \citep[e.g.][]{McLure2010,Ellis2013,Bouwens2014,Bouwens2014z10,Finkelstein2016,Bielby2016} through extremely deep optical to near-infrared (NIR) observations. However, UV-continuum selected samples using the Hubble Space Telescope ({\it HST}) are typically too faint \citep[e.g.][]{Atek2015} for extensive spectroscopic follow-up, particularly when probing distant look-back times \citep[but large area surveys can still provide ideal follow-up targets e.g.][]{Bowler2014,Bowler2016}. One alternative is to select galaxies by their Hydrogen nebular recombination lines, such as H$\alpha$ in the rest-frame optical \citep[e.g.][]{Sobral2013,Colbert2013} or Lyman-$\alpha$ (Ly$\alpha$; $\lambda_0 = 1215.67$\,\AA) in the rest-frame UV. 

Ly$\alpha$ is intrinsically the strongest emission line in the rest-frame optical and UV \citep[e.g.][]{Partridge1967,Pritchet1994} and it is routinely used to select high redshift sources ($z\sim2-7$; see e.g. \citealt{Malhotra2004}). Ly$\alpha$ is expected to be emitted by young star-forming galaxies \citep[e.g.][]{Charlot1993,Pritchet1994}, but it is also observed around AGN \citep[e.g.][]{Miley2008}. Searches for Ly$\alpha$ emitters (LAEs) have created samples of thousands of galaxies/AGN, including sources that are too faint to be detected by continuum based searches \citep[e.g.][]{Bacon2015}. The techniques used to detect LAEs include narrow-band surveys \citep[e.g.][]{Rhoads2000,Westra2006,Ouchi2008,Hu2010, Matthee2015,Konno2017,Zheng2017}, Integral Field Unit (IFU) surveys \citep[e.g.][]{vanBreukelen2005,Bacon2015,Drake2017a} and blind slit spectroscopy \citep[e.g.][]{Martin2004,Rauch2008,Cassata2011,Cassata2015}. Galaxies selected through their Ly$\alpha$ emission allow for easy spectroscopic follow-up due to their high EWs \citep[e.g.][]{Hashimoto2017} and typically probe low stellar masses \citep[see e.g.][]{Gawiser2007,Hagen2016,Oyarzun2017}. Narrow-band and/or IFU surveys have the added benefit of being truly blind, and thus allow a good assessment of the volume and selection completeness.

Unfortunately, inferring intrinsic properties of galaxies from Ly$\alpha$ observations alone can be challenging due to the highly complex resonant nature of this emission line \citep[for a review on the physics of Ly$\alpha$ radiative transfer see e.g.][]{Dijkstra2017}. A significant fraction of Ly$\alpha$ photons is scattered by the Inter-Stellar Medium (ISM), increasing the likelihood of being absorbed by dust, and in the Circum-Galactic Medium (CGM) as evidenced by the presence of extended Ly$\alpha$ halos \citep[e.g.][]{Momose2014,Wisotzki2016}. Therefore, the Ly$\alpha$ escape fraction\footnote{Throughout this study we use $\rm f_{esc}$ to quantify the escape fraction of Ly$\alpha$ photons, not Lyman-continuum photons.} \citep[$\rm f_{esc}$; see e.g.][]{Atek2008}, the ratio between the observed and the intrinsically produced Ly$\alpha$ luminosity from a galaxy, is still poorly understood quantitatively. New studies are now directly measuring $\rm f_{esc}$ of large samples of galaxies and over a range of redshifts by obtaining H$\alpha$ and Ly$\alpha$ observations simultaneously \citep[see][]{Nakajima2012,Matthee2016_CALY,Sobral2017,Harikane2017}. For example, $\rm f_{esc}$ is found to be anti-correlated with stellar mass \citep[e.g.][]{Matthee2016_CALY,Oyarzun2017}, dust attenuation \citep[e.g.][]{Verhamme2008,Hayes2011,Matthee2016_CALY} and SFR \citep[e.g.][]{Matthee2016_CALY}. Interestingly, the Ly$\alpha$ EW$_0$ seems to be the simplest empirical predictor of $\rm f_{esc}$ in LAEs with a relation that shows no evolution from $z\sim0$ to $z\sim2$ \citep[][]{Sobral2017} and that may remain the same all the way to $z\sim5$ \citep[][]{Harikane2017}.

``Typical" star-forming galaxies at $z\sim2$ have low $\rm f_{esc}$ ($\sim1-5$\%; e.g. \citealt{Oteo2015,Cassata2015}), likely because the dust present in their ISM easily absorbs Ly$\alpha$ photons \citep[e.g.][]{Ciardullo2014,Oteo2015,Oyarzun2017} and prevents most Ly$\alpha$ emission from escaping \citep[see e.g.][]{Song2014}. However, sources selected through their Ly$\alpha$ emission typically have $\sim10$ times higher escape fractions \citep[e.g.][]{Song2014,Sobral2017}, with Ly$\alpha$ escaping over $\approx2\times$ larger radii than H$\alpha$ \citep[e.g.][]{Sobral2017}. Furthermore, due to the sensitivity of $\rm f_{esc}$ to neutral Hydrogen, Ly$\alpha$ can be used as a proxy of the ISM neutral gas (HI) content \citep{Trainor2015,Konno2016} and the dust content \citep{Hayes2011}.

Statistically, the number density of LAEs as a function of luminosity (the luminosity function, LF), encodes valuable information on the global properties of LAEs and Ly$\alpha$ emission. Observations have revealed that the Ly$\alpha$ LF remains roughly constant at $z\sim3-6$ \citep[e.g.][]{Ouchi2008,Santos2016,Drake2017a}. This is in principle unexpected, as the cosmic SFRD, as traced by the UV LF, drops significantly at those redshifts \citep[e.g.][]{Bouwens2015,Finkelstein2015} and implies that intrinsic properties of galaxies may be evolving, on average, with redshift. Those may include lower dust content, leading to a higher $\rm f_{esc}$ which could compensate for a lower intrinsic production of Ly$\alpha$ photons \citep[e.g.][]{Hayes2011,Konno2016}. Another possibility is that $\xi_{\rm ion}$, which measures the ratio between ionising (LyC) and UV flux density increases with redshift \citep[e.g.][]{Duncan2015,Khostovan2016,MattheeGALEX2017}. In practice, a combined increase of both $\xi_{\rm ion}$ and $\rm f_{esc}$ is also possible, which could tell us about an evolution of both the typical stellar populations/burstiness but also on the evolving physics/ISM conditions) of the escape of Ly$\alpha$ photons.

In this work, we use 16 different narrow- and medium-band filters over the COSMOS field to select a large sample of LAEs in a total co-moving volume of $6.4\times10^7$\,Mpc$^3$ and a wide redshift range of $z\sim2-6$, addressing the current shortcomings of deep, small area surveys. Our survey can be seen as a very wide ($\approx2$ deg$^2$), low resolution IFU survey between 400-850\,nm, probing LAEs from the end of the epoch of re-ionisation at $z\sim6$ \citep[e.g.][]{Fan2006} to the peak of star-formation history at $z\sim2-3$.  

We structure this paper as follows: Section \ref{sec:data} presents the data and the extraction of sources. Section \ref{sec:criteria} presents the selection of line emitters, the criteria we applied to select LAE candidates at $z\sim2-6$ and the final SC4K sample. We present the methods in Section \ref{Methods}, including all the steps and corrections in determining Ly$\alpha$ LFs. Results are presented in Section \ref{sec:results}, including the evolution of the Ly$\alpha$ LF with redshift, comparisons with other surveys, the synergy LF (S-SC4K) and the evolution of the Ly$\alpha$ luminosity density. We discuss our results in Section \ref{Discussion}, including how $\rm f_{esc}$ and $\xi_{\rm ion}$ likely evolve with redshift. Finally, Section \ref{sec:conclusions} presents the conclusions of this paper. Throughout this work we use a $\Lambda$CDM cosmology with H$_0 = 70$\,km\,s$^{-1}$\,Mpc$^{-1}$, $\Omega _M = 0.3$ and $\Omega _\Lambda = 0.7$. All magnitudes in this paper are presented in the AB system \citep[][]{Oke1983} and we use a Salpeter \citep[][]{Salpeter1955} initial mass function (IMF).

%
%
\begin{figure*}
\centering
\includegraphics[width=14.9cm]{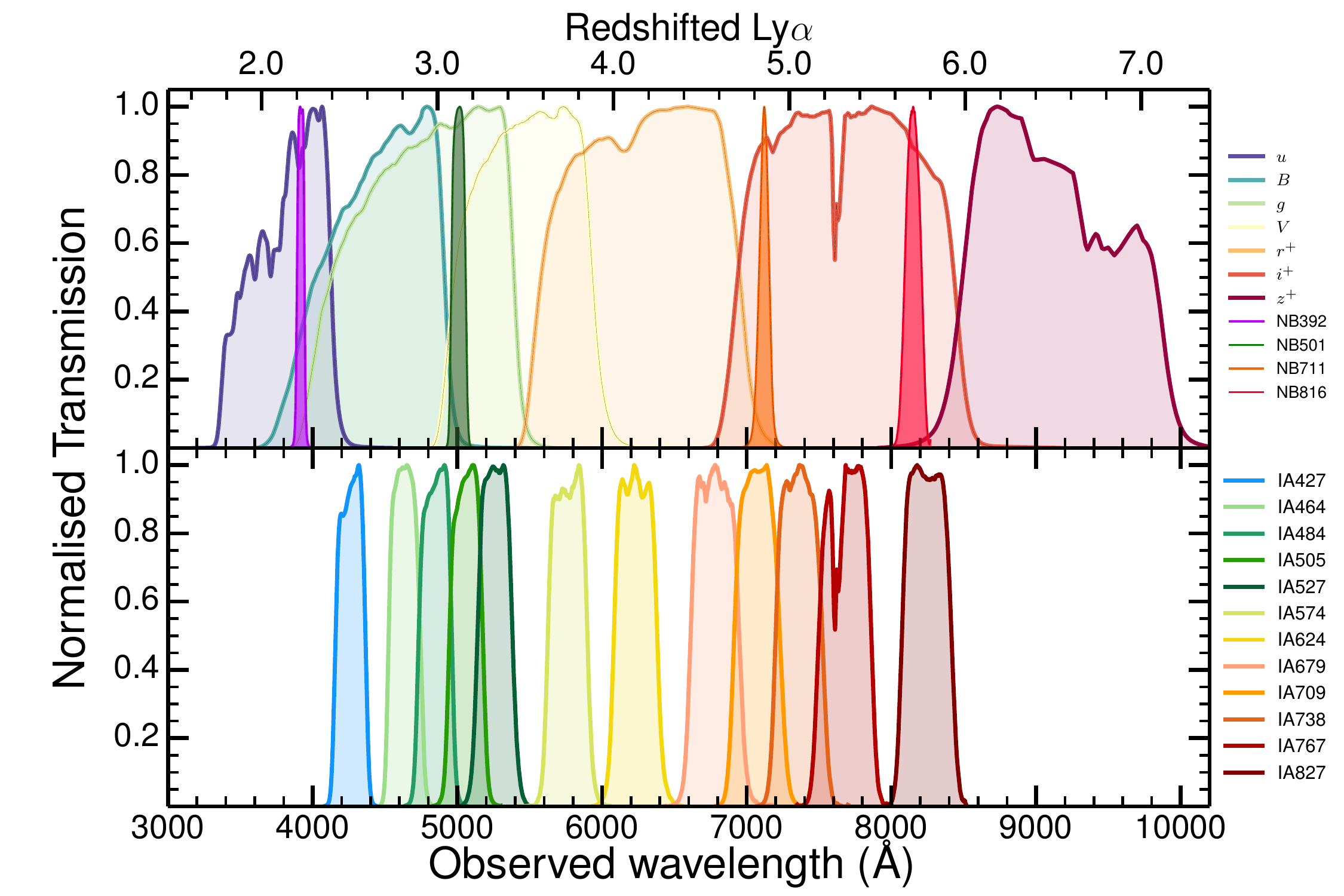}
\caption{Normalised filter profiles used in this paper. The top axis indicates the redshift of Ly$\alpha$ emission for the corresponding observed wavelength. {\it Top:} The broad-bands ($u$, $B$, $g$, $V$, $r^+$, $i^+$ and $z^+$; typical FWHM $\sim100$ nm) which we use to estimate the continuum for candidate line-emitters and four narrow-bands which we also present in our final catalogue (NB392, NB501, NB711 and NB816; typical FWHM $\sim10$ nm). {\it Bottom:} The 12 medium-bands used in this study (typical FWHM $\sim30$ nm; see Table \ref{tab:depth_phot}) which are sensitive to Ly$\alpha$ emission from $z\sim2.5$ to $z\sim5.8$. Note that some of the medium-band filters overlap slightly, which can result in some sources being detected as LAEs in two consecutive medium bands, although we note that the overlapping volume is always relatively small.}\label{fig:filters}
\end{figure*}

\section{Data and source extraction} \label{sec:data}

The COSMOS field \citep{Scoville2007,Capak2007} is one of the most widely studied regions of extragalactic sky, with a plethora of publicly available multi-wavelength coverage. Data in COSMOS include X-ray, UV, optical, NIR, FIR and radio \citep[see e.g.][]{Ilbert2009,Civano2016,Laigle2016,Smolcic2017}. We explore a range of narrow- and medium-band data \citep[][]{Capak2007,Taniguchi2007,Taniguchi2015,Santos2016,Sobral2017,MattheeBOOTES2017} over roughly the full COSMOS field in order to 1) create detection-catalogues for each band, 2) identify sources with strong excess emission in those bands relative to their broad-band counterparts and 3) obtain dual-mode photometry on all other bands available in order to further constrain the (photometric-)redshift of each source. In Figure \ref{fig:filters} we show the filter profiles of all the 12 medium-bands and the 4 narrow-bands used in this paper. These filters are capable of detecting various emission lines, particularly redshifted Ly$\alpha$ spanning a wide redshift range, from $z\sim2$ to $z\sim6$.

%
%
\begin{table}
\centering
\caption{The medium-band filters \citep[see][]{Taniguchi2015} and the depth of the data obtained with them, measured directly (3\,$\sigma$; 5\,$\sigma$ can be obtained by subtracting 0.5) in 2$''$ apertures and by \citealt{Muzzin2013} (M13). We also transform our measured 3\,$\sigma$ limiting magnitude (2\,$''$) into a flux limit (in units of erg\,s$^{-1}$\,cm$^{-2}$) in the case of the full flux within the 2$''$ medium-band aperture being from an emission line.}\label{tab:depth_phot}
\begin{tabular}{@{}ccccc@{}}
\hline
Medium & $\lambda_c$ [FWHM] & 3\,$\sigma$ Depth & 3$\sigma$ Flux & 5\,$\sigma$ (M13)  \\
 Band & ({\AA}) & (2$''$) &  ($\times10^{-17}$)   & (2.1$''$) \\
\hline
IA427 & 4263.5 [207.3] & 26.1 & 4.6  & 26.1 \\
IA464 & 4635.1 [218.1] & 26.0 & 4.5  & 25.8 \\
IA484 & 4849.2 [229.1] & 26.1 & 3.9  & 26.1 \\
IA505 & 5062.1 [231.5] & 25.8 & 4.8  & 25.9 \\
IA527 & 5261.1 [242.7] & 26.1 & 3.5  & 26.1 \\
IA574 & 5764.8 [272.8] & 25.9 & 4.0  & 25.7 \\
IA624 & 6232.9 [299.9] & 25.8 & 4.1  & 25.9 \\
IA679 & 6781.1 [335.9] & 25.7 & 4.3  & 25.6 \\
IA709 & 7073.6 [316.3] & 25.8 & 3.4  & 25.8 \\
IA738 & 7361.5 [323.8] & 25.7 & 3.5  & 25.6 \\
IA767 & 7684.9 [365.0] & 25.7 & 3.6  & 25.4 \\
IA827 & 8244.5 [342.8] & 25.7 & 3.0  & 25.4 \\
\hline
\end{tabular}
\end{table}

\subsection{Medium-band data}

We retrieve the publicly available medium-band data (see Table \ref{tab:depth_phot} and Figure \ref{fig:filters}) from the COSMOS archive \citep[see][]{Ilbert2009,Taniguchi2015}. All data were obtained with the Suprime-Cam (S-Cam) instrument on the Subaru Telescope \citep{Miyazaki2002}. The data were taken with seeing conditions varying from 0.6\,$''$ to 1.0\,$''$, with an overall FWHM of $0.8\pm0.1$\,$''$ \citep[see also][]{Muzzin2013,Taniguchi2015}.
The images have a roughly similar average depth \citep{Muzzin2013} but with some exceptions (see Table \ref{tab:depth_phot}), varying from 26.2\,mag (deepest: IA427, IA484 and IA527) to 25.4\,mag (shallowest, IA767), measured in 2.1$''$ apertures \citep[$5\sigma$, see][]{Muzzin2013}. We also obtain our own depth measurements by placing 100,000 empty/random $2''$ apertures in each of the (native) images and determining the standard deviation. The results are presented in Table \ref{tab:depth_phot} and compared with the depths measured in \cite{Muzzin2013}, who used PSF-matched data.


\subsection{Narrow-band data}

We complement our medium-band data with four narrow-band studies in the COSMOS field: the CALYMHA survey at $z=2.2$ \citep{Sobral2017} and a $z=3.1$ survey \citep[][]{MattheeBOOTES2017} using the narrow-band filters NB392 and NB501, respectively, both mounted on the 2.5\,m Isaac Newton Telescope's WFC. The NB392 data ($\lambda_c = 392$ nm, $\Delta\lambda=5.2$ nm; \citealt{Sobral2017}) have a 5$\sigma$ depth of 23.7-24.5 AB magnitude and a typical PSF-FWHM of 1.8\,$''$. The NB501 data ($\lambda_c = 501$ nm, $\Delta\lambda=10$ nm) were taken and reduced with a similar strategy and data-quality as the NB501 data described in \cite{MattheeBOOTES2017} and have a typical 5$\sigma$ depth of 24.0\,AB magnitude with 1.6$''$ PSF-FWHM. Limiting magnitudes for NB392 and NB501 data were measured with 3$''$ apertures.

In addition, we also use two narrow-band surveys exploring S-Cam data: $z=4.8$ (Perez et al. in prep) and $z=5.7$ \citep{Santos2016}; these have used the narrow-band filters NB711 and NB816, respectively (see Figure \ref{fig:filters}). We note that all NB and MB selected catalogues have been obtained in similar ways, which we describe in Section \ref{sec:extract}.

%
%
\begin{figure*}
\includegraphics[width=16.0cm]{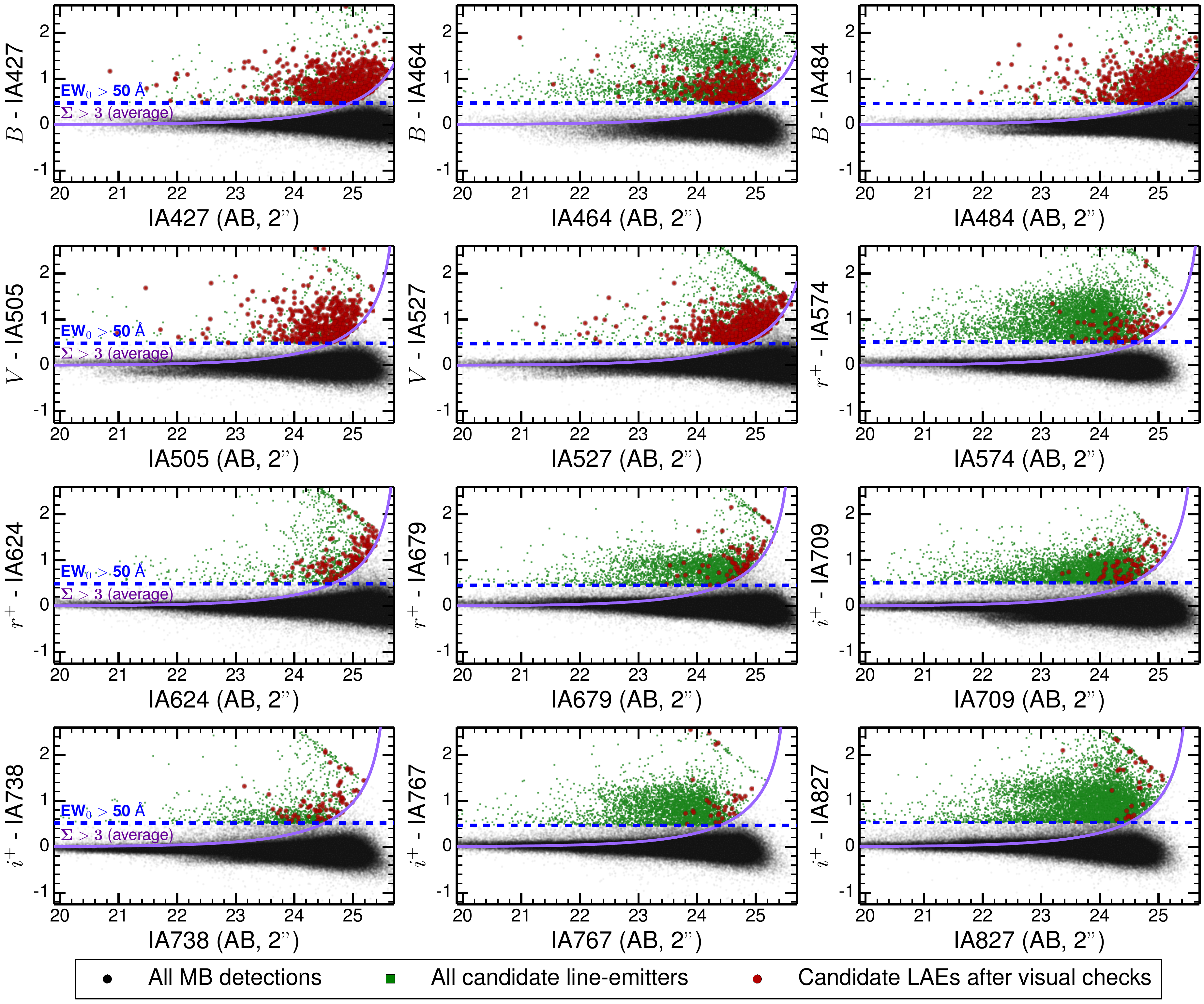}
\caption{The colour-magnitude diagrams used for the selection of line-emitters for the 12 medium-bands. Each medium-band magnitude is plotted versus the excess colour, and we identify sources with a high enough EW (corresponding to a rest-frame EW of $>50$\,{\AA} for LAEs) and with a significant excess (average $\Sigma>3$). The selection criteria of LAEs are presented in Table \ref{tab:criteria}. MB detections are shown in black, candidate line-emitters (prior to individual visual checks) are shown in green and candidate LAEs in red (after visually checking all of them). We assign the broad-band detection limit to sources with no broad-band detection. It is clear from the panels that, on average, the amount and fraction of LAEs greatly decreases from the bluer (where almost all candidate line-emitters are LAEs) to the redder filters (where only a small fraction is consistent with being a LAE).}
\label{fig:excess}
\end{figure*}

\subsection{Extraction of sources} \label{sec:extract}
 
To produce the narrow- or medium-band selected catalogues \citep[see e.g.][]{MattheeBOOTES2017}, we follow \cite{Santos2016}. Briefly, we start by registering the $u$, $B$, $g$, $V$, $r^+$, $i^+$, $z^+$, $Y$, $J$, $H$ and $K$ \citep{Taniguchi2007,Capak2007,McCracken2012} and all the medium-band (or narrow-band) images to a common astrometric reference frame using {\sc Swarp} \citep{Bertin2002}. We extract sources with a primary 2$''$ aperture\footnote{Because the NB392 and NB501 data have a broader PSF, photometry has been done with 3$''$ apertures.} (but we note we also extract them with multiple apertures, including {\sc mag-auto}, a proxy of the total magnitude) using {\sc SExtractor} \citep{Bertin1996} in dual-image mode, and with each of the medium-band images as the detection image. Therefore, for each medium-band, we create a catalogue with all the detections on that band, and with the broad-band photometry extracted at the coordinates of each detection. We thus note that our selection is purely based on the detection of a source in a medium- or narrow-band, independently of its continuum strength.

Before creating our final catalogues, we investigate the need for any significant masking to remove low quality regions and diffraction patterns around bright stars. In addition to removing such regions, we also find that there is a small area in the corner of the COSMOS field ($\approx 0.02$ deg$^2$) for which there is no $u$-band data. Given that we require blue photometry to select LAEs and reject lower redshift sources (see Tables \ref{tab:depth} and \ref{tab:criteria}), we mask/exclude this region for filters bluer than IA574. After masking, the contiguous survey area is 1.94-1.96\,deg$^2$ for the medium-band filters and 1.96\,deg$^2$ for the NB711 and NB816 filters, while the area covered by the NB392 and NB501 data is 1.21\,deg$^2$ and 0.85 deg$^2$, respectively \citep[][]{Matthee2016_CALY,Sobral2017}.

%
%
\begin{table}
\caption{The estimated depth of broad-band data used in our analysis (3$\sigma$). We measure these by placing 100,000 random 2$''$ empty apertures, and computing the standard deviation of the counts and converting it to magnitudes. The 2\,$\sigma$ and 4\,$\sigma$ limits can be obtained by adding 0.44 and subtracting 0.31, respectively.}\label{tab:depth}
\begin{center}
\begin{tabular}{ccccccc}
\hline
$u_{3\sigma}$ & $B_{3\sigma}$ & $V_{3\sigma}$ & $g_{3\sigma}$ & $r^+_{3\sigma}$ & $i^+_{3\sigma}$ & $z^+_{3\sigma}$ \\
\hline
26.81 &  27.21 &  26.50 &  26.61 &  26.55 &  26.12 &  25.23 \\
\hline
\end{tabular}
\end{center}
\end{table} 

\section{Selection criteria: SC4K} \label{sec:criteria}

\subsection{Selection of candidate line-emitters} \label{subsec:criteria_line_emit}

In order to identify sources with candidate emission lines out of all medium-band selected sources, it is necessary to estimate the continuum of each source. As the central wavelengths of medium-bands are typically offset (see Figure \ref{fig:filters}) from the central wavelengths of their overlapping broad-band, we need to investigate and apply a correction to the medium band photometry ($\rm MB_0$). This step/correction assures that a measured medium-band excess is not dependent on the intrinsic slope of the continuum (estimated with two broad-band magnitudes, $\rm BB-BB_{\rm adjacent}$), similarly to corrections applied for narrow-band surveys \citep[e.g.][]{Sobral2013,Vilella-Rojo2015}. Without such correction, sources with significant colours could mimic emission lines. In practice this requires re-calibrating either $\rm MB_0$ or $\rm BB$ photometry (or producing a new artificial $\rm BB$ magnitude) to assure that, on average, sources without an emission-line will have a zero colour excess ($\rm BB-MB\approx0$) regardless of their continuum colour ($\rm BB-BB_{\rm adjacent}$). We do this by evaluating the colour dependence of $\rm BB-MB_0$ on $\rm BB-BB_{\rm adjacent}$ and parameterising it as (calculating $m$ and $b$):
\begin{equation} \label{eq:colour_coef}
{\rm BB-MB_0}=m\times({\rm BB-BB_{\rm adjacent}}) + {b}
\end{equation}
We then use coefficients $m$ and $b$ to finally obtain:
\begin{equation} \label{eq:colour_coef}
{\rm MB=MB_0}+(m\times({\rm BB-BB_{\rm adjacent}}) + b)
\end{equation}
with the filters listed in Table \ref{tab:criteria}. The coefficients $m$ and $b$ are provided in Table \ref{tab:colour_coef}. We note that for some filter combinations both $m$ and $b$ are effectively zero. For sources without $\rm BB-BB_{\rm adjacent}$ ($<2\sigma$ detection in either band) we compute a median correction which we apply per medium-band filter. Typical median corrections are at the 0.1\,mag level and in the 0.0-0.3 range (see Table \ref{tab:colour_coef}).

%
%
\begin{figure*}
\centering
\includegraphics[width=17.7cm]{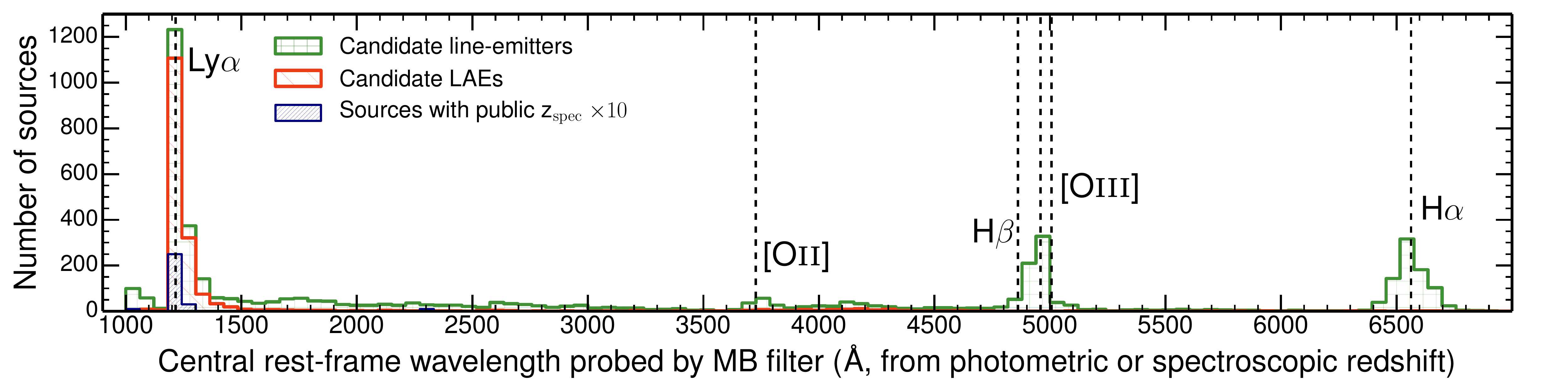}
\caption{Distribution of the central rest-frame wavelengths probed by the different medium-band filters (based on photometric or spectroscopic redshifts) of our continuum-bright (with accurate photometric redshifts) sample of line-emitters (green histogram) and its subset containing only our final LAEs (red histogram). The photometric redshifts used in this figure are taken from \citet{Laigle2016}. We convert each photometric value to a rest-frame wavelength assuming the source has an emission line at the central wavelength of the corresponding medium-band. The black dashed lines are the rest-frame wavelengths of the main emission lines probed. Our sample of continuum-bright line-emitters is clearly dominated by Ly$\alpha$ emitters, followed by a population of H$\alpha$ emitters, [O{\sc iii}]+H$\beta$ emitters, and finally [O{\sc ii}] emitters. We find that our Ly$\alpha$ selection criteria is able to remove the vast majority of lower redshift contaminants whilst maintaining the bulk of Ly$\alpha$ photometric candidates (see Section \ref{subsec:criteria_lya}).}
 \label{fig:photoz}
\end{figure*}

For the selection of line-emitters we follow the same methodology used for narrow-band surveys \citep[e.g.][]{Sobral2013,Sobral2017,MattheeBOOTES2017}, based on two main parameters: the emission-line equivalent width (EW), and the emission-line or excess significance \citep[$\Sigma$; e.g.][]{Bunker1995}; see Figure \ref{fig:excess}. $\Sigma$ quantifies how significantly above the noise a given colour excess (due to a potential emission line) is and can be written as \citep[][]{Sobral2013}:
\begin{equation}
\rm \Sigma = \frac{1-10^{-0.4(BB-MB)}}{10^{-0.4({\it ZP}-MB)}\sqrt{\rm rms_{BB}^{2}+\rm rms_{MB}^{2}}},
\end{equation}
where $\rm BB$ and $\rm MB$ are the broad- and the medium-band magnitudes and $ZP$ is the zero-point of the image. We estimate $\rm rms_{MB}$ and $\rm rms_{BB}$ by randomly placing $2''$ apertures in the appropriate images and determining the standard deviation per image. This approach takes spatially correlated noise into account. We apply an emission-line significance threshold of $\Sigma>3$, similarly to other studies \citep[e.g.][]{Matthee2015}. In addition to $\Sigma$, we also measure the observed EW ($\rm EW_{obs}$) of potential lines as:
\begin{equation}
\rm EW_{obs} = \Delta\lambda_{MB}\frac{f_{MB}-f_{BB}}{f_{BB}-f_{MB}(\Delta\lambda_{MB}/\Delta\lambda_{BB})},
\end{equation}
where $\rm \Delta\lambda_{MB}$ and $\rm \Delta\lambda_{BB}$ are the FWHM of the medium- (see Table \ref{tab:depth_phot}) and broad-band filters \citep[][]{Capak2007,Taniguchi2015}, and $\rm f_{MB}$ and $\rm f_{BB}$ are the flux densities\footnote{Note that as a consequence of the way we define/correct MB magnitudes, their flux densities ($F_{\lambda}$) need to be calculated with the same effective wavelength as the corresponding BB.} measured of the two filters.

Typical narrow-band surveys apply a Ly$\alpha$ rest-frame EW (EW$_0$) cut of $\approx25$\,{\AA} \citep[e.g.][]{Ouchi2008}, mostly to avoid contamination from other line-emitters, as Ly$\alpha$ is typically the line with the highest observed EW \citep[but see also other high EW contaminants in e.g.][]{Sobral2017}. Recent surveys also explored lowering this cut, showing that a few extra real Ly$\alpha$ sources may be recovered in those cases, which can populate the bright end \citep[][see also VUDS, e.g. \citealt{LeFevre2015}]{Sobral2017}, but also introduce many extra contaminants. Given that we are using wider filters in comparison to the typical narrow-band filters, we are forced to use a higher observed EW cut to retrieve clean samples of line-emitters. For our analysis, we find that setting an observed EW cut from 175\,{\AA} to 340\,{\AA} from the bluest (narrowest) to the reddest and broader filters is able to recover clean samples of line emitters and yields an homogeneous rest-frame Ly$\alpha$ equivalent width cut of EW$_0>50$\,{\AA} for all of our medium-bands. Note that our EW$_0$ cut (for LAEs) is about twice the typical used in narrow-band surveys (25\,\AA; see e.g. \citealt{Santos2016}), implying we are likely less contaminated by lower redshift line-emitters, but that we may be less complete. We take this into account when deriving completeness corrections, but we note that, in practice, the vast majority of LAEs at high redshift show EW$_0>50$\,{\AA}; see e.g. \cite{Ouchi2008}. For an in-depth analysis of selecting LAEs with different EW$_0$ cuts see \cite{Sobral2017} and Perez et al. (in prep.). 

The full selection procedure to search for candidate emission-line sources is illustrated in Figure $\ref{fig:excess}$, which shows the medium-band colour excess versus medium-band magnitude for each band. It can be seen that the EW threshold is well above the scatter at bright magnitudes ($\lesssim 23$). In total, we identify 40,726 potential line-emitters, with each medium-band contributing with roughly 2,000-3,000 emitters to the sample. We note, nonetheless, that we expect our full sample of $\approx40$\,k candidate line-emitters to still be contaminated by e.g. artefacts around bright stars, cosmic rays, and due to other image defects. In order to fully address this possibility, we visually inspect every single source in our final sample (see Section \ref{vis_inspec}), but we first filter out lower redshift emitters and isolate candidate LAEs.

%
%
\begin{table*}
\caption{The selection of LAEs from the sample of all line-emitters, using an observed EW threshold of EW $> 50\times(1+z)$\,{\AA} and $\Sigma>3$. The relevant LAE colour selection is given in the table. Numbers of LAEs are given after visually inspecting all candidate LAEs and rejecting interlopers. As described in the Section \ref{subsec:criteria_lya}, colour criteria are based on the Lyman break technique and removing sources with very red colours red-wards of the emission-line (which indicates that the potential Lyman break is actually a Balmer break and that the line is not Ly$\alpha$). We note that we explicitly perturb these colour selections with Markov chain Monte Carlo (MCMC) simulations and include the results in the errors when we estimate luminosity functions. We remove 20 confirmed lower redshift line emitters/contaminants, as described in Section \ref{subsec:zspecs} and we expect a $\approx10-20$\% remaining contamination. Due to a small overlap of some of the medium filters in their wings a small number (53; $\sim1$\%) of LAEs are detected as LAEs in two adjacent MB filters; these are kept in each of the filters for the full analysis. $^1$EW$_{0}$ > 5 \AA; \citet{Sobral2017} $^2$EW$_{0}$ > 25 \AA; \citet{Santos2016}; \citet{MattheeBOOTES2017}; Perez et al. in prep.}\label{tab:criteria}
\begin{center}
\begin{tabular}{ccccc}
\hline
\bf Selection & \bf Excess & \bf Ly$\alpha$ redshift & \bf LAE colour selection & \bf \# LAE\\
\bf filter & \bf filter & \bf FWHM & \bf (Section \ref{subsec:criteria_lya})  & \bf candidates\\
\hline
IA427 & $B$ ($u$) & $2.42-2.59$ & $(u>u_{3\sigma}$ $\vee$ $u-B>0.4)$ \& $(B-r^+<0.5)$ & 741\\
IA464 & $B$ ($V$)& $2.72-2.90$& $(u>u_{3\sigma}$ $\vee$ $u-B>0.5)$ \& ($B-r^+<0.8)$ & 311\\
IA484 & $B$ ($V$)& $2.89-3.08$ & $(u>u_{3\sigma}$ $\vee$ $u-B>0.5)$ \& $(B-r^+<0.75)$& 711\\
IA505 & $V$ ($B$)&$3.07-3.26$ & $(u>u_{3\sigma}$ $\vee$ $u-V>1.3)$ \& $(B-r^+<0.5)$& 483\\
IA527 & $V$ ($B$)& $3.23-3.43$& $(u>u_{3\sigma}$ $\vee$ $u-V>1.5)$ \& $(V-i^+<1.0)$& 641\\
IA574 & $r^+$ ($V$)& $3.63-3.85$& $(u>u_{3\sigma})$ \& ($B>B_{3\sigma}$ $\vee$ $B-r^+>1.0$) \& $(V-i^+<0.5)$& 98\\
IA624 & $r^+$ ($i^+$)& $4.00-4.25$& $(B>B_{3\sigma})$ \& ($V>V_{3\sigma}$ $\vee$ $V-r^+>0.5$) \& $(r^+-i^+<1.0)$& 142\\
IA679 & $r^+$ ($i^+$)& $4.44-4.72$& $(B>B_{3\sigma})$ \& ($V>V_{3\sigma}$ $\vee$ $V-r^+>0.5$) \& $(r^+-i^+<1.0)$ & 79\\
IA709 & $i^+$ ($r^+$)& $4.69-4.95$& $(B>B_{3\sigma})$ \& ($V>V_{3\sigma})$ \& ($r^+>r^+_{3\sigma}$ $\vee$ $r^+-i^+>0.8$) \& $(i^+-z^+<1.0)$& 81\\
IA738 & $i^+$ ($r^+$)& $4.92-5.19$& $(B>B_{3\sigma})$ \& ($V>V_{3\sigma})$ \& ($r^+>r^+_{3\sigma}$ $\vee$ $r^+-i^+>0.5$) \& $(i^+-z^+<1.0)$& 79\\
IA767 & $i^+$ ($z^+$)& $5.17-5.47$& $(B>B_{3\sigma})$ \& ($V>V_{3\sigma})$ \& ($r^+>r^+_{3\sigma}$ $\vee$ $r^+-i^+>0.5$) \& $(i^+-z^+<1.0)$ & 33\\
IA827 & $i^+$ ($z^+$)& $5.64-5.92$& $(B>B_{3\sigma})$ \& ($V>V_{3\sigma})$ \& ($r^+>r^+_{3\sigma}$ $\vee$ $r^+-i^+>0.5$) \& $(i^+-z^+<1.0)$& 35\\  
\hline
NB392$^1$ &$u$ ($B$)&2.20-2.24& $(z - K) > (B - z)$ $\vee$ $z_{phot}=1.7-2.8$ $\vee$ $z_{spec}=2.20-2.24$ & 159\\
NB501$^2$ &$g^{+}$$^2$ &3.08-3.16& $(u>u_{3\sigma}$ $\vee$ $u-g^{+}>1)$ \& $(g^{+}-i^+<1.5)$ &45\\
NB711$^2$ &$i^+$ ($z^+$)&4.83-4.89& $(B>B_{2\sigma})$ \& $(V>V_{2\sigma})$ \& $[(r^+>r^+_{2\sigma} \vee (r^+<r^+_{2\sigma} \wedge r^+-i^+>1.0)]$&78\\
NB816$^2$ &$i^+$ ($z^+$)&5.65-5.75& $(B>B_{2\sigma})$ \& $(V>V_{2\sigma})$ \& $[(r^+>r^+_{2\sigma} \vee (r^+<r^+_{2\sigma} \wedge r^+-i^+>1.0)]$&192\\
\hline
\multicolumn{4}{l}{Full SC4K sample (This study, 12 medium-band + 4 narrow-band)}&{\bf Total 3908}\\
\end{tabular}
\end{center}
\end{table*} 

\subsection{Photometric and spectroscopic redshifts} \label{subsec:photo}

In order to test how robust our emission-line selection criteria are, we use a large compilation of photometric and spectroscopic redshifts \citep[e.g.][]{Lilly2007,Ilbert2009,Laigle2016}. We use the distribution of photometric redshifts to identify the likely rough rest-frame wavelength of each emission-line picked up by the medium-bands, and show the results in Figure \ref{fig:photoz}. We find evidence for the presence of a large population of H$\beta$+[O{\sc iii}]$_{5007}$ and H$\alpha$ emitters, although the sample is dominated by candidate LAEs. Excluding the dominating LAEs, the most common remaining sources are H$\alpha$ emitters, followed by [O{\sc iii}]+H$\beta$. [O{\sc ii}]$_{3727}$ emitters represent a less frequent population among all the candidate line-emitters, and we also find evidence for some 4000\,{\AA} and Lyman break sources making it to the sample of potential line emitters. The relative proportion of sources is not surprising, given the combination of volume and observed EW distributions of all these lines \citep[see e.g.][]{Sobral2014,Khostovan2016,Hayashi2017}.

While Figure \ref{fig:photoz} shows that our sample of high EW candidate line-emitters is dominated by LAE candidates, it also reveals that many other line-emitters are expected to be in the sample. This is confirmed by spectroscopic redshifts of the full sample and stresses the importance of excluding lower redshift emitters in order to produce relatively clean and complete samples of LAEs.

\subsection{Selection of LAEs at $\bf z\sim2.5-6$} \label{subsec:criteria_lya}

In order to isolate LAEs from lower redshift line-emitters (see Figure \ref{fig:photoz}), we apply two criteria. First, we identify the presence of a colour break blue-ward of the medium-band excess emission and no significant emission bluer of that (see Table \ref{tab:depth}). Secondly, we remove sources that have red colours (e.g. $B-r>0.5$ for $z\sim2.5$; $i-z>1.0$ for $z\sim5.5$); see Table \ref{tab:criteria}. The first step selects the Lyman break, while the second criterion removes sources likely to be stars or red galaxies with a strong Balmer break (at a rest-frame wavelength $\sim400$\,nm) that mimics the Lyman break \citep[see e.g.][]{Matthee2014,Matthee2017followup}. Narrow-band surveys for LAEs typically apply the same/similar standard criteria \citep[e.g.][]{Ouchi2008,Matthee2015,Bielby2016,Santos2016}, with the difference being how strict the criteria/flexible the cuts are and what bands are available to trace/identify the Lyman break. Some surveys conducted in the blue bands rely mostly on a high EW$_{0}$ cut \citep[e.g.][]{Ciardullo2014,Konno2016}, but as discussed in e.g. \cite{Sobral2017}, even in the bluest bands it is crucial to filter lower redshift contaminants out of the sample of line-emitters due to bright, high EW lines such as C{\sc iii}] and C{\sc iv} \citep[see][]{Stroe2017,Stroe2017b}, particularly in wide-field surveys. We note that our colour criteria to exclude very red sources only removes extreme red sources and is based on current spectroscopic samples that show that essentially no real LAE will be removed by our cuts. However, it is possible that a handful of even more extremely red LAEs \citep[which are interesting on themselves; see e.g.][]{Ono2010,Taniguchi2015_APJ,Matthee2016_CALY} may be rejected in this way.

We apply our LAE selection by taking full advantage of the deep available broad-band photometry (see Table \ref{tab:depth}), which covers the wavelengths of the Lyman break and the Lyman continuum for our entire redshift range (see Figure \ref{fig:filters}). Our colour selection criteria (see Table \ref{tab:criteria}) are defined such that a candidate LAE is required to either have no detection blue-ward of the medium-band (i.e. being a drop-out galaxy), or, if the continuum is bright enough, to have a strong colour break between the two broad bands adjacent to the Ly$\alpha$ break expected wavelength. By not applying too strict colour-criteria, we ensure that sources with Lyman-Werner radiation or Lyman continuum leakage are not removed from our sample, as long as they have a Lyman break. We note that such sources are typically AGN, with high spectroscopic completeness in currently available spectroscopic surveys in COSMOS \citep[e.g.][]{Lilly2007}.

The exact values for each criterion are determined empirically using the large compilation of spectroscopic and reliable photometric redshifts discussed in Section \ref{subsec:photo}, but we also perturb these in Section \ref{perturb_LAE_sel}. Our LAE selection criteria selects up to $\sim50$\,\% of line emitters as LAEs for the lower redshift slices ($z\sim2-3$) but only $\sim2-5$\% of line emitters as LAEs for the highest redshift slices ($z\sim5-6$). This is a consequence of the differences in luminosity depth in Ly$\alpha$, but even more so due to the volumes and redshifts of the other main emission lines such as H$\alpha$, [O{\sc iii}]+H$\beta$ and [O{\sc ii}] which become more prominent for redder filters. Our results show that even with a high EW cut, we expect that about 50\% of sources will not be Ly$\alpha$ in the bluest bands, while about 95-97\% of sources in the red bands will be lower redshift line emitters\footnote{Due to the Lyman break criteria, our survey (and all similar Ly$\alpha$ surveys) is strongly biased against galaxy-galaxy lensed LAEs, as any lower redshift galaxy lensing a distant LAE will be classed as a lower-redshift interloper and the lensing system rejected.} (see Figure \ref{fig:photoz}). After the LAE selection, we retrieve a total of 6,156 potential Ly$\alpha$ emitters out of the full 40,726 potential line emitters (15\%).

\subsection{Visual inspection of all LAE candidates}\label{vis_inspec}

In order to obtain a clean sample of LAE candidates, we visually inspect all the candidates for spurious detections in their corresponding medium-band. In practice, we remove i) fake sources due to diffraction patterns, ii) fake sources which are selected close to the borders of images where the local noise is higher, iii) sources that are clear artefacts and iv) sources which are real but that clearly have their fluxes boosted in the medium-bands due to bright halos or diffraction of nearby stars. This is the same approach taken in the large-area narrow-band surveys which we also explore, namely \cite{Santos2016} and \cite{Sobral2017}. From a total of 6,156 LAE candidates, we conservatively reject/exclude 2,703 sources, and end up with a sample of 3,453 LAEs. We note that due to very different local noise properties, artefacts and image quality/depth, some bands (e.g. IA574 and IA827) have very high spurious fractions of $\approx90$\,\% in the initial LAE candidate sample, while other bands such as IA427 and IA679 have lower spurious fractions of $\sim15-25$\%. It is worth noting that due to the strict selection criteria in terms of non-detection in the optical in many bands, along with the high excess observed in the medium-bands, we easily select every single spurious/artefact in the full COSMOS images/catalogue. We thus stress the importance of visually checking all sources for such wide area surveys or, at least, to visually check a representative sub-sample and apply statistical corrections. We take into account the removal of spurious sources when computing the total areas and volumes, but we note that these only remove up to $\approx0.03$\% of the total area and thus they are completely negligible.

\subsection{Spectroscopic completeness, contamination and the final sample of LAEs} \label{subsec:zspecs}

Figure \ref{fig:photoz} reveals that our sample of line-emitters is greatly dominated by LAEs, and we expect that our photometric selection will further remove contaminants. This can nonetheless be quantified/investigated by using a relatively large number of spectroscopic redshifts of i) the full set of line emitters and ii) our samples of LAEs. Ideally, a sample that is highly complete will show that essentially all spectroscopically confirmed LAEs in i) will be contained in sample ii), while a highly clean sample will see most of contaminants in i) not be selected for ii).

We compile a large sample of spectroscopic redshifts in the COSMOS field \citep[e.g.][]{Lilly2007,Shioya2009,LeFevre2015,Kriek2015,Cassata2015} to find that 132 sources within our sample of LAE candidates have a spectroscopic redshift. Out of the 132 sources, we confirm 112 as LAEs in the appropriate band. This suggests a contamination of about 15\%, well within the range of what is typically found for large area Ly$\alpha$ narrow-band surveys at similar redshifts \citep[e.g.][]{Santos2016,Sobral2017,Shibuya2017,Harikane2017}. We also investigate whether there is any significant dependence of this contamination rate on redshift, Ly$\alpha$ luminosity or EW$_0$. We find that within the Poissonian errors the contamination is found to be relatively constant and to be between 10-20\%, similar to those found for narrow-band surveys of LAEs \citep[e.g.][]{Ouchi2008,Bielby2016}. In Appendix \ref{HK_colours} we provide further evidence of low contamination in typical $H-K_{\rm s}$ colours of $z\sim3$ LAEs. There are only mild indications that the higher redshift and the highest luminosity samples may be slightly more contaminated \citep[similarly to what has been found/discussed in e.g.][]{Matthee2015,Matthee2017followup,Harikane2017}, but such trends require further spectroscopic follow-up of our sample (Santos et al. in prep.).

Reliable redshift identifications can also be obtained through the dual narrow-band technique \citep[e.g.][]{Sobral2012,Nakajima2012,MattheeBOOTES2017}, where multiple unique combinations of strong emission lines can be observed in specific combinations of narrow- or medium-band filters\footnote{Here we use line-emitters identified in NB711 (Perez et al. in prep), NB816 \citep{Santos2016}, NB921 \citep{Sobral2013,Matthee2015}, NBJ, NBH and NBK \citep{Sobral2013,Khostovan2015} to search for another line, in addition to Ly$\alpha$ detected in our MBs.}. Within the SC4K sample of LAEs, we have already identified 27 Ly$\alpha$-C{\sc iii}] emitters at $z=2.7-3.3$, one Ly$\alpha$-C{\sc iv} emitter at $z=4.3$ (an X-Ray AGN) and 22 Ly$\alpha$-[O{\sc iii}] emitters at $z=3.3$ (three of these [O{\sc iii}] emitters are also C{\sc iii}] emitters). One dual-emitter already had a spectroscopic redshift. Hence, we obtain 51 additional reliable redshifts confirming all these sources as LAEs.

We also note that some of the contaminants are not easy to isolate by using broad-band colours. For example, SC4K-IA767-43371, with a redshift of $z=5.441$, is selected as a LAE candidate in both IA767 and IA827. While this source is a confirmed LAE (in IA767), the emission line in IA827 is N{\sc v} (1240\,\AA). As such, we remove this source from being a IA827 LAE. There are further 19 LAE candidates which are lower redshift interlopers and thus are removed from the final sample, either due to archival redshifts or from follow-up with AF2/WHT (Santos et al. in prep). We find that the confirmed interlopers/contaminants have a diverse nature. At lower redshift most are C{\sc iii}] and C{\sc iv} \citep[][]{Sobral2017,Stroe2017}, while at higher redshift there is a mix of [O{\sc iii}]+H$\beta$ and [O{\sc ii}]. We stress that neither of these class of sources could easily be removed by adjusting our selection criteria and certainly not without compromising our completeness, which we currently estimate to be at the level of $\sim85-90$\,\%. After removing the 19 spectroscopically confirmed interlopers, our final sample of medium-band selected sources contains 3434 candidate LAEs.


\subsection{UV continuum properties of SC4K LAEs}

In the process of selecting LAEs we find sources which have no continuum counterpart in the COSMOS data. These are typically found in very deep narrow-band or IFU studies \citep[e.g.][]{Ouchi2008,Wisotzki2016,Oyarzun2017}, but here we also find them in shallower data. In our samples, $\approx10$\% of LAEs have no continuum detection around the narrow- or medium-band. These LAEs likely occupy the lower stellar mass range of our sample and may have higher escape fractions due to their very high EWs \citep[see e.g.][]{Verhamme2017,Sobral2017}. We note that due to the fixed broad-band depths, the fraction of candidate LAEs without rest-frame UV detections becomes larger with redshift, from just $\sim1-2$\% at $z\sim2.5$ to $\sim10$\,\% at $z\sim5$ and reaching 30\% for our highest redshift sources. For sources without a rest-frame UV detection, we assume that the continuum flux is an upper limit based on the measured rms$_{\rm BB}$ and derive lower limits for their EWs. We note that by stacking our LAEs in the rest-frame UV (F814W, {\it HST}), \cite{Paulino-Afonso2017} find they have a typical rest-frame UV luminosity of $M_{UV}\sim-20$, which ranges from $M_{UV}=-19.2\pm0.2$ for our lowest redshift sample (the deepest in Ly$\alpha$) to up to $M_{UV}\sim-21$ at higher redshift \citep[see][]{Paulino-Afonso2017}.

%
%
\begin{figure}
\centering
\includegraphics[width=8.8cm]{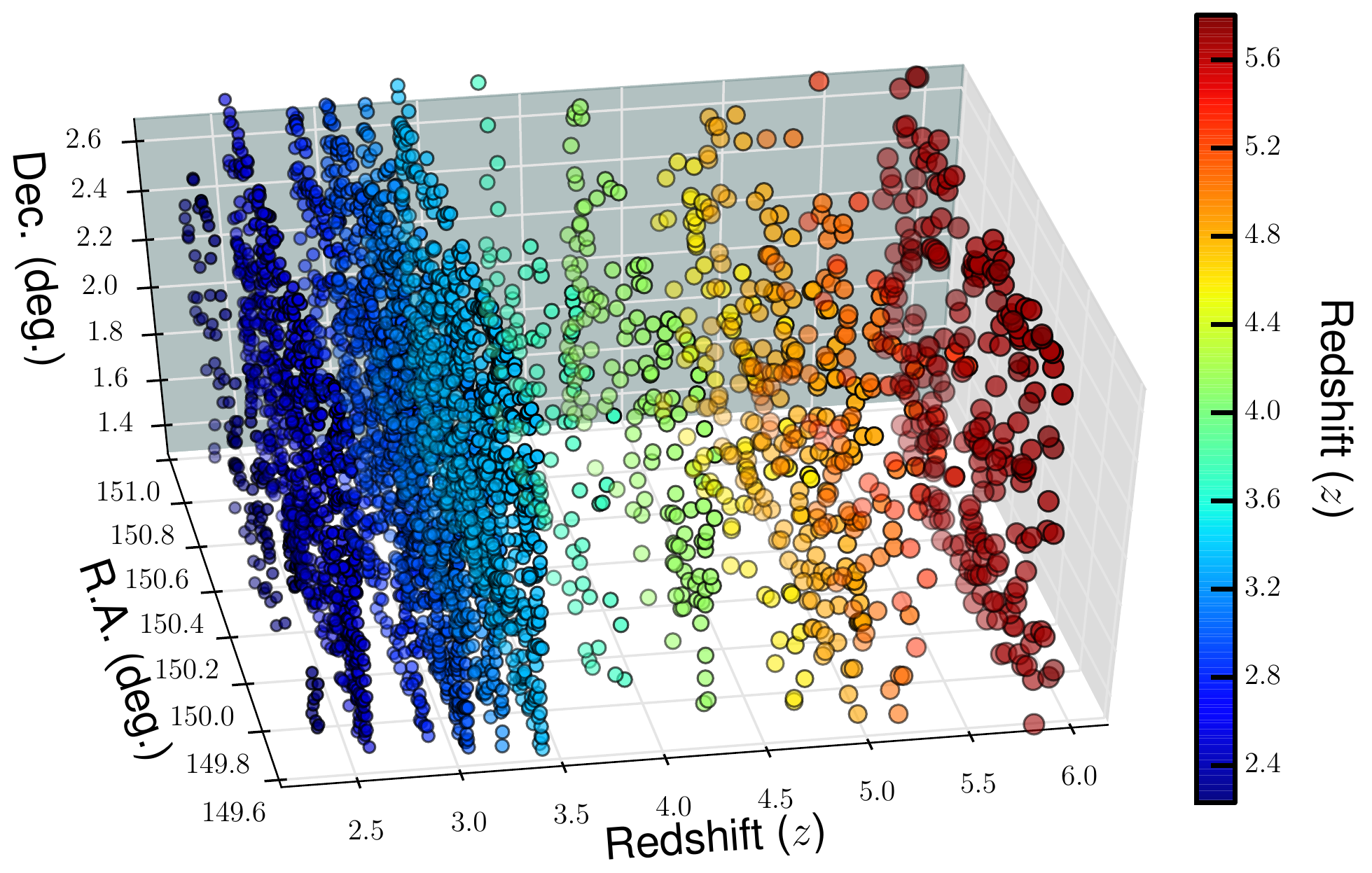}
\caption{The 3D distribution of the SC4K sample presented in this paper in the full 2\,deg$^2$ COSMOS field (see Table \ref{catalogue_appendix}), showing all LAEs from the 16 different redshift slices, colour coded by redshift (blue to red from lower to higher redshift). The redshift is computed using the central wavelength of the medium- or narrow-band filter. SC4K probes roughly 4,000 LAEs selected over a total volume close to $\sim10^{8}$\,Mpc$^{3}$ (see Table \ref{tab:volumes} for volumes probed per filter).}
\label{fig:3D}
\end{figure}

\subsection{Final sample: SC4K} \label{subsec:final_sample}

Our sample of medium-band selected LAEs consists of 3434 sources (see Table \ref{tab:criteria}), visually inspected for spurious detections. We complement our medium-band LAEs with four narrow-band studies (Table \ref{tab:criteria}) in the COSMOS field which follow the same methodology as in this paper. We add 159 LAEs at $z\sim2.23$ \citep[CALYMHA survey;][]{Sobral2017} and 45 sources at $z\sim3.1$ \citep[][]{MattheeBOOTES2017}, selected with narrow-bands NB392 and NB501. In addition, we also include 78 LAEs at $z\sim4.8$ (Perez et al. in prep) and 192 LAEs at $z\sim5.7$ \citep{Santos2016}, selected with the narrow-bands NB711 and NB816, respectively. Our final sample of LAEs contains 3908 sources. We name this sample of $\sim4,000$ (4k) LAEs, obtained by ``slicing'' the COSMOS field (Figure \ref{fig:3D}), as SC4K. For an example and description of the catalogue, see Table \ref{catalogue_appendix}. Our survey is roughly equivalent to a very wide, low resolution $(R\sim20-80)$ IFU Ly$\alpha$ survey covering all the way from $z\sim2.2$ to $z\sim6$. A 3D view (showing the full COSMOS field and redshift as a depth dimension) of SC4K is shown in Figure \ref{fig:3D}.

\section{Methods and corrections}\label{Methods}

%
%
\begin{table}
\caption{The Ly$\alpha$ survey co-moving volumes per redshift/filter slice assuming top-hat filter profiles for medium- and narrow-band filters. We provide the filter name and the Ly$\alpha$ volume corresponding to the 50\% transmission points in the normalised filter profile. The two final columns on the right present the limiting luminosity limit ($\log_{10}$\,L$_{\rm Ly\alpha}$/erg\,s$^{-1}$) for each slice, by using the formal flux limits from Table \ref{tab:depth_phot} and the 30\% completeness limit that we measure with out methodology (see Section \ref{sec:compcor}).}\label{tab:volumes}
\begin{center}
\begin{tabular}{ccccc}
\hline
Filter  & Area & Volume & L$_{\rm Ly\alpha,3\sigma}$   & L$_{\rm Ly\alpha}$ 30\%  \\
  & (deg$^2$) & (10$^6$\,Mpc$^3$) & (log$_{10}$) & (log$_{10}$) \\
\hline
IA427 &  1.94 &  4.0 &  42.4 & 42.5 \\
IA464 &  1.94 &  4.2 &  42.5 & 42.9 \\
IA484 &  1.94 &  4.3 &  42.5 & 42.7 \\
IA505 &  1.94 &  4.3 &  42.6 & 42.7 \\
IA527 &  1.94 &  4.5 &  42.5 & 42.7 \\
IA574 &  1.96 &  4.9 &  42.7 & 43.0 \\
IA624 &  1.96 &  5.2 &  42.8 & 42.9 \\
IA679 &  1.96 &  5.5 &  43.0 & 43.1 \\
IA709 &  1.96 &  5.1 &  42.9 & 43.1 \\
IA738 &  1.96 &  5.1 &  43.0 & 43.3 \\
IA767 &  1.96 &  5.5 &  43.0 & 43.4 \\
IA827 &  1.96 &  4.9 &  43.0 & 43.4 \\ 
\hline
NB392 &  1.21 &  0.6 &  42.3 & 42.3  \\  
NB501 &  0.85 &  0.9 &  42.9 & 43.0  \\ 
NB711 &  1.96 &  1.2 &  42.6 &  42.9   \\  
NB816 &  1.96 &  1.8 &  42.5 & 42.5   \\  
\hline  
Total &  1.96 &  61.5 &  42.4-43 & 42.5-43.4 \\
\hline
\end{tabular}
\end{center}
\end{table} 

\subsection{Ly$\alpha$ luminosities and survey volumes}

We compute Ly$\alpha$ luminosities for each of our LAE candidates per filter/redshift slice by using i) their estimated Ly$\alpha$ fluxes in $2''$ apertures ($F_{\rm Ly\alpha}$; see e.g. \citealt{Sobral2017}) and ii) the luminosity distance ($D_L$) corresponding to Ly$\alpha$ lines detected at the central wavelength of each filter. Luminosity distances (D$_L$) range from $20\times10^3$\,Mpc at $z\approx2.5$ to $55\times10^3$\,Mpc at $z\approx5.8$. Ly$\alpha$ luminosities are then calculated as L$_{\rm Ly\alpha}=4\pi F_{\rm Ly\alpha}D_{L}^2$. We find that our ``formal'' 3\,$\sigma$ limit MB detections correspond to Ly$\alpha$ luminosity limits ranging from 10$^{42.4}$\,erg\,s$^{-1}$ at $z=2.5$ to 10$^{43.0}$\,erg\,s$^{-1}$ at $z=5.8$ (see Table \ref{tab:volumes} for luminosity limits per filter).

\cite{Paulino-Afonso2017} measured the rest-frame UV sizes of our LAEs, concluding they have half-light-radii in the range $\approx0.1-0.2''$ ($\approx0.7-1.3$\,kpc), and thus significantly smaller than our $2''$ apertures. However, due to the use of ground-based imaging (with a larger PSF) and the fact that we are tracing Ly$\alpha$ and not the rest-frame UV, the $2''$ apertures may be missing some flux. We thus study how the fluxes computed in 2\,$''$ apertures compare with fluxes derived from using an estimate of the full flux using e.g. {\sc mag-auto}. We find an average ratio (Flux$_{\rm \sc [mag-auto]}$/Flux$_{\rm [2'']}$) of $\approx1.03\pm0.26$ (median of 1.02). There is no systematic difference in our sample as a whole nor any significant trend with redshift. Therefore, in this study we do not apply any further aperture correction and base our measurements on our directly measured $2''$ aperture quantities \citep[see discussion in][]{Drake2017b}.

We compute the co-moving volumes probed by each of the medium-bands by approximating them to top-hat filters (using the measured FWHM; Table \ref{tab:depth_phot}). We find co-moving volumes within $(4.0-5.5)\times10^6$\,Mpc$^3$ per medium-band and a total co-moving volume of $5.7\times10^7$\,Mpc$^3$ over all 12 medium-bands; see Table \ref{tab:volumes}. The sum of all narrow-band volumes contributes with a modest volume of $4.5\times10^6$\,Mpc$^3$, but allows to probe fainter Ly$\alpha$ luminosities (see Table \ref{tab:volumes}). The full Ly$\alpha$ survey volume in SC4K is therefore dominated by the medium-band filter survey and amounts to $6.2\times10^7$\,Mpc$^3$. We note that while our survey is only sensitive to the more typical and bright Ly$\alpha$ emitters, it provides a unique opportunity to explore the bright end of the Ly$\alpha$ luminosity function mostly for the first time, being fully complementary to other previous surveys. For example, we probe a volume $\approx$\,50,000 times larger than MUSE \citep[][]{Drake2017a} and $\approx50-60$ times the volumes of typical 1\,deg$^2$ narrow-band surveys \citep[][]{Ouchi2008} and still a factor of a $\sim2-3$ larger than current $\sim10-20$\,deg$^2$ surveys with Hyper-Suprimecam \citep[e.g.][]{Konno2017}.

\subsection{Corrections to the Ly$\alpha$ luminosity function} \label{corrections_errs_LF}

\subsubsection{Completeness correction} \label{sec:compcor}

Sources with weak emission lines or with low EWs may be missed by our selection criteria, causing the measured number density of sources to be underestimated. To estimate the line flux completeness we follow \cite{Sobral2013}, adapted for Ly$\alpha$ studies by \cite{Matthee2015}. Briefly, for each medium-band we obtain a sample of non-emitters at the redshift we intend to study from the appropriate MB catalogue. To do so, we use the sources which are not classified as line-emitters (we exclude the line-emitters) and, from these, we select sources which are consistent with being at a redshift $\pm0.2$ of the Ly$\alpha$ redshift for a given filter. We do this by i) applying the same Lyman break selection as we did for the sample of line-emitters and ii) by selecting sources with photometric redshifts within $\pm0.2$ \citep{Laigle2016} of the redshift window shown in Table \ref{tab:criteria}. We check that our method leads to a distribution in MB magnitudes of non-emitters in agreement with that of the LAEs, with a tail of $\sim2-5$\% brighter sources. Overall, our empirical approach leads to a sample of non-line emitters that is slightly brighter than that of LAEs, and thus can be seen as a conservative approach in estimating completeness corrections that does not require making any assumptions to create fake/mock sources.

Our procedure results in samples of non-line-emitters per MB filter that are at roughly the same redshift as our LAEs and allow us to estimate our line-flux completeness with an empirical/data approach. To do so, we add emission line flux to sources in steps of 10$^{-18}$\,erg\,s$^{-1}$\,cm$^{-2}$, which results in increasing the flux of the medium- and broad-bands depending on the filter's FWHM. For each step in flux added, we apply our emission-line selection criteria and identify those that, with the flux added, now make it into a new sample of line emitters and compare those with the total sample that was flux-boosted. By determining the fraction that we retrieve (after applying our $\Sigma$ and EW cuts; see Section \ref{subsec:criteria_line_emit}) as a function of added line-flux in comparison with the full sample, we obtain a completeness estimation for each flux, which we apply to our luminosity functions. We only calculate the Ly$\alpha$ luminosity function for luminosity bins in which we find a completeness of $30$\% or higher at the lowest luminosity limit of the bin; these are in the range L$_{\rm Ly\alpha}=10^{42.5-43.4}$\,erg\,s$^{-1}$ (see Table \ref{tab:volumes}). Our lowest luminosity bin is the one affected by the largest incompleteness and thus the one with the highest completeness correction being applied, which is typically a factor of $\approx2$. We find that the completeness functions strongly depend on line flux, with completeness typically growing from 30\% to 90\% in $\approx0.4-0.5$\,dex increases in Ly$\alpha$ luminosity, and reaching $\approx100$\% with a further $\sim0.5$\,dex increase.

%
%
\begin{figure}
\centering
\includegraphics[width=8cm]{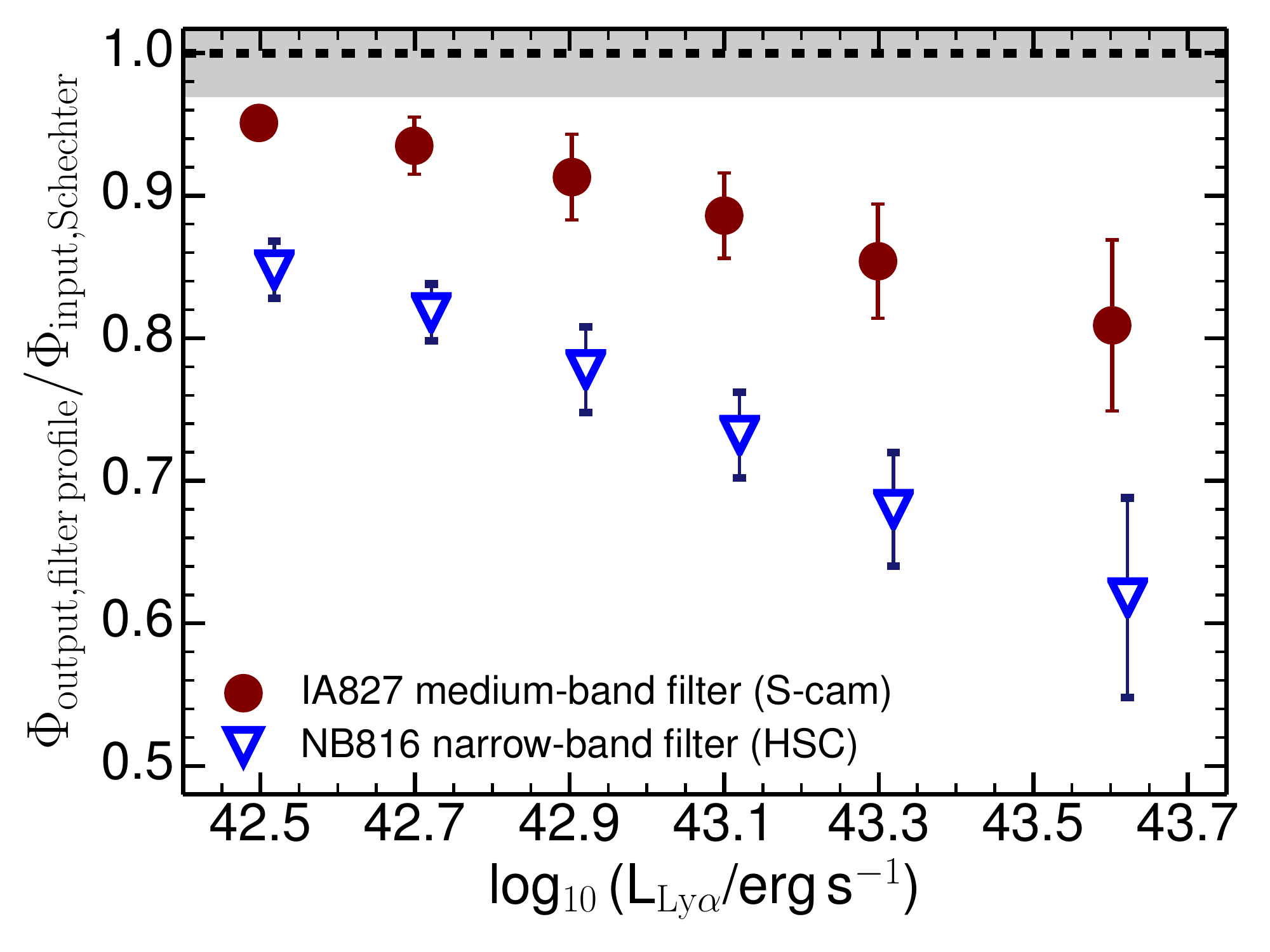}
\caption{The observed ratio between 10,000 observed Ly$\alpha$ luminosity functions using the real filter profiles and a Schechter input simulated sample of LAEs assuming tophat filters. Our results highlight the need to correct for filter profile effects which pushes sources from intrinsically bright to observed fainter bins, and highlights that the corrections are particularly important for narrow-band surveys, but are still relevant for medium-band surveys.}
\label{fig:filter_cor_816_827}
\end{figure}

\subsubsection{Filter profile effects and corrections} \label{sec:filtercor}

As discussed in detail in e.g. \citet{Sobral2013} and \citet{Matthee2015}, due to the non-top-hat shape of narrow-band filters, sources can be observed at a low transmission (almost no source is observed at full transmission when a filter is well described by a Gaussian function), particularly once survey volumes are large. As a result, assuming a top-hat filter will cause a complex underestimation of the flux, which is manifested in the luminosity function as a transfer of intrinsically bright sources towards observed fainter sources. For an intrinsic Schechter distribution, and particularly for the exponential regime (bright end), this effect results in an underestimation of the number density of the brightest emitters (as they can only be detected as bright over a small redshift range corresponding to the filter's peak transmission), and sometimes an overestimation of the faintest sources (as brighter sources detected away from peak transmission will look fainter). However, the necessary corrections depend on i) the filter profile, ii) the intrinsic shape of the luminosity function and iii) the depth and survey volumes.

While medium-bands are broader than narrow-bands and in general better fitted by a top-hat, a full investigation of the role of the filter profiles is still required. We estimate potential corrections for each filter by simulating ten million sources with an input random redshift distribution\footnote{Note that the output distribution is not random and follows closely the filter profile; this is what is used to study the effect of the filter profile.} which is wide enough to cover down to zero transmission by each filter on the blue and red wings. We generate these ten million sources with a luminosity distribution given by the observed (completeness corrected) luminosity function, following \cite{Sobral2013}. By convolving the full population with i) the real filter profile and ii) the top-hat approximation we can determine the number density ratio between i) and ii) per luminosity and derive corrections based on the filter profile; an example for IA827 and the NB816 (from HSC) filters is shown in Figure \ref{fig:filter_cor_816_827} (see also Figure \ref{fig:filter_cor_all}).

Our results show that the use of medium-band filters results in smaller corrections (see also Appendix \ref{sec:apdx_filtercor}) than those derived for typical narrow-band filters (Figure \ref{fig:filter_cor_816_827}). This is because fluxes are only significantly underestimated at the wings of the medium-band filters, which correspond to a much smaller fractional volume than for narrow-band filters. We also note that the input shape of the luminosity function is crucial for the estimated filter profile effect: while an observed Schechter function leads to a large correction in the exponential part, a bright end which is observationally described by a much slower decline with luminosity (e.g. a power-law with a shallow slope) results in smaller corrections (see full discussion in Appendix \ref{sec:apdx_filtercor}). Our results thus mean that while for previous deep surveys mostly tracing the faint-end power-law component of the Schechter function the corrections could be relatively small, for the bright end (under a Schechter assumption) the corrections can be large, close to a factor of 2-3 for narrow-band filters at the highest luminosities, while they can be a factor of 1.2-1.3 for medium-bands (see Figure \ref{fig:filter_cor_all}).

While the filter profile effects can be small for medium-bands, we still take them into account by applying a statistical correction to each luminosity bin. This produces our final luminosity function (LF). We provide a more detailed analysis and discussion of the effects, assumptions and corrections due to the various filter profiles when contrasted to top-hat approximations in Appendix \ref{sec:apdx_filtercor}. We also note that indirect statistical tests for our corrections can be obtained when comparing our results with e.g. MUSE \citep[][]{Drake2017b} and other IFU surveys which are not affected by filter profile effects (see Section \ref{global_LF_Sec}).

\subsection{Flux robustness and errors: random and systematic}

\begin{figure}
\centering
\includegraphics[width=8.cm]{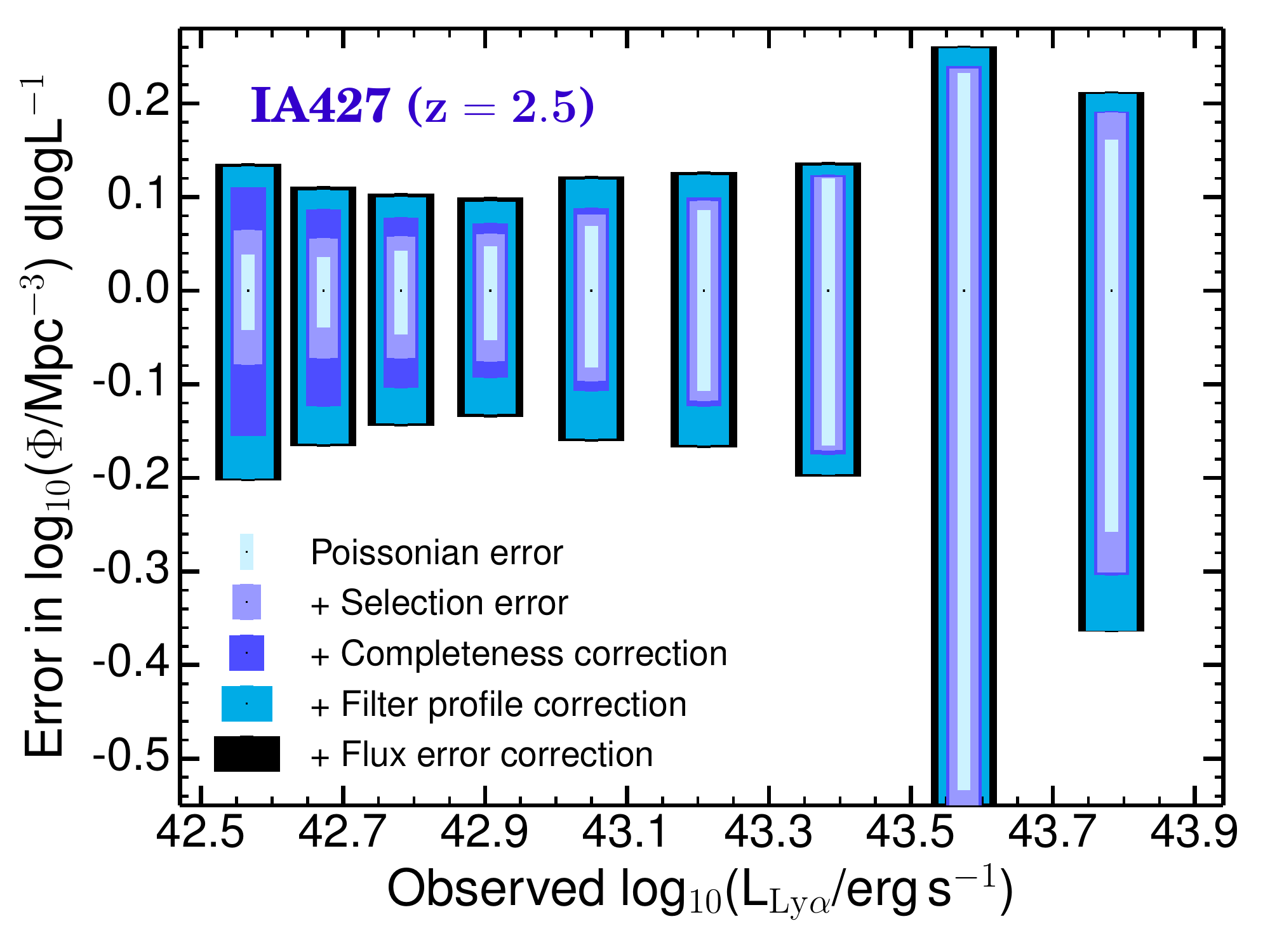}
\caption{An example of the different error contributions to the Ly$\alpha$ luminosity bins from different sources of uncertainty which are taken in series in our analysis. We show bins with bin widths that increase with increasing Ly$\alpha$ luminosity. These include: Poissonian, perturbations to the selection criteria (from line emitters to LAEs; see Table \ref{tab:criteria}), line flux completeness, filter profile corrections and flux errors. We find that selection perturbation errors are most important at the faintest luminosities, but they still contribute to the brightest bins.}
\label{fig:perturbation_errors}
\end{figure}

Due to errors in the photometry, both in the MB and BB magnitudes, estimated Ly$\alpha$ fluxes will be subject to errors and, in some cases, also prone to potential systematic effects. We study these errors and their potential role in the derivation of the Ly$\alpha$ LF. Briefly, we assume that each MB and BB magnitudes and their uncertainties are described by normal distributions centred on the measured value and with the width that is equal to its associated 1\,$\sigma$ error. We then perturb each galaxy magnitude 10,000 times by randomly picking values from their individual probability distributions. We use these perturbed magnitudes to compute the Ly$\alpha$ LF. We do not find any systematic difference, showing that the errors on the MB and BB photometry have no systematic effect in our methodology. We use the difference between the median value and the 16th and 84th percentiles of the perturbed number density distribution as the lower and upper errors on the number density for each luminosity bin. We find small variations due to this effect, with a median error of $\approx0.03$ error in $\log_{10}(\Phi)$. This is particularly sub-dominant when compared to other sources of uncertainty, but we still add it (0.03\,dex) at the end in quadrature (see Figure \ref{fig:perturbation_errors}).

%
%

\subsection{Completeness-contamination errors in the LAE selection and final errors}\label{perturb_LAE_sel}

While the flux and EW selection/limits can be taken into account for corrections and accounted for in errors (see Section \ref{sec:compcor}), there are other sources of uncertainties that are linked with the photometric or colour-colour criteria applied to select LAEs/filter lower redshift sources (Section \ref{subsec:criteria_lya}). While no single cut is perfect (even more so due to photometric errors), it is possible to perturb the selection and conduct a Markov chain Monte Carlo (MCMC) analysis in order to estimate the effects of varying the selection in the derivation of the Ly$\alpha$ LF and propagated quantities. Here we implement such an analysis. We perturb the LAE selection criteria described in Table \ref{tab:criteria} in a i) $\pm$0.2 dex interval around each colour-colour and photometric cut, independently and ii) by randomly varying by $^{+0.31}_{-0.44}$ the 3\,$\sigma$ magnitudes (corresponding to varying non-detection limits in the range $2-4$\,$\sigma$, from the least to the most conservative cuts) of bands tracing bluer of the Lyman-limit, used to reject interlopers. We run a MCMC simulation, with 10,000 iterations for each filter, randomly picking sets of values inside the full explored range, assuming all have an equal probability (flat prior). We then calculate the selection criteria errors as the difference between the median value and the 16th and 84th percentiles within each luminosity bin for all realisations.

An example of the estimation of the full errors affecting log$_{10}$($\Phi$) can be found in Figure \ref{fig:perturbation_errors} for $z=2.5$ (IA427). We find that the perturbations result in standard deviations of 0.03 to 0.1\,dex per luminosity bin at $z\sim2.5$, representing up to $50\%$ of the total error. The perturbation error is larger in absolute terms at the brightest bins, but it becomes a much more significant fractional contribution to the faintest bins where the Poissonian errors are very small (see Figure \ref{fig:perturbation_errors}). The errors from perturbing the selection criteria are roughly a factor of up to 2.5 the Poissonian error per bin at the bins probing the faintest luminosities (with the largest number of sources), while they can be as low as 0.2-0.8 of the Poissonian errors if the bin is populated with only 5-10 sources (where the Poissonian error is already large). We also find that perturbation errors are more important (larger) at $z\sim2.5-2.8$ and $z>4$ than they are at $z\sim3-3.3$. This roughly coincides with jumps in the selection criteria and whenever different colours/bands are used (see Table \ref{tab:criteria}).

We add our estimated perturbation errors in quadrature to the Poissonian errors, noting that they are particularly important for the faint end where the Poissonian errors are an underestimation of the full uncertainties. We then scale the errors by the line flux completeness correction and the filter profile correction, which we assume we know with 30\% accuracy (and thus add an extra 30\% of such corrections to the errors, taking a conservative approach). We note that we do not add any errors due to cosmic variance, but that given the very large volumes and the multiple redshift slices, we expect these to be just a small fraction of our full errors that are much larger than the formal Poissonian errors. Finally, even though our samples are expected to be contaminated by interlopers at the 10-15\% level, similarly to other narrow-band surveys, our LAE selection-completeness implies we may be missing 10-15\% of real LAEs (when we transform the sample of line emitters into candidate LAEs), and thus in our analysis we do not apply any corrections for this contamination or completeness, as they should roughly cancel out.

\subsection{Redshift binning} \label{sec:binning}

Our multiple redshift slices allow to trace LAEs across well defined cosmic times from $z\sim2$ to $z\sim6$. We can also combine the slices to produce a global Ly$\alpha$ LF or obtain slightly broader redshift slices which are populated by a much larger number of sources, and that overcome even more cosmic variance. We bin all our $z\sim2.5-6$ slices (IA427 through to IA827) in order to produce a global high redshift LF and compare it with similar measurement made with the MUSE instrument \citep[e.g][]{Drake2017b} or with slit observations \citep[e.g.][]{Cassata2011}. We also split the sample into 5 different redshift bins in the following way:
\begin{itemize}
\item $z\sim2.2$ ($z=2.22\pm0.02$; NB392)
\item $z\sim2.5$ ($z=2.5\pm0.1$; IA427)
\item $z\sim3.1$ ($z=3.1\pm0.4$; IA464, IA484, IA505, IA527)
\item $z\sim3.9$ ($z=3.9\pm0.3$; IA574, IA624)
\item $z\sim4.7$ ($z=4.7\pm0.2$; IA679, IA709)
\item $z\sim5.4$ ($z=5.4\pm0.5$; IA738, IA767, IA827)
\end{itemize}

When producing redshift binned LFs, we only use the volumes associated with a given medium-band if that specific filter provides the necessary depth for a completeness above 30\%.

\subsection{Schechter, power-law and combined fits}

The Schechter function \citep{Schechter1976} is a widely used parametrization of the LF, defined by three parameters: the power-law slope $\alpha$, the characteristic number density $\Phi^\star$ and the characteristic luminosity L$^\star$. Observations up to extremely low luminosities are necessary to accurately constrain the power-law slope $\alpha$ \citep[e.g][]{Dressler2015,Drake2017b}. Our medium-bands cover ``typical" luminosities and higher, thus not probing much fainter than L$^{\star}$, and do not allow to measure $\alpha$ on their own. However, several studies have been able to obtain good constraints on $\alpha$ from $z\sim2$ to $z\sim6$ \citep[e.g.][]{Dressler2015,Drake2017b, Santos2016,Konno2016}, which has been shown to be very steep ($<-1.5$) and potentially varying from $\alpha\approx-1.7$ at $z=2.2$ \citep{Konno2016,Sobral2017} to $\alpha\approx-2$ (or even steeper; see \citealt{Drake2017b}) by $z\sim6$ \citep{Dressler2015,Santos2016,Drake2017b}. We therefore fit Schechter functions by fixing/varying $\alpha$ between $-1.6$ and $-2.0$, but we also explore fits with $\alpha$ fixed to $-1.8$ at all redshifts in order to investigate the potential redshift evolution of $\rm L^{\star}_{Ly\alpha}$ and $\rm \Phi^*_{Ly\alpha}$ at fixed $\alpha$. Finally, we also fit $\alpha$ explicitly by combining our results with ultra-deep observations.

In addition to fitting Schechter functions, we also fit power-laws of the form log$_{10}\Phi = A\,\log_{10}{\rm L_{Ly\alpha}} + B$ to the full LFs and compare these with Schechter fits. Finally, we also explore combinations of a Schechter for lower luminosities and a power-law at higher luminosities when a single function yields a very bad fit on its own (see Section \ref{sec:results}). For all LFs, we follow a MCMC approach for the fits, perturbing each bin independently within its asymmetric Gaussian error probability distribution and re-fitting for 10,000 realisations per LF. We take the median of all the best fits as the most likely combination of parameters and estimate the errors by computing the 16th and 84th percentiles for all 10,000 realisations per LF estimation. We note that due to degeneracies in the parameters, these errors can sometimes exaggerate the errors on individual parameters (i.e., parameters are linked and only allowed to vary within some specific relation and not independently), but they are generally well constrained. For each best-fit we also compute the corresponding $\rm \chi^2_{red}$ and use these to obtain the median $\rm \chi^2_{red}$ and the 16th and 84th percentiles of all realisations.

%
%
\begin{figure*}
\centering
\includegraphics[width=13.8cm]{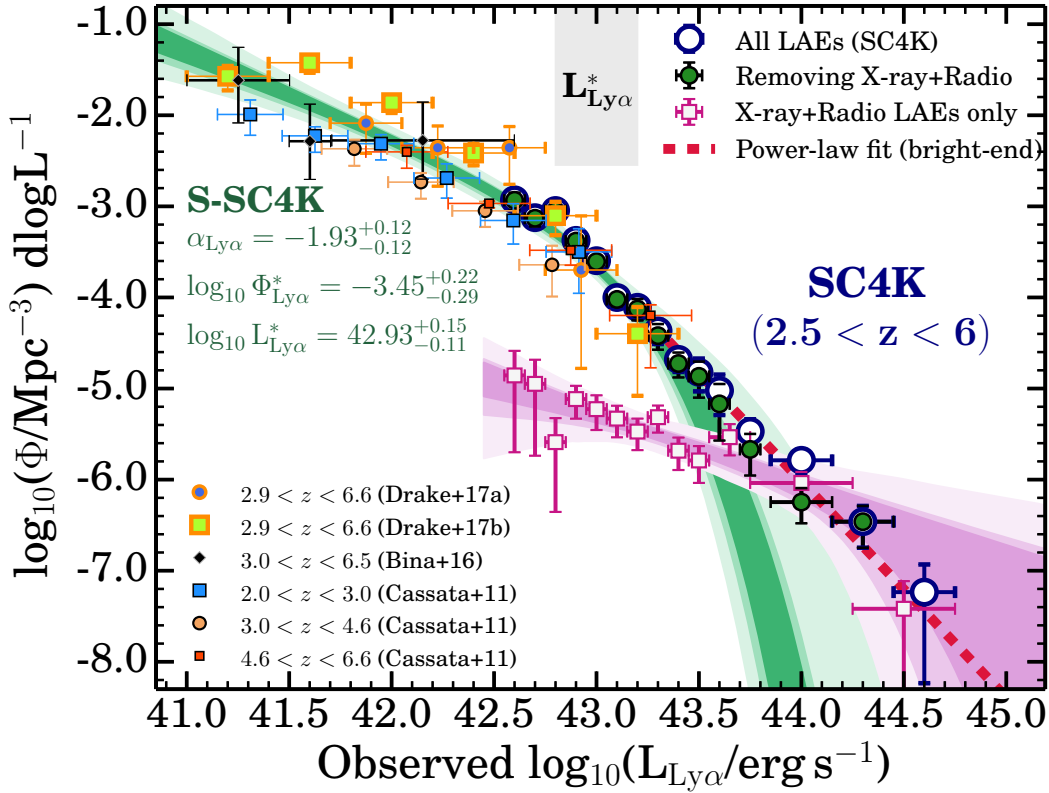}
\caption{The combined global Ly$\alpha$ LF at $2.5<z<6$. Our large samples of luminous LAEs, obtained over a co-moving volume of $0.6\times10^{8}$\,Mpc$^{3}$ are able to constrain the bright end of the Ly$\alpha$ LF for the first time down to number densities of $\sim10^{-7}$\,Mpc$^{-3}$ and Ly$\alpha$ luminosities of $\sim10^{44.5}$\,erg\,s$^{-1}$. We find a significant excess of bright LAEs at the highest luminosities when compared to a single Shechter function and show that is likely driven by a population where Ly$\alpha$ is AGN-driven. We compute a proxy for the AGN Ly$\alpha$ LF with X-ray and radio AGN among our sample (1 and 2\,$\sigma$ contours shown for Schechter function fits). We also compare our results with recent MUSE/VLT \citep{Bina2016,Drake2017a,Drake2017b} and VIMOS/VLT observations \citep[e.g.][]{Cassata2011}, showing a very good agreement in the L$^*_{\rm Ly\alpha}$ range where all studies overlap. Deeper, smaller volume studies from the literature allow to cover the sub-L$^\star_{\rm Ly\alpha}$ luminosity regime, being perfectly complementary to our approach. Overall, we show the Ly$\alpha$ LF determined over 4 orders of magnitude in Ly$\alpha$ luminosity and 6 orders of magnitude in number densities at $z\sim2.5-6$ resulting in the `synergy' Ly$\alpha$ LF (S-SC4K; $2.5<z<6$) and the 1, 2 and 3\,$\sigma$ confidence levels when fitting a Schechter function up to $10^{43.3}$\,erg\,s$^{-1}$ (we also show the power-law fit done for higher luminosities). Results are provided in Table \ref{LF_fits_global}.}
\label{Global_LF}
\end{figure*}

\subsection{X-ray and radio properties: AGN candidates within our LAEs} \label{sec:nature_lya}

We explore {\it Chandra} X-ray \citep[e.g.][]{Civano2016} and VLA radio data \citep[e.g.][]{Smolcic2017} within COSMOS to identify AGN in our sample. Full details are given in \cite{Calhau2018}. Briefly, we use the publicly available {\it Chandra} Cosmos Legacy survey \citep[][]{Elvis2009,Puccetti2009,Civano2016} to select sources with X-ray counterparts, within the overlap region with SC4K (1.86\,deg$^2$). Out of the full SC4K sample of 3908 LAEs presented in this paper, 3707 have {\it Chandra} X-ray coverage. From those, we identify 109 sources with X-ray emission in the \cite{Civano2016} catalogue, making them strong candidates of being X-ray AGN (L$_X>10^{42.5}$\,erg\,s$^{-1}$). \cite{Calhau2018} presents a detailed analysis on the X-ray activity of our full sample of LAEs. We find a global X-ray AGN fraction among our SC4K LAEs of $2.9\pm0.3$\% for $z=2-6$. The AGN fraction shows evidence for a decline with increasing redshift for typical to bright LAEs. At $z\sim2.2-2.7$ the X-ray AGN fraction is $3.9\pm0.6$\%, declining to $3.5\pm0.4$\% and $0.4\pm0.2$\% for redshifts $2.7<z<3.5$ and $3.5<z<6$, respectively. We also identify a clear relation between the X-ray AGN fraction and Ly$\alpha$ luminosity of LAEs \citep[see][]{Calhau2018}, qualitatively similar to what has been found at lower redshift \citep{Wold2014,Wold2017} and also found and discussed in \cite{Ouchi2008} and \cite{MattheeBOOTES2017}. 
 
In the VLA radio data \citep[1.4\,GHz and 3\,GHz; see][]{Schinnerer2010,Smolcic2017} we identify 62 individual sources, with these being dominated by the 3\,GHz detections (61), and we class these as AGN. Out of these, 30/62 are also X-ray sources. We therefore find a total of 141 AGN sources among the SC4K sample of LAEs, yielding a total AGN fraction of $3.6\pm0.3$\%. If we split the sample in three redshift ranges, we find that the AGN fraction slowly declines from $z\sim2.2-2.7$ ($4.7\pm0.7$\%) to $z\sim2.7-3.5$ ($4.4\pm0.4$\%) and then drops significantly at $z\sim3.5-6$ ($1.2\pm0.4$\%). Concentrating on radio AGN within our sample of LAEs, we find that the (radio) AGN fraction is relatively constant ($1.9\pm0.4$\%) at $z\sim2.2-3.5$ and then drops to $0.9\pm0.2$\% at $z\sim3.5-6$.

%
%
\begin{table*}
\caption{The global Ly$\alpha$ LF at $z\sim2.5-6$ from SC4K only, when combined with the latest MUSE results \citep[][]{Drake2017b} and when using the derived consensus global Ly$\alpha$ LF, S-SC4K \citep[SC4K and][]{Cassata2011,Bina2016,Drake2017a,Drake2017b}. We also show the results when explicitly removing radio and X-ray AGN from the sample (see Section \ref{sec:nature_lya}). The corresponding $\rho_{\rm Ly\alpha}$ have been computed by integrating Schechter functions down to $1.75\times10^{41}$\,erg\,s$^{-1}$, corresponding to 0.04\,$L^\star_{z=3}$ from \citet{Gronwall2007}; see Section \ref{sec:Lya_rho}. All errors are the 16th and 84th percentiles for all 10,000 realisations per LF estimation which, due to degeneracies in the parameters, can sometimes exaggerate the errors on individual parameters. We also provide a comparison (ratio) between reduced $\chi^2$ for Schechter and power-law fits ($\rm \chi^2_{Sch}/\chi^2_{PL}$) fitted over the same luminosity range for a fair comparison; values below 1 indicate that a Schechter fit performs significantly better, while a large value indicates that a simple power-law fit provides a relatively lower reduced $\chi^2$. * Note that fits to the full LF are given for completeness and comparison, but that they fail to fit the data as a whole, as the combined faint and bright ends are not accurately described by a single Schechter or power-law functions.}\label{LF_fits_global}
\begin{tabular}{@{}lcccccc@{}}
\hline
Global Ly$\alpha$ sample &  $\alpha$ &  $\rm \log_{10}\,L^*_{Ly\alpha}$  & $\log_{10}\,\Phi^*_{\rm Ly\alpha}$ &  $\rho_{\rm Ly\alpha}/10^{40}$ Sch & Power-law (PL)   & $\rm \chi^2_{Sch}/$ \\
 ($2.5<z<5.8$) &    &  (erg\,s$^{-1}$)  & (Mpc$^{-3}$)  &   (erg\,s$^{-1}$\,Mpc$^{-3}$) &  (A\,$\log_{10}\,$L+B)  &  $\rm \chi^2_{PL}$  \\
\hline 
SC4K ($\rm \log_{10}L_{Ly\alpha}<43.3$) & $-1.8\pm0.2$ (fix) &  $42.81^{+0.07}_{-0.06}$  & $-3.16^{+0.13}_{-0.14}$  &  $0.98^{+0.22}_{-0.17}$ & $-2.09^{+0.17}_{-0.17}$, $86.1^{+7.3}_{-7.1}$  & 0.6  \\
SC4K+MUSE ($\rm \log_{10}L_{Ly\alpha}<43.3$) & $-1.80^{+0.11}_{-0.11}$ &  $42.72^{+0.07}_{-0.06}$  & $-2.92^{+0.14}_{-0.16}$  &  $1.32^{+0.12}_{-0.12}$ & $-1.36^{+0.05}_{-0.05}$, $55.1^{+2.2}_{-2.4}$  & 0.4   \\
S-SC4K ($\rm \log_{10}L_{Ly\alpha}<43.3$) & $-1.93^{+0.12}_{-0.12}$ &  $42.93^{+0.15}_{-0.11}$  & $-3.45^{+0.22}_{-0.29}$  &  $0.88^{+0.09}_{-0.09}$ & $-1.29^{+0.06}_{-0.06}$, $52.0^{+2.6}_{-2.7}$  & 0.8   \\
\hline
 SC4K$^*$ (All LAEs) & $-1.8\pm0.2$ (fix) &  $43.59^{+0.06}_{-0.06}$  & $-4.53^{+0.13}_{-0.16}$  &  $0.33^{+0.07}_{-0.05}$ & $-2.22^{+0.08}_{-0.10}$, $91.7^{+4.1}_{-3.6}$  & 8.0   \\
SC4K+MUSE$^*$ (All LAEs) & $-2.55^{+0.06}_{-0.06}$ &  $43.92^{+0.12}_{-0.11}$  & $-5.47^{+0.24}_{-0.26}$  &  $1.40^{+0.17}_{-0.15}$ & $-1.78^{+0.04}_{-0.05}$, $72.7^{+2.0}_{-1.9}$  & 0.7   \\
S-SC4K$^*$ (All LAEs) & $-2.45^{+0.06}_{-0.06}$ &  $43.87^{+0.10}_{-0.10}$  & $-5.32^{+0.21}_{-0.23}$  &  $1.04^{+0.12}_{-0.12}$ & $-1.69^{+0.05}_{-0.05}$, $68.6^{+2.0}_{-2.0}$  & 0.7   \\
\hline
X-ray + radio AGN only & $-1.7^{+0.3}_{-0.2}$ &  $51.3^{+1.2}_{-7.3}$  & $-11.0^{+5.0}_{-2.6}$  &  $0.027^{+0.013}_{-0.013}$ & $-0.75^{+0.17}_{-0.17}$, $27.1^{+7.3}_{-7.2}$  & 1.3   \\
\hline
 SC4K$^*$ (w/o X-ray+radio) & $-1.8\pm0.2$ (fix) &  $43.56^{+0.06}_{-0.05}$  & $-4.56^{+0.12}_{-0.14}$  &  $0.29^{+0.06}_{-0.05}$ & $-2.38^{+0.09}_{-0.10}$, $98.7^{+4.4}_{-4.1}$  & 8.2   \\
SC4K+MUSE$^*$ (w/o X-ray+radio) & $-2.63^{+0.06}_{-0.06}$ &  $43.90^{+0.12}_{-0.10}$  & $-5.59^{+0.25}_{-0.28}$  &  $1.48^{+0.18}_{-0.17}$ & $-1.86^{+0.05}_{-0.05}$, $76.2^{+2.2}_{-2.1}$  & 0.7   \\
S-SC4K$^*$ (w/o X-ray+radio) & $-2.52^{+0.07}_{-0.07}$ &  $43.84^{+0.11}_{-0.09}$  & $-5.40^{+0.21}_{-0.25}$  &  $1.09^{+0.14}_{-0.13}$ & $-1.77^{+0.05}_{-0.05}$, $72.0^{+2.2}_{-2.1}$  & 0.7   \\
\hline
\end{tabular}
\end{table*}

\section{Results} \label{sec:results}

\subsection{The global Ly$\alpha$ LF at $\bf z\sim2.5-6$}\label{global_LF_Sec}

In Figure \ref{Global_LF} we present the global Ly$\alpha$ LF at $z\sim2.5-6$, determined with a total volume of close to $\sim10^8$\,Mpc$^{3}$. Our results probe Ly$\alpha$ luminosities from $\sim10^{42.5}$\,erg\,s$^{-1}$ to $\sim10^{44.5}$\,erg\,s$^{-1}$, covering 2 orders of magnitude in Ly$\alpha$ luminosity with a single survey. Down to our observational limits, we find that the global Ly$\alpha$ LF at $z\sim2.5-6$ resembles a single or double power-law (or a double Schechter, but not a single Schechter function) and extends to luminosities and number densities that reach into what is expected from the quasar luminosity function \citep[e.g.][]{Richards2006} and follow-up of quasars in Ly$\alpha$ \citep[e.g.][]{Borisova2016}. Fitting the global SC4K Ly$\alpha$ LF leads to a power-law of $\rm -2.22^{+0.08}_{-0.10}\,\log_{10}(L_{Ly\alpha})+91.7^{+4.1}_{-3.6}$ (see Table \ref{LF_fits_global}), which describes the data significantly better than a single Schechter function ($\rm \chi^2_{Sch}/\chi^2_{PL}\approx8$; see Table \ref{LF_fits_global}). If we exclude X-ray AGN and radio AGN, we find that the global Ly$\alpha$ LF becomes steeper at the bright end. We can also derive a X-ray+radio AGN Ly$\alpha$ LF, which we also present in Figure \ref{Global_LF}, together with the range of Schechter fits encompassing 1 and 2$\sigma$ ranges of all realisations. We find evidence for the AGN population being responsible for the `bump' at high Ly$\alpha$ luminosities, which can be parameterised by a Schechter function\footnote{It can also be relatively well parameterised by a simpler power-law function, see Table \ref{LF_fits_global}.} with a higher characteristic luminosity, a lower characteristic number density and potentially a shallower slope than the global population (see Table \ref{LF_fits_global}). The full cross-over between the likely two populations happens at $\approx10^{43.5}$\,erg\,s$^{-1}$, although given that X-ray and radio only provide a partial view of all the AGN, this transition may happen at slightly lower luminosities $\approx10^{43.2-43.3}$\,erg\,s$^{-1}$ \citep[see e.g.][]{Sobral2018}. It is worth noting that the typical characteristic number density of the AGN component of the Ly$\alpha$ LF is close to $\approx10^{-6}$\,Mpc$^{-3}$, similar to the number densities of clusters in the Universe, and that may provide a natural link between bright LAEs at $z>2.5$ (typically seen as very extended and thus called Ly$\alpha$ `blobs') and the physical environments they inhabit (potential `proto-clusters').

In Figure \ref{Global_LF} we also show results obtained with much deeper surveys, including MUSE \citep[][]{Bina2016,Drake2017a,Drake2017b} and results from slit spectroscopy using VIMOS/VLT \citep{Cassata2011}; see also Table \ref{SSC4K_refs_IDs}. We find excellent agreement within the error bars with the MUSE results presented by \cite{Drake2017a,Drake2017b}, although we note that the agreement is only possible to be tested around L$_{\rm Ly\alpha}^*$, where all studies overlap. Future results from the MUSE-wide project \citep[see][]{Herenz2017,Caruana2018}, or a compilation of extra-galactic MUSE archival observations, may be able to extend the volume covered by MUSE and further increase the overlap, allowing for more detailed comparisons and to evaluate any systematics/differences. Extremely deep MUSE data allow to not only blindly find faint LAEs, but even more importantly to measure the full Ly$\alpha$ luminosity of each source without effects from narrow-band filter profiles \citep[see][]{Drake2017b,Leclercq2017}. The comparison thus provides statistical evidence that our corrections are able to recover the full Ly$\alpha$ LF.

Comparing our results with \cite{Cassata2011}, we find a very good agreement with their $z\sim2-3$ and $z\sim4.6-6.6$ Ly$\alpha$ LFs. The $z\sim3.0-4.6$ LF from \cite{Cassata2011} is slightly below ours in the small luminosity range overlap (we can only use one of their bins to directly compare with ours), but we note that their results are also below those from MUSE \citep[see][]{Drake2017a,Drake2017b}. Apart from cosmic variance and the large differences in selection (our selection is directly on Ly$\alpha$, more similar to MUSE), there could also be some cosmic evolution. Furthermore, we note that the use of slits and potential underestimation of slit corrections may further explain the differences. Both narrow-band surveys and MUSE have established that Ly$\alpha$ emission is significantly extended \citep[e.g.][]{Momose2014,Wisotzki2016,Sobral2017,Leclercq2017}, thus making slit spectroscopy hard to correct. Slit corrections can be particularly challenging as they are often based on the UV continuum, but there is no simple relation between the Ly$\alpha$ extent and the UV extent \cite[see e.g.][]{Leclercq2017}.

Ly$\alpha$ surveys from deeper (necessarily smaller) volumes are needed to cover the sub-L$^\star_{\rm Ly\alpha}$ luminosity regime \citep[][]{Bina2016,Drake2017a,Drake2017b}, as highlighted in Figure \ref{Global_LF}. Overall, we can now determine the Ly$\alpha$ LF over 4 orders of magnitude in Ly$\alpha$ luminosity at $z\sim2.5-6$. Figure \ref{Global_LF} also reveals how complementary ultra-deep MUSE and slit observations are to very wide narrow- and medium-band surveys \citep[e.g. SC4K and][]{Konno2017}, allowing unique synergies and providing the first combined view all the way from the faintest Ly$\alpha$ sources to the brightest. We fully explore the combined strength of deep surveys\footnote{In order to account for potential systematic differences between surveys, cosmic variance and due to the way we compute errors, we add errors of $^{+0.05}_{-0.08}$ to data bins determined with deeper observations/by other studies, as they are able to explain current differences between surveys and methods. We note, nonetheless, that these errors are very uncertain on themselves and depend on which surveys/methods are being compared.} (to probe the faint end) and SC4K (to probe the bright end) and derive a combined Ly$\alpha$ LF (S-SC4K; see Section \ref{consensus}) presented in Figure \ref{Global_LF} and Table \ref{LF_fits_global}. We obtain two cases: when combining SC4K with the latest MUSE results \citep[][]{Drake2017b} and when combining all studies with SC4K \citep{Bina2016,Drake2017a,Drake2017b,Cassata2011}. While we note that a single Schechter function is simply not an appropriate fit to the full LF, we still provide those for completeness and for comparison of parameters. Restricting the fit to L$_{\rm Ly\alpha}<10^{43.3}$\,erg\,s$^{-1}$ allows to fit a single Schechter which is likely tracing an overall population of SF-dominated galaxies, showing a steep slope, $\alpha=-1.93^{+0.12}_{-0.12}$ (greatly improved when using MUSE only see \citealt{Drake2017b}), with $\rm L^*_{\rm Ly\alpha}=10^{42.93^{+0.15}_{-0.11}}$\,erg\,s$^{-1}$ and $\rm \Phi^*_{\rm Ly\alpha}=10^{-3.45^{+0.22}_{-0.29}}$\,Mpc$^{-3}$ (see Table \ref{LF_fits_global}).

%
%
\begin{figure*}
\centering
\includegraphics[width=17.1cm]{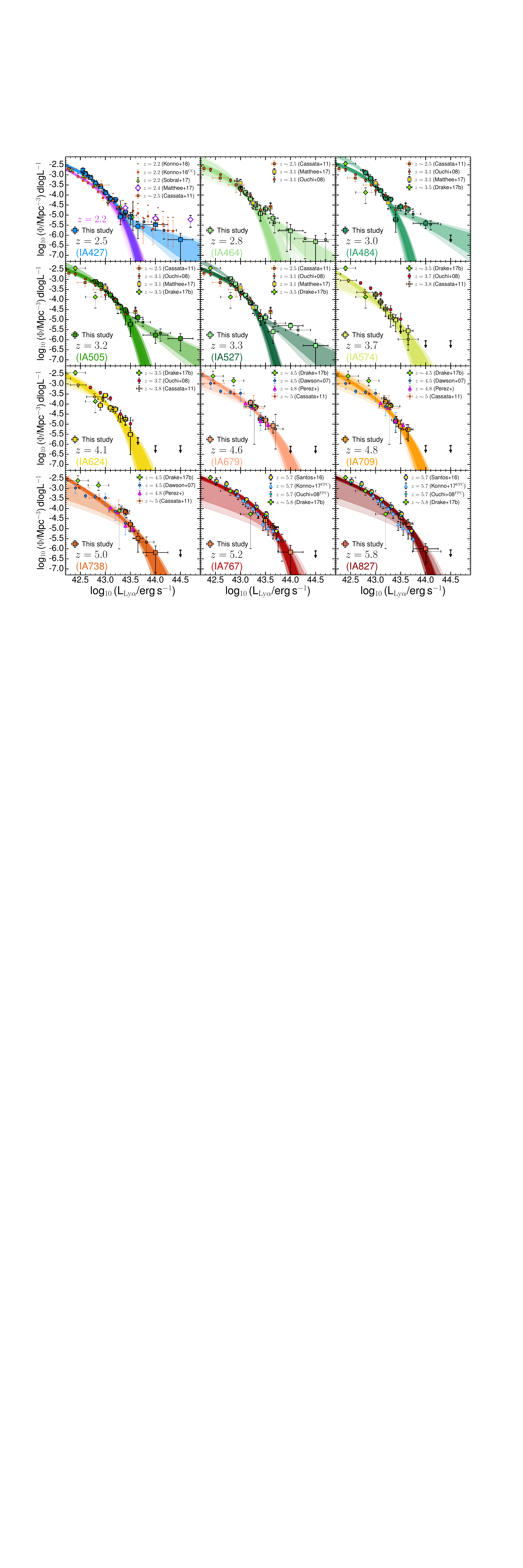}
\caption{The evolution of the (bright end of the) Ly$\alpha$ LF from $z\sim2.2$ to $z\sim6$ in 12 (+1) redshift slices and comparison with a variety of surveys at roughly the same redshift as each slice. Our results reveal a significant evolution at the bright end, with the number counts falling down as a steepening potential power-law at $z\sim2.2-3.3$ which can be described as a single Schechter function at $z>3.5$. We show two bin realisations for visualisation of binning effects, but also the much more representative range of Schechter and power-law (percentiles, corresponding to 1, 2$\sigma$) fits from perturbing the data. In addition, we also show the fits and uncertainties when exploring synergies with deeper surveys (S-SC4K), which greatly reduces the uncertainties (darker contours). Note that we show both the original \citet{Konno2016} $z=2.2$ LF in small points, and after correcting for potential contamination \citep[see][]{Sobral2017}. At $z\sim5.8$ we compare our measurement with NB surveys \citep[e.g.][]{Ouchi2008,Konno2017} corrected for filter profile effects as in \citet{Santos2016}.}
\label{fig:lfindividual}
\end{figure*}

Due to the different Ly$\alpha$ luminosity limits, our global LF presented in Figure \ref{Global_LF} is inevitably dominated by sources at different redshifts as a function of luminosity, with the lower luminosity bins being dominated by the (deeper) lower redshift data, while at higher luminosities all redshifts contribute roughly equally. This is not a problem in the case of a slow or negligible evolution in the Ly$\alpha$ LF with redshift from $z\sim2.5$ to $z\sim6$ \citep[e.g.][]{Ouchi2008}, but this has not been fully established yet, particularly for the bright end \citep[for the evolution of the faint-end, see][]{Drake2017b}. Our large sample of typical to bright LAEs is ideal to investigate whether that is the case and to quantify any potential evolution with redshift.

\subsection{The evolution of the Ly$\bf\alpha$ luminosity function from $\bf z\sim2$ to $\bf z\sim6$ in 12 redshift slices} \label{Evolution_LF_12slices}

After presenting the global Ly$\alpha$ LF for our full sample in Section \ref{global_LF_Sec}, we now explore the multiple redshift slices in SC4K (see Table \ref{S-SC4K_full_results}). In Figure \ref{fig:lfindividual} we present the Ly$\alpha$ LF per redshift slice all the way from $z\sim2.2$ to $z\sim5.8$ by deriving them per filter/redshift. We find a mild but noticeable evolution of the bright end of the Ly$\alpha$ LF with redshift from $z\sim2.2$ to $z\sim6$. This evolution seems to be mostly visible in terms of i) an evolution of the shape and ii) an evolution in luminosity. At lower redshift ($z\sim2.2-3.3$) there is a significant extra component (in addition to a single Schechter) to the Ly$\alpha$ LF above luminosities of $\approx10^{43.3}$\,erg\,s$^{-1}$, while such component seems to completely disappear by $z\sim3.7$ and to not show up in any of the Ly$\alpha$ LFs towards higher redshift. Interestingly, when considering only the major Schechter component of the Ly$\alpha$ LF, we find evidence for $\rm L^*_{Ly\alpha}$ to be evolving with redshift towards $z\sim6$ (see Table \ref{tab:integrals} and Figure \ref{fig:all_stacks}).

In order to quantify the potential redshift evolution and its significance, we use our best-fits of single power-laws, single Schechter functions and combinations of both and compare the resulting reduced $\chi^2$ (see Table \ref{S-SC4K_full_results}). We find that a single Schechter function is a particularly bad fit when including the bright end of the Ly$\alpha$ LF ($\rm \chi^2_{red}\sim10-30$) from $z=2.5$ to $z=3.3$. A single power-law fits better ($\rm \chi^2_{red}\sim3-7$), while a combination of a Schechter at lower luminosities and a power-law at higher luminosities with a transition around 10$^{43.3}$\,erg\,s$^{-1}$ provides the best fits (see Figure \ref{fig:lfindividual}). The combined fit are similar to the ones applied in recent large volume Ly$\alpha$ studies at a variety of redshifts \citep[e.g.][]{Konno2016,Sobral2017,MattheeBOOTES2017,Wold2017}. Interestingly, the Schechter component of the Ly$\alpha$ LF shows little evolution in L$^*_{\rm Ly\alpha}$ from $z\sim2.5$ to $z\sim3.1$, but reveals an important $\Phi^*_{\rm Ly\alpha}$ evolution from $z\sim2.5$ to $z\sim3.3$ (see Figure \ref{fig:all_stacks}), which may be consistent with an `extended' period of peak activity in the Universe \citep[][]{Madau2014}. For $z>3.5$, a single Schechter fit provides very good fits, although a single power-law could in principle also describe the bright end of the Ly$\alpha$ LF. From $z\sim2.2$ to $z\sim3.3$ the Ly$\alpha$ LF reveals a rise in $\Phi^*_{\rm Ly\alpha}$ by a factor $\approx4$, along with a potential steepening of the power-law component at the bright end of the LF. For $z>3.5$, where the power-law component is not seen anymore, our results reveal a fall of $\Phi^*_{\rm Ly\alpha}$ and a rise of L$^*_{\rm Ly\alpha}$ up to $z\approx5.8$ (see also Table \ref{S-SC4K_full_results}).

%
%
\begin{table*}
\caption{The results of fitting different Ly$\alpha$ LFs with a Schechter function at the appropriate luminosity range. For SC4K only we do not fit $\alpha$, but instead fix it from $-1.6$ to $-2.0$ in steps of 0.05 with a uniform prior; the $\pm0.2$ shown therefore reflects the variation we impose on $\alpha$, and not an uncertainty in fitting $\alpha$. For S-SC4K we explicitly fit all three parameters. For each fit we also integrate the Ly$\alpha$ LF to obtain $\rho_{\rm Ly\alpha}$, derived for different redshift bins, down to $0.04L^\star$. All errors are the 16 and 84 percentiles of all the fits, derived from our 10,000 realisations per LF. We also convert $\rho_{\rm Ly\alpha}$ to a star formation rate density by using Equation \ref{eq:SFR} \citep{Kennicutt1998} assuming a Salpeter IMF between $0.1-100\,{\rm M}_{\sun}$ and affected by $\rm f_{esc}$ (note that any correction for dust extinction will also be included in the $\rm f_{esc}$ term).}\label{tab:integrals}
\begin{tabular}{@{}ccccccc@{}}
\hline
Redshift bin &  $\alpha$ &  $\rm \log_{10}\,L^*_{\rm Ly\alpha}$  & $\log_{10}\,\Phi^*_{\rm Ly\alpha}$ &  $\rho_{\rm Ly\alpha}$\,/\,$10^{40}$ & SFRD$_{\rm Ly\alpha}\times{\rm f_{esc}}$/\,$10^{-2}$  & Reference(s)   \\
 (SC4K only) &    &  (erg\,s$^{-1}$)  & (Mpc$^{-3}$)  &  (erg\,s$^{-1}$\,Mpc$^{-3}$) &  (M$_{\odot}$\,yr$^{-1}$\,Mpc$^{-3}$)  & (Table \ref{SSC4K_refs_IDs})   \\
\hline 
 $z=2.2\pm0.1$ ($\rm L<10^{43.3}$) & $-1.8\pm0.2$ (fix) &  $42.69^{+0.14}_{-0.11}$  & $-3.33^{+0.21}_{-0.26}$  &  $0.48^{+0.04}_{-0.04}$ & $0.44^{+0.04}_{-0.04}$  &  12 \\
 $z=2.5\pm0.1$ ($\rm L<10^{43.3}$) & $-1.8\pm0.2$ (fix) &  $42.76^{+0.08}_{-0.07}$  & $-3.23^{+0.15}_{-0.15}$  &  $0.73^{+0.18}_{-0.14}$ & $0.67^{+0.16}_{-0.13}$   &  SC4K \\
 $z=3.1\pm0.3$ ($\rm L<10^{43.3}$) & $-1.8\pm0.2$ (fix) &  $42.69^{+0.05}_{-0.04}$  & $-2.73^{+0.11}_{-0.12}$  &  $1.90^{+0.56}_{-0.39}$ & $1.73^{+0.51}_{-0.36}$   &  SC4K \\
 $z=3.9\pm0.2$  & $-1.8\pm0.2$ (fix) &  $42.89^{+0.11}_{-0.10}$  & $-3.71^{+0.30}_{-0.28}$  &  $0.34^{+0.21}_{-0.12}$ & $0.31^{+0.19}_{-0.11}$   &  SC4K \\
 $z=4.7\pm0.1$  & $-1.8\pm0.2$ (fix) &  $43.10^{+0.13}_{-0.12}$  & $-3.82^{+0.33}_{-0.32}$  &  $0.48^{+0.33}_{-0.18}$ & $0.43^{+0.30}_{-0.16}$    &  SC4K \\
 $z=5.4\pm0.4$  & $-1.8\pm0.2$ (fix) &  $43.35^{+0.12}_{-0.11}$  & $-4.18^{+0.31}_{-0.30}$  &  $0.41^{+0.28}_{-0.16}$ & $0.37^{+0.26}_{-0.15}$   &  SC4K \\
\hline 
 S-SC4K: synergy Ly$\alpha$ LF \\
\hline 
 $z=2.2\pm0.1$ ($\rm L<10^{43.3}$) & $-2.00^{+0.15}_{-0.14}$  &  $42.82^{+0.13}_{-0.10}$  & $-3.59^{+0.22}_{-0.28}$  &  $0.52^{+0.05}_{-0.05}$ & $0.47^{+0.04}_{-0.04}$   &  2.1, 6.1, 12 \\
 $z=2.5\pm0.1$ ($\rm L<10^{43.3}$) & $-1.72^{+0.15}_{-0.15}$  &  $42.71^{+0.09}_{-0.08}$  & $-3.10^{+0.17}_{-0.21}$  &  $0.74^{+0.08}_{-0.07}$ & $0.67^{+0.07}_{-0.07}$    &  2.1, 5.1 \\
 $z=3.1\pm0.3$ ($\rm L<10^{43.3}$) & $-1.63^{+0.17}_{-0.16}$  &  $42.77^{+0.12}_{-0.09}$  & $-3.06^{+0.21}_{-0.26}$  &  $0.86^{+0.10}_{-0.09}$ & $0.78^{+0.09}_{-0.08}$    &  2.1, 5.1  \\
 $z=3.9\pm0.2$ & $-2.26^{+0.18}_{-0.17}$  &  $42.93^{+0.13}_{-0.11}$  & $-3.66^{+0.30}_{-0.35}$  &  $1.11^{+0.19}_{-0.16}$ & $1.00^{+0.17}_{-0.14}$    &  2.2, 5.1, 5.2 \\
 $z=4.7\pm0.1$ & $-2.35^{+0.19}_{-0.19}$  &  $43.28^{+0.20}_{-0.14}$  & $-4.25^{+0.34}_{-0.49}$  &  $1.16^{+0.40}_{-0.27}$ & $1.05^{+0.36}_{-0.25}$    &  2.3, 3, 5.2, 10  \\
 $z=5.4\pm0.4$ & $-1.98^{+0.14}_{-0.14}$  &  $43.28^{+0.09}_{-0.09}$  & $-3.83^{+0.21}_{-0.22}$  &  $1.11^{+0.21}_{-0.17}$ & $1.01^{+0.19}_{-0.16}$    &  5.3, 9.4, 11 \\
\hline
\end{tabular}
\end{table*}

Using the redshift bins defined in Section \ref{sec:binning} we show the overall redshift evolution of the Ly$\alpha$ LF in Figure \ref{fig:all_stacks} (see Table \ref{tab:integrals}). We also use other/different filter combinations to obtain different redshift bins, and find that the results are all consistent within the error-bars, and thus not dependent on the choice of binning. The increased statistical sample from the redshift bins provides stronger constraints on the Ly$\alpha$ LF, and further reinforces the results already mentioned when looking at each of the individual 12 redshift slices, including the presence of a potential power-law (or extra Schechter) component at high luminosities at $z\sim2-3.5$, which seems to disappear or be at too low number densities for even our survey to detect beyond $z\sim3.5$. Focusing on the Schechter components (fitting a Schechter only up to 10$^{43.3}$\,erg\,s$^{-1}$ at $z<3.3$ where a clear excess at the bright end is found), and for a fixed $\alpha=-1.8$, we find that L$_{\rm Ly\alpha}^*$ may evolve in a relatively continuous way from 10$^{42.69^{+0.05}_{-0.04}}$\,erg\,s$^{-1}$ at $z\sim3.1$ to 10$^{43.35^{+0.12}_{-0.11}}$\,erg\,s$^{-1}$ at $z\sim5.4$, which would imply a factor $\sim4-5$ increase in the typical luminosity. This is accompanied by a strong decline of $\Phi^*_{\rm Ly\alpha}$ of $\sim10-30$ times from $z\sim3.1$ to $z\sim5.4$ (see Table \ref{tab:integrals} and Figure \ref{fig:all_stacks}).

The apparent decline in $\Phi^*_{\rm Ly\alpha}$, accompanied by a positive L$_{\rm Ly\alpha}^*$ evolution may be linked to an evolution of the nature of the sources or changes in the conditions of the ISM and CGM, but potentially also with an evolution of the AGN population. We discuss possible explanations in Section \ref{Discussion}. We note that by excluding X-ray and radio AGN we find a reduction of the number densities of LAEs at the highest luminosities, lowering and steepening the potential power-law component, but without removing it. This means that if the power-law component is fully AGN driven \citep[][]{Wold2017,Sobral2018} there is still a significant component of the AGN population that is simply not detectable in the X-rays or radio \citep[][]{Sobral2018}, potentially because these AGN are very young and/or of very low black hole mass, but highly efficient in the production of Ly$\alpha$ photons which might easily escape, or due to the timescales involved in the AGN turning on and off. Our results thus highlight two potentially important/different physical mechanisms contributing to the Ly$\alpha$ LF at $z\sim2-6$.

\subsection{Comparison with other studies at $\bf z\sim2-6$} \label{sec:comparison}

A wide range of Ly$\alpha$ surveys using narrow-bands, slits or IFUs have derived Ly$\alpha$ LFs at $z\sim2-6$, mostly probing at and below L$^*_{\rm Ly\alpha}$ \citep[e.g.][]{Shimasaku2006,Westra2006,Dawson2007,Gronwall2007,Murayama2007,Rauch2008,Ouchi2008,Shioya2009,Cassata2011,Drake2017b}; see Table \ref{SSC4K_refs_IDs}. These are both perfect comparisons to our results and useful extensions to fainter luminosities.

\begin{figure*}
\centering
\includegraphics[width=18.0cm]{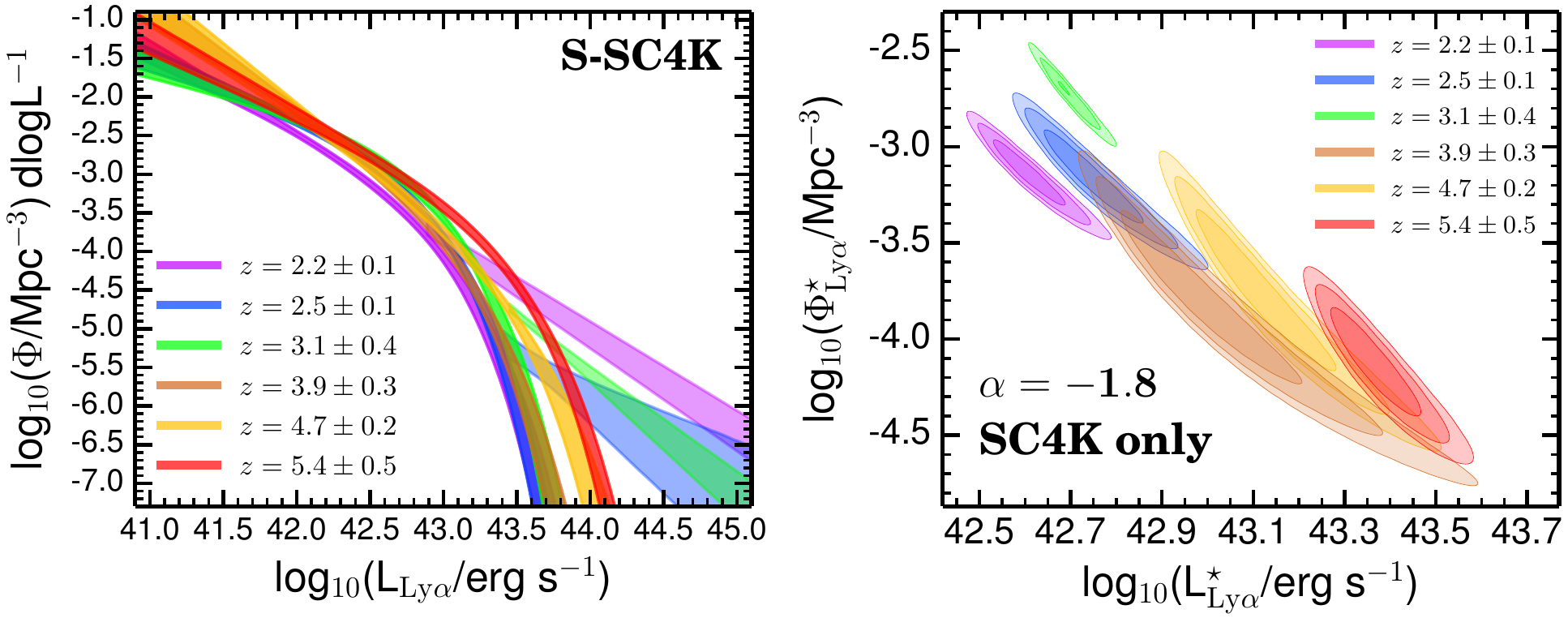}
\caption{{\it Left:} The evolution of the Ly$\alpha$ LF with redshift from $z\sim2$ to $z\sim6$ from this study, exploring our synergy approach (S-SC4K), showing the 16th and 84th percentiles of all realisations/fits. We find a mild $\rm L^*_{Ly\alpha}$ rise with increasing redshift, at the same time that $\rm \Phi^*_{Ly\alpha}$ declines. This leads to a mild evolution in the Schechter-like component with redshift. We find that the extra power-law/Schechter component at $\rm L_{Ly\alpha}>10^{43.3}$\,erg\,s$^{-1}$ declines with increasing redshift, mostly by becoming steeper and with a lower normalisation, which may be linked with the decline in the AGN population. By $z\sim3.9$ the extra component is no longer seen at the current observational limits. {\it Right:} The $\rm L^*_{Ly\alpha}$-$\rm \Phi^*_{Ly\alpha}$ contours for the Schechter fits by fixing $\alpha=-1.8$ (without any perturbation) by using the SC4K MBs only. The lines are the 1$\sigma$, 2$\sigma$ and 3$\sigma$ contours for $\rm L^\star_{Ly\alpha}$ and $\rm \Phi^\star_{Ly\alpha}$ for each redshift bin. This shows the mild but significant evolution of both $\rm L^\star_{Ly\alpha}$ and $\rm \Phi^\star_{Ly\alpha}$ with redshift.}
\label{fig:all_stacks}
\end{figure*}

A comparison between the Ly$\alpha$ LFs from this work and other studies at similar redshifts from the literature is shown in Figure \ref{fig:lfindividual}. We find that the $z=2.2$ Ly$\alpha$ LF from \citet{Sobral2017} is in good agreement with our $z=2.5$ measurements at the bright end, but the comparison reveals a positive $\rm \Phi^*_{Ly\alpha}$ evolution from $z\sim2.2$ to $z\sim2.5$ (see also Figure \ref{fig:all_stacks}). The $z\sim2.2$ Ly$\alpha$ LF presented by \cite{Konno2016} is in reasonable agreement with ours, and also implies evolution from $z=2.2$ to $z\sim2.5$, but implies higher number densities of bright sources \citep[see discussion on the importance of filtering out lower redshift interlopers and how they can easily account for 50\% of high EW sources in the bright end at $z\sim2$; see][]{Sobral2017}. We also show the \cite{Konno2016} results when removing likely contaminants in Figure \ref{fig:lfindividual} as in \cite{Sobral2017}, which results in an even better agreement with our results at the bright end. The $z=2.4$ LF from \citet{MattheeBOOTES2017} is also in good agreement with our measurement at $z\sim2.5$. We note that the number densities observed for the brightest bin in \citet{MattheeBOOTES2017} are marginally higher than ours (Figure \ref{fig:lfindividual}), and that those high luminosity sources have now all been spectroscopically confirmed \citep[see][]{Sobral2018}, and thus contamination is not able to explain the small discrepancy. The observed lower number densities for our results based on the medium bands when compared with \citet{MattheeBOOTES2017} may be explained by some of the brightest sources having lower EWs and thus being missed by our relatively high EW cut, even after applying our completeness corrections \citep[see full discussion in][and also in Perez et al. in prep.]{Sobral2017}. Cosmic variance is another possibility. We also compare our results to \cite{Cassata2011} and find a good agreement.

%
%

The `mild' increase from $z\sim2$ up to $z\sim3.3-3.7$ of the number density of LAEs (factor of $\approx4$) across the entire luminosity range is consistent with measurements from several studies, where a similar rise of the Schechter function is seen by comparing e.g. \citet{Sobral2017} at $z\sim2.2$ with \cite{Ouchi2008} at $z\sim3.1$ and $z\sim3.7$ (Figure \ref{fig:lfindividual}). In fact, at $z=3.7$, \cite{Ouchi2008} finds higher number densities at all luminosities than ours, although by $z=4.1$ our measurements agree very well with \cite{Ouchi2008}. At $z=4.8$, SC4K provides a unique opportunity to directly compare results from a MB and NB at roughly the same central wavelength (Perez et al. in prep), and we find a very good agreement at all luminosities probed by both bands, with the NB data allowing to go deeper, while the MB allows to probe a wider volume.

As we move to even higher redshifts ($z\sim5-6$), there is tentative evidence for a `boost' in luminosity (accompanied by a decline in number density and a potential steepening of the Ly$\alpha$ LF; \citealt{Drake2017b}), which agrees with results from \cite{Santos2016}, and with those at $z\sim5.7$ from \cite{Ouchi2008} when corrected in the same way as our results \citep[see discussion in e.g.][and also Section \ref{sec:filtercor}]{Matthee2015,Santos2016}. Recent results from HSC \citep[][]{Konno2017} reach volumes similar to ours at $z=5.7$ and hint for an overall lower number density of sources than those found by \cite{Ouchi2008} or \cite{Santos2016}. This difference is mitigated when we apply the filter profile corrections (see e.g. Figure \ref{fig:lfindividual}), but still suggests an overall lower number density of sources or systematic differences in estimating/measuring fluxes.

\subsection{S-SC4K: the synergy Ly$\alpha$ LF($\bf z$)} \label{consensus}

Overall, our results show very good agreement with the literature for the range of luminosities where surveys can be directly compared. Our results also extend previous surveys not only to higher luminosities, but also to a much higher number of redshift slices, allowing to investigate the fine redshift evolution of the Ly$\alpha$ LF in terms of the apparent shape change in the bright end and its positive luminosity evolution (of the main Schechter component) by a factor of about $\approx5$ from $z\sim3$ to $z\sim6$ and a decline in the number density of sources by a factor $\approx10$ or more. Interestingly, recent results from MUSE \citep[][]{Drake2017b} provide strong evidence for $\alpha$ being steep and tentative evidence for it steepening with increasing redshift. However, ultra-deep MUSE data on their own still suffer from an important short-coming: the uncertainty in determining the characteristic luminosity and/or number density of sources \citep[e.g. errors on $\alpha$ up to $^{+1.4}_{-\infty}$ at $z\sim3-6.6$ due to poor constraints on the bright end; see][]{Drake2017b}. Our SC4K survey is exactly what is needed (see Figure \ref{Global_LF}) to provide the extra constraints on the bright end and break the degeneracies.

We combine our SC4K results with other surveys probing to fainter luminosities than SC4K, to derive a synergy/consensus Ly$\alpha$ LF (S-SC4K) from the peak of star-formation into the end of re-ionisation. We present the results in Figures \ref{Global_LF}, \ref{fig:lfindividual}, \ref{fig:all_stacks} and Tables \ref{LF_fits_global} and \ref{tab:integrals}. We find evidence for a steepening of the faint-end slope (see Table \ref{S-SC4K_full_results}) from $z\sim2.5$ ($\alpha=-1.7\pm0.2$) to $z\sim5$ ($\alpha=-2.5\pm0.2$). Most importantly, we find that $\alpha$ is always very steep and close to $\alpha=-2$ at all redshifts probed. The synergy LF (S-SC4K; Figure \ref{fig:all_stacks}) also shows a roughly continuous increase in $\rm L^*_{Ly\alpha}$ by a factor of $\approx3-4$ from $z=2.5$ to $z\sim5-6$ (for the main Schechter component; note that at $z<3.3$ the Ly$\alpha$ LF requires an extra bright component to be properly modelled). In addition, we also find evidence of a decline in the typical number density at $\rm L^*_{Ly\alpha}$, with $\rm \Phi^*_{Ly\alpha}$ continuously reducing by a factor of $\approx5$. Overall, we show that there is evolution in the Ly$\alpha$ LF from $z\sim2.5$ to $z\sim6$, driven by an apparent steepening of the faint end slope, together with both a decline in number density and a positive luminosity evolution (factors of $\sim3-5$). It is also worth highlighting that a single Schechter function is not capable to encompass the full evolution of the Ly$\alpha$ LF at $z\sim2-3.3$, due to the significant power-law or extra brighter Schechter component. We also note that it is possible that the extra population of likely AGN dominating the bright end at lower redshift (see Figure \ref{fig:all_stacks}) may still contribute at higher redshift and may in principle be partially responsible for the luminosity evolution. However, as Section \ref{sec:Lya_rho} shows, due the very steep faint end slope of the Ly$\alpha$ LF, the Ly$\alpha$ luminosity density is dominated by the faintest sources and thus the evolution of the bright end by itself does not dominate the luminosity budget, though it may be very important to understand the physics of sources contributing to it. We also stress that while the bright sources are not the dominant sources of Ly$\alpha$ luminosity density in the Universe, only the combination of ultra-deep and large volume surveys can provide the full constraints necessary to fully measure the evolution of the Ly$\alpha$ LF and the population of sources that contributes to it.

\subsection{The redshift evolution of $\bf \rho_{Ly\alpha}$} \label{sec:Lya_rho}

\begin{figure}
\centering
\includegraphics[width=8.4cm]{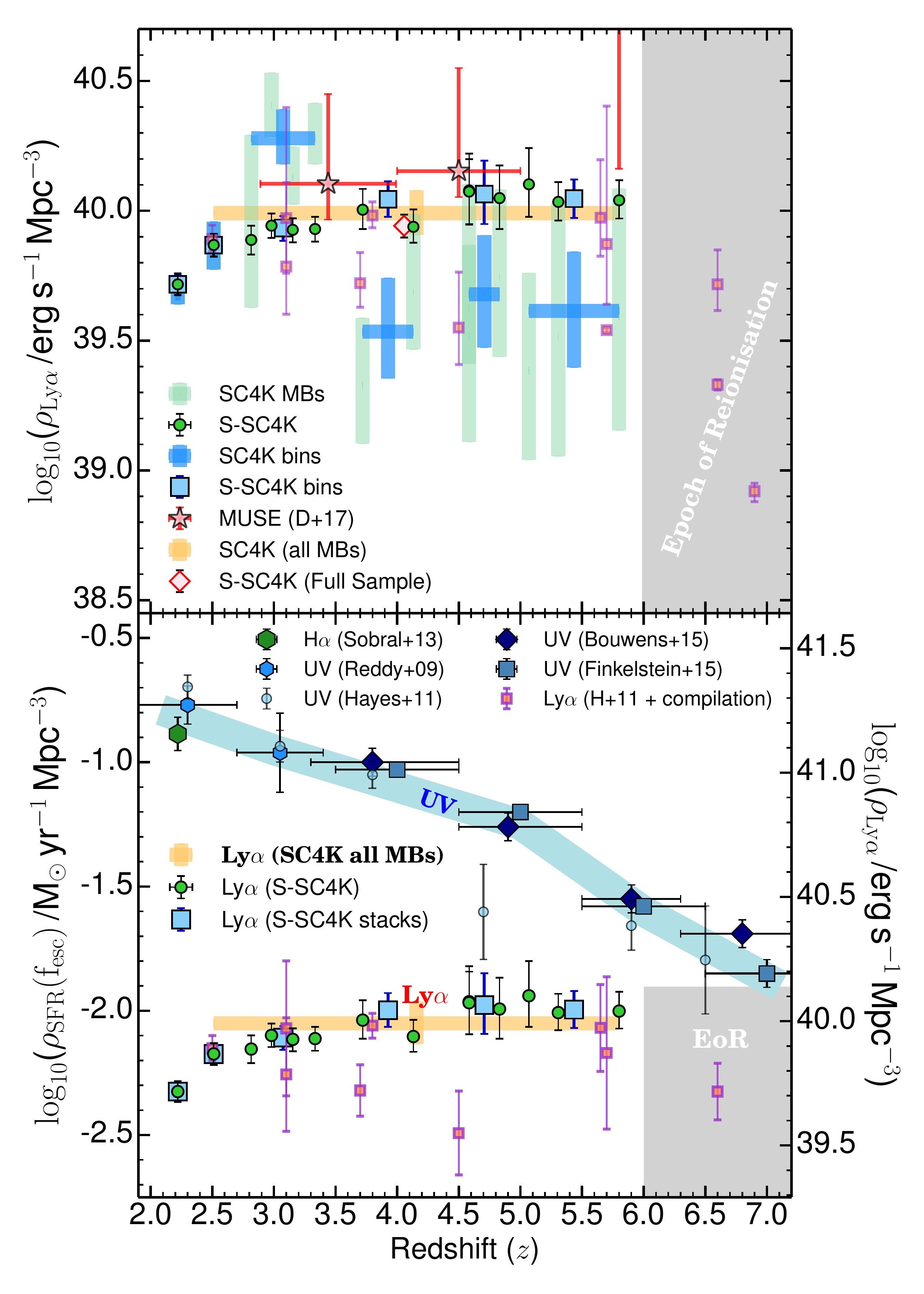}
\caption{{\it Top:} The evolution of the Ly$\alpha$ luminosity density ($\rho_{\rm Ly\alpha}$). We show the measurements using SC4K only, including per filter and also redshift stacks. We find a relatively constant $\rho_{\rm Ly\alpha}$ across redshift, with the MB filters on their own suggesting a slight decline in $\rho_{\rm Ly\alpha}$ from $z\sim3$ to $z\sim4-6$. Combining SC4K with deep surveys (S-SC4K) reveals the importance of probing both the faint and bright ends. The combined constraints show that $\rho_{\rm Ly\alpha}$ rises from $z\sim2$ to $z\sim3.5$ and then stays constant with redshift all the way to $z\sim6$. {\it Bottom:} We compare our results with surveys measuring the UV \citep[][]{Hayes2011,Bouwens2015,Finkelstein2015} and H$\alpha$ \citep[][]{Sobral2013} luminosity densities transformed to SFRDs. While the global star formation rate density (UV luminosity density) of the Universe is falling sharply from $z\sim2$ to $z\sim6$ by a factor of $\approx5$, the contribution from Ly$\alpha$ selected sources is rising, particularly due to the steepening of the Ly$\alpha$ LF, accompanied by a higher typical luminosity and despite the lower typical number density.}
\label{fig:sfrd}
\end{figure}

We explore SC4K and S-SC4K to measure the evolution of the Ly$\alpha$ luminosity density ($\rho_{\rm{Ly}\alpha}$) from $z\approx2.2$ to $z\sim6$, in multiple redshift slices, with unprecedented detail. We compute $\rho_{\rm{Ly}\alpha}$ by integrating the LF down to different limits. For a direct comparison with \cite{Hayes2011}, we integrate LFs down to $1.75\times10^{41}$\,erg\,s$^{-1}$, corresponding to 0.04\,$L^\star_{z=3}$\footnote{This corresponds to integrating down to $\rm \approx0.16$\,M$_{\odot}$\,yr$^{-1}$ for a Salpeter IMF and $\rm f_{esc}=1.0$; see Section \ref{fesc_xion}.} from \cite{Gronwall2007}. For each LF, we calculate 10,000 integrals, each perturbing individual data-points within their asymmetric Gaussian distributions, fitting the LF and computing the integral. For SC4K-only LFs we vary $\alpha$ with a uniform probability distribution between $-1.6$ and $-2.0$ for a more conservative error estimation (errors are the 16 and 84 percentiles of all the integrals). The results are shown in Figure \ref{fig:sfrd} and Tables \ref{LF_fits_global} and \ref{tab:integrals}.

We find evidence for $\rm \rho_{Ly\alpha}$ to increase with redshift, with a rise from $z\sim2$ to $z\sim3$ and then a tentative decline at $z\sim4$ and remaining constant at $z\sim4-6$ (Figure \ref{fig:sfrd}). These results are clear using both the individual redshift slices and also the redshift bins. We note that the decline in $\rm \rho_{Ly\alpha}$ seen from $z\sim3$ to $z\sim4$ with SC4K coincides with the disappearance of the bright-end excess of the Ly$\alpha$ LF, although we note that the potential power-law component at the highest luminosities, by itself, only represents $\sim1-5$\% of the Schechter luminosity density\footnote{In our analysis we do not include the integral of the power-law component.}. The evolution from $z\sim3.3$ to $z\sim4$ may be linked with a significant evolution in the nature of Ly$\alpha$ emitters.

%
%

When using S-SC4K, we obtain far superior constraints on $\rm \rho_{Ly\alpha}$ (much better than e.g. MUSE or SC4K on their own; see Figure \ref{fig:sfrd}). We still find that $\rm \rho_{Ly\alpha}$ increases from $z\sim2.2$ to $z\sim3-4$ by a factor of $\approx2$, and clear evidence for $\rm \rho_{Ly\alpha}$ to be relatively constant with redshift from $z\sim4$ to $z\sim6$. Our results thus show that despite the clear evolution of the Ly$\alpha$ LF from $z\sim4$ to $z\sim6$, its integral remains roughly constant. Interestingly,we note that the global SC4K LF on its own ($2.5<z<6$) yields a value of $\rm \rho_{Ly\alpha}$ which is actually representative of the majority of the individual measurements at $z\sim3-6$. We find that the relative constancy of $\rm \rho_{Ly\alpha}$ with increasing redshift is driven by a steepening of the faint-end slope $\alpha$ with increasing redshift, together with an increase in $\rm L^*_{Ly\alpha}$, which counter-balances the significant reduction in $\rm \Phi^*_{Ly\alpha}$ with increasing redshift. Therefore, our results show that whilst $\rm \rho_{Ly\alpha}$ stays relatively constant with redshift, there is a strong shift towards fainter LAEs becoming more and more dominant in the global $\rm \rho_{Ly\alpha}$ towards re-ionisation.

We compare our results with the literature \citep[see e.g.][and references therein]{Ouchi2008,Hayes2011,Matthee2015,Santos2016,Zheng2017,Drake2017b} and find good agreement with our measurements within the errors. The scatter of individual measurements and previous studies done on single fields and/or just probing either the bright or faint regimes is also very clear in Figure \ref{fig:sfrd}. For example, MUSE data on their own suggest a potential increase in $\rm \rho_{Ly\alpha}$, while SC4K on its own would suggest a reduction. Our results highlight the importance of combining the strengths of each approach/instrument/measurement in order to truly reveal the behaviour of $\rm \rho_{Ly\alpha}$ with redshift.

On the bottom panel of Figure \ref{fig:sfrd} we convert $\rm \rho_{Ly\alpha}$ to a star-formation rate density (SFRD; see full assumptions in Section \ref{fesc_xion}) so we can more directly compare it with the UV luminosity density also converted to SFRD \citep[e.g.][]{Bouwens2015,Finkelstein2015}. Our results reveal the striking difference between the evolution of the UV and Ly$\alpha$ SFRDs with increasing redshift. While the SFRD traced by Lyman break galaxies (and H$\alpha$ emitters at $z=2.2$) is strongly declining (by a factor of about 5 from $z\sim2.2$ to $z\sim6$), the Ly$\alpha$ SFRD is increasing to $z\sim3-4$ and then remaining constant all the way to the end of the epoch of re-ionisation at $z\sim6$. Therefore, our results re-enforce the increasing importance of LAEs at higher redshift in the global SFRD, and hint for global evolution in the properties of galaxies for this to happen, including the Ly$\alpha$ escape fraction (which would result in a higher Ly$\alpha$ luminosity density for a fixed UV luminosity density) and/or the typical ionisation efficiency (which can also lead to a higher production of Ly$\alpha$ photons). Furthermore, the Ly$\alpha$ escape fraction is sensitive to a number of galaxy properties such as the dust content \citep[e.g.][]{Atek2008,Hayes2010,Shibuya2014,Matthee2016_CALY,Oyarzun2017} and covering fraction of neutral Hydrogen \citep{Henry2015}, and thus any of these may be evolving. The production efficiency of ionising photons is related to the nature of stellar populations, such as the metallicity and initial mass function \citep[e.g.][]{Schaerer2003,Erb2014,Reddy2017}. In Section \ref{fesc_xion} we explore these possibilities in detail.

\section{Discussion}\label{Discussion}

%
%
\begin{figure*}
\centering
\includegraphics[width=14.0cm]{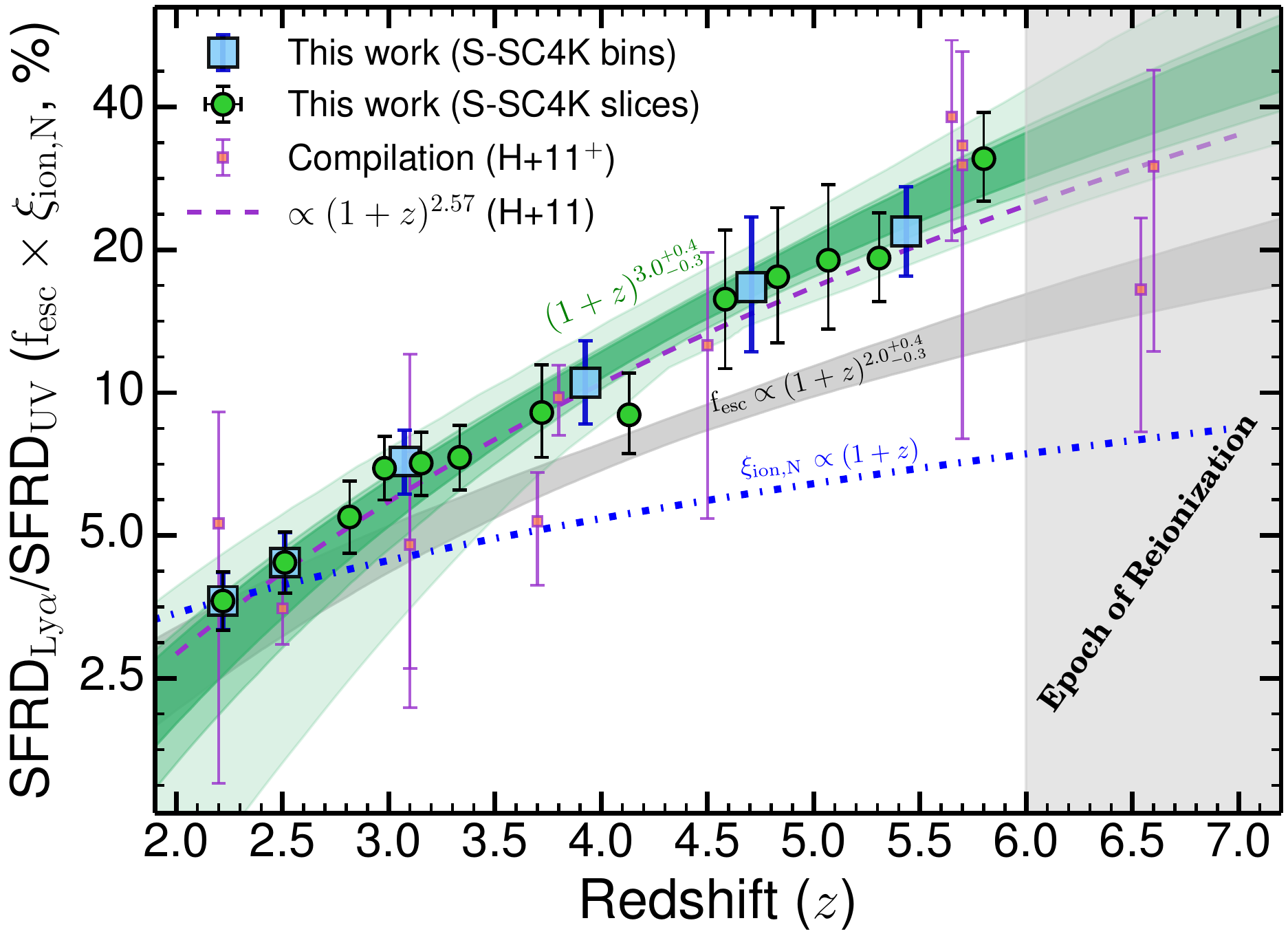}
\caption{The evolution of SFRD$_{\rm{Ly}\alpha}$/SFRD$_{UV}$ from $z=2.2$ to $z\sim6$ with S-SC4K. We find the ratio to increase from $\approx4$\% at $z=2.2$ to $\approx30\%$ at $z\sim6$, implying a very high Ly$\alpha$ to UV luminosity density ratio in the early Universe. We parameterise the rise with redshift as a power-law and find $\propto(1+z)^{3.0^{+0.4}_{-0.3}}$ (we show the 1, 2 and 3\,$\sigma$ range of all fits), a slightly steeper relation than in \citealt{Hayes2011} (we also include more recent measurements from the literature). Furthermore, by modelling the rise of $\rm \xi_{ion}$ as $\propto(1+z)$ \citep[see][]{MattheeGALEX2017}, we infer that $\rm f_{esc}$ is rising as $(1+z)^{2.0^{+0.4}_{-0.3}}$. Our results suggest a significant evolution in the typical burstiness/stellar populations ($\rm \xi_{ion}$) by a factor of $\approx2$ and an even stronger evolution in the typical ISM conditions leading to an inferred $\rm f_{esc}$ increase of a factor $\approx4$ from $z\sim2.2$ to $z\sim6$.}
\label{fig:ratio_uv_lya}
\end{figure*}

\subsection{The evolution of the cosmic Ly$\alpha$/UV ratio} \label{sec:uv_lya}

Based on our results, $\rho_{\rm Ly\alpha}$ rises by a factor of about $\sim2$ from $z\sim2.2$ to $z\sim3$ and is then relatively constant up to $z\sim6$. However, as shown in Figure \ref{fig:sfrd}, the UV luminosity density\footnote{UV luminosity densities are integrated down to 0.04\,$L^\star_{UV,z=3}$ following \cite{Hayes2011}; see also discussions in \cite{Hayes2011} and e.g. \cite{Sobral2017} on integration limits.} decreases by a factor $\approx 5$ over the same redshift range \citep[e.g.][]{Reddy2009,Bouwens2015,Finkelstein2015}. Figure \ref{fig:ratio_uv_lya} shows that the cosmic SFRD$_{\rm{Ly}\alpha}$/SFRD$_{UV}$ increases significantly with redshift by a factor of $\sim7-8$ from $z\sim2$ to $z\sim6$, driven by the mild positive evolution of $\rho_{\rm{Ly}\alpha}$ with redshift and the sharp decline in $\rho_{UV}$ (Figure \ref{fig:sfrd}). Our measurements follow a similar trend estimated by \cite{Hayes2011}, but provide significantly better sampling in terms of redshift and further constraining both the bright (SC4K) and faint ends (S-SC4K); see Figure \ref{fig:ratio_uv_lya}.

Observationally, our results mean that from $z\sim2$ to $z\sim6$ there is a systematic increase in the luminosity density of Ly$\alpha$ photons in the Universe relative to 1500\,{\AA} UV photons. Such increase should be vastly dominated by the large number of faint LAEs that likely become more dominant towards higher redshift, but there is also independent evidence for a higher Ly$\alpha$/UV ratio for fixed UV luminosities towards $z\sim6$, including at high UV luminosities \citep[see][and references therein]{Curtis-Lake2012,Schenker2014,Stark2017}. We explore and discuss potential explanations and interpretations for the rise of the cosmic Ly$\alpha$/UV ratio in Section \ref{fesc_xion}.

\subsection{The redshift evolution of $\bf f_{esc}$ and $\bf \xi_{ion}$}\label{fesc_xion}

The $\rho_{\rm{Ly}\alpha}$/$\rho_{UV}$ ratio is a tracer for the relative strength of Ly$\alpha$ to the UV. In Figure \ref{fig:ratio_uv_lya} we show that SFRD$_{\rm{Ly}\alpha}$/SFRD$_{UV}$ ($\sim\rho_{\rm{Ly}\alpha}/\rho_{UV}$) rises with redshift significantly. In order to fully interpret and discuss the redshift evolution of the $\rho_{\rm{Ly}\alpha}$/$\rho_{UV}$ ratio, it is necessary to derive how it depends on the Ly$\alpha$ escape fraction and production efficiency of ionising photons. We follow \cite{Bouwens2016} and \cite{MattheeGALEX2017}, and define $\rm \xi_{ion}$ \citep[see discussions in][]{Shivaei2017}, the production efficiency of Hydrogen ionising photons (Lyman continuum, LyC), as:
\begin{equation}
\rm \xi_{ion} = \frac{Q_{ion}}{L_{UV}}\times({1-f_{esc,LyC}})\,\,{\rm (Hz\,erg^{-1})},
\end{equation}
where $\rm L_{UV}$ is the dust-corrected UV luminosity in erg\,s$^{-1}$\,Hz$^{-1}$ at a wavelength of 1500 {\AA}, and assuming a $\approx0$\,\% escape fraction of LyC photons ($\rm f_{esc,LyC}$). Q$_{\rm ion}$, the number of emitted ionising (LyC) photons per second, is related to the dust-corrected H$\alpha$ luminosity ($\rm L_{H\alpha}$) as:
\begin{equation}
\rm Q_{ion} =  \frac{L_{\rm H\alpha}}{c_{H\alpha}}\,\,{\rm (s^{-1})},
\end{equation}
where $c_{\rm H\alpha} =1.37\times10^{-12}$ erg \citep[e.g.][]{Kennicutt1998,Schaerer2003} is the recombination coefficient. Under the assumption of case B recombination, a temperature of 10$^4$\,K, an electron density 350\,cm$^{-3}$ and a 0\% escape fraction of ionising LyC photons, the H$\alpha$ luminosity is related to Ly$\alpha$ (with f$_{\rm esc}$ being the Ly$\alpha$ escape fraction) as:
\begin{equation}
\rm L_{\rm H\alpha} =\frac{L_{\rm Ly\alpha}}{8.7 f_{\rm esc}}\,\,{\rm (erg\,s^{-1})}.
\end{equation}

With our assumptions so far, we can use L$_{\rm H\alpha}$ to estimate the SFR\footnote{For continuous SF over 10\,Myr timescales.}, following \citet{Kennicutt1998} for a Salpeter IMF $(0.1-100 \,{\rm M}_{\sun})$:
\begin{equation} \label{eq:SFR}
\rm {\rm SFR_{H\alpha}}=7.9\times10^{-42} \, \, L_{\rm H\alpha}\,\,{\rm (M_{\odot}\,yr^{-1})}.
\end{equation}

We combine these equations to derive an expression for the relation between the Ly$\alpha$ and UV luminosities:
\begin{equation}
\rm \xi_{ion}\times f_{\rm esc} = \frac{L_{\rm Ly\alpha}}{8.7 c_{H\alpha} L_{UV}} \,\,{\rm (Hz\,erg^{-1})}. \label{xi_fesc}
\end{equation}

Quantitatively, both UV and Ly$\alpha$ luminosities are related to the SFR. The (dust-corrected) UV luminosity through direct continuum emission from young stars, and Ly$\alpha$ luminosity through the recombination radiation in H{\sc ii} regions from LyC photons originating from young stars. Following \cite{Kennicutt1998}, $\rm \xi_{ion}$ is related to the H$\alpha$ and UV SFR as:
\begin{equation}
\rm \xi_{ion} = 1.3\times10^{25} \frac{SFR_{\rm H\alpha}}{SFR_{UV}} \, \rm (Hz\,erg^{-1}).
\end{equation}
In this equation, the constant $1.3\times10^{25}$\,Hz\,erg$^{-1}$ is dependent on the IMF and stellar spectral synthesis models. The ratio between the H$\alpha$ and UV SFRs is a measure of burstiness of SF \citep[see also][]{Smit2016} and is equal to 1 if there is a continuous SF history for the last 100\,Myr. Therefore, an increasing $\rm \xi_{ion}$ could trace both the nature of stellar populations (i.e. the hardness of the ionising spectrum) and/or the burstiness of star formation. This degeneracy can be resolved with photo-ionisation modelling when multiple emission-lines with a range of ionisation energies are observed (for example using the Helium Balmer lines). If we define $\rm \xi_{ion,N}=\xi_{ion}/(1.3\times10^{25}\,Hz\,erg^{-1})$, we can write:
\begin{equation}
\rm \xi_{ion,N}\times f_{\rm esc} = \frac{SFR_{\rm Ly\alpha}}{SFR_{UV}},
\end{equation}
allowing us to more directly interpret the ratio between $\rm SFR_{\rm Ly\alpha}$ and $\rm SFR_{UV}$. \cite{MattheeGALEX2017} discusses how $\rm \xi_{ion}$ correlates with H$\alpha$ EW, and how the widely agreed rise of typical H$\alpha$ EWs\footnote{See also results showing a rise in typical EWs of other rest-frame optical lines such as [O{\sc iii}] in \citet{Khostovan2016}.} with redshift \citep[e.g.][]{Fumagalli2012,Sobral2014,Faisst2016} suggests that $\rm \xi_{ion}$ rises by a factor of about $\sim2$ from $z\sim2$ to $z\sim6$ as $\propto(1+z)$, in agreement with e.g. \cite{Nakajima2016} and \cite{Harikane2017}. Assuming $\rm \xi_{ion,N}\approx1$ at $z=2.2$ \citep[see][]{MattheeGALEX2017,Shivaei2017}, we can then measure $\rm f_{esc}$ directly for $2<z<6$ by using:
\begin{equation}
\rm f_{\rm esc} = \frac{3.2}{(1+z)}\frac{SFR_{\rm Ly\alpha}}{SFR_{UV}} \, \, (2<z<6).
\end{equation}

We check with \cite{Sobral2017} that the above approach is able to roughly recover $\rm f_{esc}$ at $z=2.2$ measured directly with H$\alpha$ ($4\%$ with the integration limits we use and without using the power-law component of the Ly$\alpha$ LF). By comparing with our observations in Figure \ref{fig:ratio_uv_lya}, we infer an evolution of $\rm f_{esc}$ of a factor $\approx4$ (from $\approx3.8$\% at $z\sim2.2$ to $\approx15$\% at $z\sim6$), with an increase roughly proportional to $(1+z)^{2.0\pm0.3}$ for $\rm f_{esc}$ (see Figure \ref{fig:ratio_uv_lya}). Our results thus suggest that the strong evolution in the SFRD$_{\rm{Ly}\alpha}$/SFRD$_{UV}$ ratio with redshift is driven by an increase in $\rm \xi_{ion}$ (tracing high burstiness and/or an average change in stellar populations which) by a factor of $\sim2$, rising as $1+z$ and $\rm f_{esc}$ by a factor of $\approx4-5$ from $z\sim2$ to $z\sim6$, rising as $(z+1)^2$. Overall, this explains the rise of SFRD$_{\rm{Ly}\alpha}$/SFRD$_{UV}$ as $(1+z)^{3.0\pm0.3}$. Our results thus imply evolution in both ISM conditions and on the burstiness/nature of the stellar populations with increasing redshift.

\subsection{The compact nature of LAEs and relation to the global increase in $\rm \bf f_{esc}$} \label{discuss_morph_fesc}

\cite{Paulino-Afonso2017} presents the full visual and automated morphological and structural analysis in the rest-frame UV of the SC4K sample presented in this paper. They find that LAEs are systematically smaller in the rest-frame UV than the global population of star-forming galaxies, presenting sizes which are roughly constant with redshift of $\approx1$\,kpc \citep[see also][]{Bond2011,Malhotra2012,Guaita2015}. \cite{Paulino-Afonso2017} also points out that while ``typical'' star-forming galaxies at $z<2$ are $\sim2-4\times$ larger than LAEs \citep[][]{vanderWel2014,Ribeiro2016,PA_2017}, the differences in typical sizes become smaller with increasing redshift. By $z\sim6$, the general population of SFGs presents basically the same morphological properties as LAEs have across all redshifts.

Furthermore, \cite{Paulino-Afonso2017} also discusses how the sizes and compactness of LAEs depend on rest-frame Ly$\alpha$ EW$_0$. The EW$_0$ of the Ly$\alpha$ line has recently been shown to be the simplest/most robust empirical predictor of $\rm f_{esc}$ \citep[][]{Sobral2017}, with the relation between EW$_0$ and $\rm f_{esc}$ showing no significant evolution at $z\sim0-5$, \citep[see][]{Sobral2017,Harikane2017}. \cite{Paulino-Afonso2017} find that LAEs with the highest EWs are the smallest and most compact at all redshifts. This suggests a relation between compactness and/or size and $\rm f_{esc}$, and may be one of the physical reasons why we find that globally $\rm f_{esc}$ seems to rise with increasing redshift. In this case, it would be because the general population of galaxies are, as a whole, compact and small enough, for Ly$\alpha$ photons to more easily escape. However, we note that smaller and more compact galaxies will typically be also less evolved, potentially more bursty and with lower metallicity stellar populations, which can also lead to boosting Ly$\alpha$ through a higher $\rm \xi_{ion}$. The potentially higher $\rm f_{esc}$ at higher redshift could also be caused more directly by e.g. lower dust content and/or a more porous CGM due to strong stellar winds \citep[e.g.][]{Geach2014Nat} produced in compact and highly star-forming regions, which would allow the escape of more Ly$\alpha$ photons.

The morphological information may be important to potentially explain the increase in $\rm f_{esc}$ with redshift, but in principle it does not tell us anything about the burstiness or the stellar populations and/or AGN activity that may be happening within LAEs across cosmic time. This is important to understand the potential evolution in $\rm \xi_{ion}$ \citep[][]{MattheeGALEX2017}, even more so as our results provide evidence that both $\rm \xi_{ion}$ and $\rm f_{esc}$ evolve with redshift. Further physical insight may be obtained by studying local analogues like `green peas' or `blueberry' galaxies \citep[e.g.][]{Yang2017a,Yang2017b,Izotov2017Eion}. Such analogue galaxies allow for detailed studies to be performed to make crucial measurements and test hypothesis/modelling results \citep[e.g.][]{Verhamme2006,Verhamme2015,Izotov2016,Yang2017a} regarding the connection between $\rm f_{esc}$ and the Ly$\alpha$ emission line peak separation, width and other properties \citep[see e.g.][]{Verhamme2017}. Furthermore, these low redshift sources, showing essentially the same properties as SC4K galaxies at higher redshift, are ideal to further explore and test the link between LyC and Ly$\alpha$ photon escape \citep[e.g.][]{Verhamme2015,Verhamme2017,DijkstraGronke2016,Izotov2017_high_Lyc} and their relation with size/compactness and other physical properties.

\subsection{The bright end of the Ly$\bf \alpha$ LF: AGN?}\label{bright_end_discussion}

Previous studies \citep[e.g.][]{Konno2016,Sobral2017,MattheeBOOTES2017,Wold2017} have found evidence for a relation between the potential power-law component of the bright end of the Ly$\alpha$ LF and the AGN nature of sources populating it. Such evidence has been primarily driven by the detection of many of those sources in the X-rays \citep[e.g.][]{Konno2016,Sobral2017}. With the availability of deep {\it Chandra} and VLA data, we have identified that $3.6\pm0.3$\% of all our sources are likely AGN, with 109 ($2.9\pm0.3$\%) being X-ray AGN, 62 ($1.7\pm0.2$\%) being radio AGN and 30 ($0.8\pm0.1$\%) being both. While these are a very small fraction overall, as shown in \cite{Calhau2018} and \cite{Sobral2018}, AGN LAEs become more significant at the brightest Ly$\alpha$ luminosities, a consequence of their relatively flat Ly$\alpha$ LF which we have found, with a potential high L$^*_{\rm Ly\alpha}$. \cite{Calhau2018} finds a significant correlation between the X-ray AGN fraction of LAEs and the Ly$\alpha$ luminosity; this fraction is consistent with $0.7\pm0.3$\% below L$^*_{\rm Ly\alpha}$, but it grows towards 100\% at the highest Ly$\alpha$ luminosities \citep[see also][]{MattheeBOOTES2017,Sobral2018}. We thus find that removing the X-ray and radio AGN leads to removing sources from the bright-end of the LF, but an excess relative to a Schechter persists at $z\sim2-3$ even after removing X-ray and radio sources. We argue that there is still a significant population of AGN sources that is undetected in the radio and X-rays, even after stacking. X-ray or radio-detected AGN only provide a lower constrain on the total number of AGN, as not all AGN have strong X-ray or radio emission. As shown in \cite{Sobral2018}, virtually all the spectroscopically confirmed LAEs at $z\sim2-3$ with $>10^{43.2}$\,erg\,s$^{-1}$ are AGN. Such AGN are revealed by deep rest-frame UV spectroscopy, even though the majority does not show any detectable X-ray or radio emission. These results indicate that the most luminous LAEs at $z\sim2-3$ are powerful AGN that emit copious amounts of Ly$\alpha$ photons, boosting the bright end of the Ly$\alpha$ LF. Further evidence comes from the relation between X-ray and Ly$\alpha$ luminosities which suggests that Ly$\alpha$ is tracing the accretion rate for those sources, and not SF processes. AGN LAEs have X-ray luminosities in the range L$_{\rm X-ray}=10^{43.4-45.1}$\,erg\,s$^{-1}$, implying high black hole accretion rates of 0.1-4\,M$_\odot$\,yr$^{-1}$. AGN LAEs have radio luminosities of $\approx10^{30.7}$\,erg\,s$^{-1}$\,Hz$^{-1}$, but little relation with Ly$\alpha$, probing down to lower Ly$\alpha$ luminosities, and potentially indicating `bursty' AGN accretion.

\cite{Calhau2018} discusses how the relation between AGN fraction and Ly$\alpha$ luminosity evolves with redshift, consistent with a decline in the normalisation or an evolution towards much higher Ly$\alpha$ luminosities. For $3.5<z<6$ (where we fail to detect the power-law component), the X-ray+radio AGN fraction of LAEs remains relatively low for the entire luminosity range, although it still rises with Ly$\alpha$ luminosity from $0.9\pm0.4$\% at the lowest Ly$\alpha$ luminosities to $11\pm7$\% at $\approx10^{44}$\,erg\,s$^{-1}$. These results are consistent with those from \citet{Wold2014,Wold2017} at $z\sim0-1$, but provide evidence for the AGN fraction evolving (declining) with redshift. While we find no convincing evidence of a significant population of AGN LAEs beyond $z>3.5$, and no detectable power-law component in the LF, it is possible that it continues to exist at $z>3.5$, but just with number densities below our surveyed volumes and/or with a LF that is more similar to the fainter population of LAEs, thus making it indistinguishable from those. If these sources occupy the faint-end of the quasar luminosity function, one would potentially expect number densities of 10$^{-9}-10^{-10}$\,Mpc$^{-3}$ \citep{McGreer2013} for the most luminous $z=5$ quasars, which would be easily below our detection limit. It is also possible that the bright end still contains AGN sources even towards $z\sim6$, but that they are just not X-ray or radio luminous enough to be detected either individually or by stacking \citep[see][]{Calhau2018}. Such potential ``hidden" AGN activity in luminous LAEs at higher redshift could still be driving the apparent L$^*_{\rm Ly\alpha}$ rise towards $z\sim6$ and might be tentatively showing up in deep spectroscopic observations of some of the most luminous LAEs at $z\sim6-7$ with potential detections of He{\sc ii} and/or N{\sc v} \citep[e.g.][]{Laporte2017,Sobral2017_CR7}. In addition, we also note that while high accretion rates and relatively high black hole masses in fainter LAEs are excluded, faint LAEs may still contain young, low mass AGN that would make them currently undetectable in the X-rays and radio.

\subsection{The nature and evolution of faint to bright LAEs across $\bf z\sim2-6$: progenitors of sub-L$^*$ galaxies to proto-cluster tracers}\label{discuss_nature_global}

Clustering analysis \citep[][]{Khostovan2018} of the SC4K sample shows a clear dependence of the clustering length and the inferred dark matter halo mass on both the Ly$\alpha$ luminosity and the UV luminosity or SFR. At the highest Ly$\alpha$ luminosities, LAEs are likely hosted by quite massive dark matter haloes of 10$^{13-14}$\,M$_{\odot}$, where one expects AGN activity to be prominent. These observational results are in good agreement with modelling from e.g. \cite{Garel2016} who finds that the brightest LAEs at high redshift should reside in more massive dark matter haloes and be the progenitors of more massive haloes today, while the super faint LAEs now being found by MUSE \citep[][]{Drake2017b} are likely the progenitors of sub-L$^*$ galaxies today. \cite{Khostovan2018} finds similar results, with the dark matter haloes and the clustering strength of the faintest LAEs from the narrow-band selected surveys being closer to $\sim10^{11}$\,M$_{\odot}$, similar to results from e.g. \cite{Ouchi2010} and other clustering studies focusing on very faint LAEs \citep[e.g.][]{Kusakabe2017}. The high number densities of faint LAEs at high redshift, driven by the steep ($\alpha\approx-2$) faint-end slope of the Ly$\alpha$ LF \citep[S-SC4K and e.g.][]{Dressler2015,Drake2017b} reveal that a very large number of sources are emitting Ly$\alpha$ photons that can escape in the early Universe. These numerous LAEs \citep[this study and e.g.][]{Drake2017b} with high Ly$\alpha$ escape fractions and high EWs \citep[e.g.][]{Sobral2017,Hashimoto2017}, highly ionising \citep[][]{Nakajima2016}, compact/small sources \citep[][]{Malhotra2012,Paulino-Afonso2017} may play a crucial role in the early Universe. For example, our results imply that by $z\sim6$, LAEs are likely key contributors to the global LyC photons produced in the Universe.

Overall, LAEs have low UV luminosities (which can easily make them undetected even in very deep continuum surveys), but high production of LyC photons (expressed as a high ionisation efficiency; \citealt{Nakajima2016,MattheeGALEX2017,Harikane2017}). Thus, our results strongly add to current observations by pointing towards LAEs being exactly the sources that ultra-deep continuum surveys strive to detect using gravitational lensing \citep[e.g.][]{Atek2015}. Due to the strength and high EWs of the Ly$\alpha$ emission line at high redshift, LAE surveys are simply much more efficient at picking the numerous, UV-faint, compact and highly ionising sources in spite of their ultra-faint UV magnitudes. Examples of such faint, strongly Ly$\alpha$ emitting galaxies have recently been found in e.g. \cite{Vanzella2016}. Furthermore, recent results of local galaxies showing the same properties as SC4K sources \citep[including $\rm M_{UV}$, Ly$\alpha$ EWs and sizes e.g.][]{Izotov2016,Izotov2017_high_Lyc} provide even more evidence for the importance of LAEs in the early Universe in terms of their contribution to both the SFRD and as the sources that likely re-ionised the Universe.

SC4K is also able to find some of the rarest, brightest LAEs across cosmic time which are likely powered by AGN. Most importantly, the brightest LAEs seem to be highly clustered, and there is convincing evidence that they trace, on average, some of the densest regions of the Universe usually classed as `proto-clusters' \cite[e.g.][]{Franck2016}. This is because the brightest LAEs within SC4K across the COSMOS field are hosted by dark matter haloes of $\sim10^{13-14}$\,M$_{\odot}$ at $z>2.5$ \citep[][]{Khostovan2018}, which will easily result in massive clusters of $\sim10^{14-15}$\,M$_{\odot}$ in the local Universe when extrapolating to the present day using halo mass accretion growth. The number densities of these sources also agrees with our findings, being below $10^{-6}$\,Mpc$^{-3}$. The results thus bring further context into the findings of bright LAEs in or around some of the most over-dense regions in the Universe at $z\sim2-6$ \citep[][]{Venemans2007,Yamada2012}, including e.g. Ly$\alpha$ `blobs' \citep[e.g.][]{Matsuda2004,Kubo2013} and point towards large volume Ly$\alpha$ surveys as ideal ways to find these extremely over-dense regions. Given the high fraction of AGN among the population of these very high luminosity LAEs, it is not surprising that many studies also find those sources (e.g. X-ray or radio detected; see e.g. \citealt{Venemans2007}) to be good tracers of over-densities throughout the Universe \citep[see][and references therein]{Lehmer2009,Matsuda2011,Kubo2013,Overzier2016}.

\section{Conclusions} \label{sec:conclusions}

We have presented a new sample of $\sim4,000$ typical ($\gtrsim$\,L$^{\star}_{\rm Ly\alpha}$) LAEs (SC4K; Table \ref{catalogue_appendix}), selected through 12 medium- and 4 narrow-band filters in the full $\sim2$\,deg$^2$ COSMOS field, covering a wide redshift range ($z\sim2-6$). We use our large sample to construct Ly$\alpha$ LFs for the different redshift slices and investigate the evolution across cosmic time. We also combine SC4K with results from the literature to obtain a powerful consensus/synergy Ly$\alpha$ survey (S-SC4K) that spans over 4 orders of magnitude in Ly$\alpha$ luminosity across $z\sim2-6$. Our main results are:
\begin{itemize}

\item SC4K extensively complements ultra-deep surveys, jointly covering over 4\,dex in Ly$\alpha$ luminosity and revealing a global ($2.5<z<6$) synergy LF with a steep faint end slope $\alpha=-1.93^{+0.12}_{-0.12}$, a characteristic luminosity of $\rm \log_{10}L^*_{\rm Ly\alpha}={42.93^{+0.15}_{-0.11}}$\,erg\,s$^{-1}$ and a characteristic number density of $\rm \log_{10}\Phi^*_{\rm Ly\alpha}={-3.45^{+0.22}_{-0.29}}$\,Mpc$^{-3}$.

\item The Schechter component of the Ly$\alpha$ LF shows a factor $\sim5$ rise in $\rm L^*_{\rm Ly\alpha}$, from $\approx10^{42.7}$\,erg\,s$^{-1}$ at $z\sim2$ to $\approx10^{43.35}$\,erg\,s$^{-1}$ at $z\sim6$ and a $\sim7\times$ decline in $\Phi^*_{\rm Ly\alpha}$ from $z\sim2$ to $z\sim6$. We also find evidence for the faint-end slope to steepen from $\alpha=-1.7\pm0.2$ at $z\sim2.5$ to $\alpha=-2.5\pm0.2$ at $z\sim5$. Most importantly, $\alpha$ is always very steep and close to $\alpha=-2$ at all redshifts probed.

\item A Schechter function provides a good fit to the LF up to luminosities of $\sim10^{43.3}$\,erg\,s$^{-1}$, but we find a significant extra power-law (or Schechter) component above L$_{\rm Ly\alpha}=10^{43.3}$\,erg\,s$^{-1}$. We show that the extra component is partially driven by X-ray and radio AGN, as their Ly$\alpha$ LF resembles the excess. This extra component is found to decline (steepen) significantly with redshift and/or becomes mixed with the main Schechter component beyond $z\sim3.5$, likely linked with the evolution of the AGN population. This means that above $z\sim3.5$ a single Schechter function becomes a good description of the Ly$\alpha$ luminosity function from the lowest to the highest Ly$\alpha$ luminosities.

\item The Ly$\alpha$ luminosity density rises by a factor $\sim2$ from $z\sim2$ to $z\sim3$ but is then found to be roughly constant ($1.1^{+0.2}_{-0.2}\times10^{40}$\,erg\,s$^{-1}$\,Mpc$^{-3}$) to $z\sim6$, despite the $\sim0.7$\,dex drop in UV luminosity density. As a consequence, the SFRD$_{\rm{Ly}\alpha}/\rm{SFRD}_{UV}$ ratio rises from $4\pm1$\% to $30\pm6$\% from $z\sim2.2$ to $z\sim6$. LAEs become increasingly important as SFRD contributors into the epoch of re-ionisation, and not simply a relatively minor/rare population.

\item  Our results are consistent with a rise of a factor of $\approx2$ in the cosmic ionisation efficiency ($\xi_{\rm ion}$) and imply a factor $\approx4\pm1$ increase in the cosmic $\rm f_{esc}$ from $z\sim2$ to $z\sim6$. We find that an increase of $\rm f_{esc}$ with redshift as $(1+z)^{2.0\pm0.3}$ and a further increase of $\xi_{\rm ion}$ as $(1+z)$ can successfully model the global increase of SFRD$_{\rm{Ly}\alpha}/\rm{SFRD}_{UV}$ as $(1+z)^{3.0\pm0.3}$.

\item Our results hint for evolution in both the typical burstiness/stellar populations and even more so in the typical ISM conditions for Ly$\alpha$ photons to escape more efficiently at higher redshift. These trends may well be connected with the typically younger and more metal-poor galaxies becoming more dominant -- explaining the higher typical $\rm \xi_{ion}$ -- and also typically smaller/more compact morphologies, likely linked with the rise of $\rm f_{esc}$. SC4K LAEs are ideal follow-up candidates for these scenarios to be tested with current state-of-the-art and upcoming instruments/telescopes.

\end{itemize}

\section*{Acknowledgements}

We thank the anonymous referee for their constructive comments that helped us improve the manuscript. DS acknowledges the hospitality of the IAC and a Severo Ochoa visiting grant. SS and JC acknowledge studentships from the Lancaster University. JM acknowledges a Huygens PhD fellowship from Leiden University. APA acknowledge financial support from the Science and Technology Foundation (FCT, Portugal) through research grants UID/FIS/04434/2013 and fellowship PD/BD/52706/2014. The authors thank Alyssa Drake, Kimihiko Nakajima, Yuichi Harikane, Max Gronke, Irene Shivaei, Helmut Dannerbauer, Huub R\"{o}ttgering, Marius Eide and Masami Ouchi for many engaging and stimulating discussions. We also thank Sara Perez, Alex Bennett and Tom Rose for their involvement in the early stages of this project. Based on data products from observations made with ESO Telescopes at the La Silla Paranal Observatory under ESO programme IDs 294.A-5018, 097.A-0943, 098.A-0819, 099.A-0254 and 179.A-2005 and on data products produced by TERAPIX and the Cambridge Astronomy Survey Unit on behalf of the UltraVISTA consortium. Based on observations using the WFC on the 2.5\,m Isaac Newton Telescope, as part of programs 2013AN002, 2013BN008, 2014AC88, 2014AN002, 2014BN006, 2014BC118 and 2016AN001. The INT is operated on the island of La Palma by the Isaac Newton Group in the Spanish Observatorio del Roque de los Muchachos of the Instituto de Astrofisica de Canarias. This work is based in part on data products produced at Terapix available at the Canadian Astronomy Data Centre as part of the Canada-France-Hawaii Telescope Legacy Survey, a collaborative project of NRC and CNRS.

We are grateful to the CFHTLS, COSMOS-UltraVISTA and COSMOS survey teams. We are also unmeasurably thankful to the pioneering and continuous work from previous Ly$\alpha$ surveys' teams. Without these previous Ly$\alpha$ and the wider-reach legacy surveys, this research would have been impossible. We also thank the VUDS team for making available spectroscopic redshifts from data obtained with VIMOS at the European Southern Observatory Very Large Telescope, Paranal, Chile, under Large Program 185.A-0791. Finally, the authors acknowledge the unique value of the publicly available programming language {\sc Python}, including the {\sc NumPy} \& {\sc SciPy} \citep[][]{van2011numpy,jones}, {\sc Matplotlib} \citep[][]{Hunter:2007}, {\sc Astropy} \citep[][]{Astropy2013} and the {\sc Topcat} analysis program \citep{TOPCAT2005}. We publicly release a catalogue with all LAEs using in this paper (SC4K), so it can be freely explored by the community (see five example entries in Table \ref{catalogue_appendix}). 

\bibliographystyle{mnras}
\bibliography{myBib}

\appendix

\section{Catalogue of Lyman-$\alpha$ emitters (SC4K)}

%
%
\begin{table*}
\caption{Our full SC4K catalogue of candidate LAEs which we release with this paper. The SC4K catalogue contains the samples obtained with the 12 COSMOS medium-bands, together with 4 narrow-band samples from \citet{Santos2016}, \citet{Sobral2017}, \citet{MattheeBOOTES2017} and Perez et al. (in prep.). We provide five example entries. The full catalogue is available in electronic format ({\sc fits} table) with the final refereed paper. Errors on EW$_0$, Flux and L$_{\rm Ly\alpha}$ are computed by independently perturbing the MB and BB magnitudes along their Gaussian uncertainties 10,000 times per source and computing the 16th and 84th percentiles of each computed quantity. Note that for faint sources EWs are affected by large uncertainties; see e.g. IA427-141. The AGN flag in the catalogue provides information on the matches with public X-ray (including coverage) and radio catalogues (see Section \ref{sec:nature_lya}): 0 -- no match/no coverage; 1 -- X-ray detected; 2 -- radio detected.} \label{catalogue_appendix}
\begin{center}
\begin{tabular}{@{}ccccccccc@{}}
\hline
ID & R.A. & Dec. & MB or NB  & BB & EW$_0$ & Flux/$10^{-17}$ & $\log_{10}\,$L$_{\rm Ly\alpha}$ & AGN flag \\
 (SC4K-) & (J2000) & (J2000) & (AB) & (AB) & ({\AA}) & (erg\,s$^{-1}$\,cm$^{-2}$) & (erg\,s$^{-1}$) & (X-ray or radio) \\
\hline
IA427-141 & 10\,03\,20.01 & +02\,13\,38.8 & $24.83\pm0.06$ & $26.47\pm0.28$ & $2000^{+2000}_{-1400}$ & $13.4^{+0.7}_{-0.7}$ & $42.83^{+0.02}_{-0.02}$ & 0 \\
IA427-446 & 10\,02\,38.96 & +02\,14\,16.3 & $24.87\pm0.07$ & $25.97\pm0.18$ & $289^{+270}_{-124}$ & $10.6^{+0.9}_{-1.0}$ & $42.73^{+0.03}_{-0.04}$ & 0 \\
IA427-865 & 10\,02\,17.97 & +02\,15\,03.2 & $24.84\pm0.07$ & $25.82\pm0.15$ & $205^{+136}_{-98}$ & $10.2^{+1.0}_{-1.8}$ & $42.71^{+0.04}_{-0.08}$ & 0 \\
IA427-1169 & 10\,03\,10.85 & +02\,15\,37.6 & $24.28\pm0.04$ & $25.25\pm0.09$ & $193^{+126}_{-38}$ & $16.8^{+2.4}_{-1.0}$ & $42.92^{+0.06}_{-0.03}$ & 0 \\
IA427-1559 & 10\,02\,13.65 & +02\,16\,28.9 & $24.90\pm0.07$ & $25.46\pm0.11$ & $65^{+7}_{-21}$ & $6.5^{+0.3}_{-1.6}$ & $42.51^{+0.02}_{-0.13}$ & 0 \\
\hline
\end{tabular}
\end{center}
\end{table*}

We publicly release the full SC4K catalogue of 3,908 LAEs at $z\sim2-6$ derived and used in this paper, based on data obtained with 16 different medium- and narrow-band filters over the full COSMOS field. We show 5 example entries of the catalogue in Table \ref{catalogue_appendix}. The full electronic version of the catalogue will be available with the refereed paper in a {\sc fits} table. Table \ref{tab:colour_coef} presents the colour terms used to correct medium-band magnitudes and to compute emission line fluxes.

%
%
\begin{table}
\caption{The colour coefficients for each medium-band, used to correct the observed medium-band magnitudes ($\rm MB_0$) into $\rm MB$, as defined in Equation \ref{eq:colour_coef}: ${\rm MB=MB_0}-(m\times({\rm BB-BB_{adjacent}}) + b)$. For sources without a colour determination ($\rm BB-BB_{adjacent}$) we add the median correction listed in the table. Note that we use the MB magnitudes (and not $\rm MB_0$) for our SC4K catalogue (see Table \ref{catalogue_appendix}) and all derived quantities.}\label{tab:colour_coef}
\begin{center}
\begin{tabular}{ccccc}
\hline
MB & $\rm BB-BB_{\rm adjacent}$ & $m$ & $b$ & Median \\
 &   &  & & correction  \\
\hline
IA427 & $B-U$ & 0.33  & -0.11 & 0.01 \\
IA464 & $B-V$ & 0.0  & 0.0 & 0.0 \\
IA484 & $B-V$ & 0.0 & 0.0 & 0.0 \\
IA505 & $V-B$ & 0.0 & 0.0 & 0.0 \\
IA527 & $V-B$ & 0.0 & 0.0 & 0.0 \\
IA574 & $r^{+}-V$ & 0.0 & 0.0 & 0.0 \\
IA624 & $r^{+}-i^{+}$ & 0.0 & 0.0 & 0.0\\
IA679 & $r^{+}-i^{+}$ & -0.30 & -0.18 & 0.31 \\
IA709 & $r^{+}-i^{+}$ & -0.31 & 0.0 & -0.13 \\
IA738 & $r^{+}-i^{+}$ & -0.14 & 0.08 & -0.14 \\
IA767 & $i^{+}-z$ & 0.0 & 0.25 & -0.25 \\
IA827 & $i^{+}-z$ & -0.49 & 0.34 & -0.20 \\
\hline
\end{tabular}
\end{center}
\end{table}

\section{[OIII]+H$\beta$ excess in the $K_s$ band at $z\approx3$}\label{HK_colours}

%
%
\begin{figure}
\centering
\includegraphics[width=8.4cm]{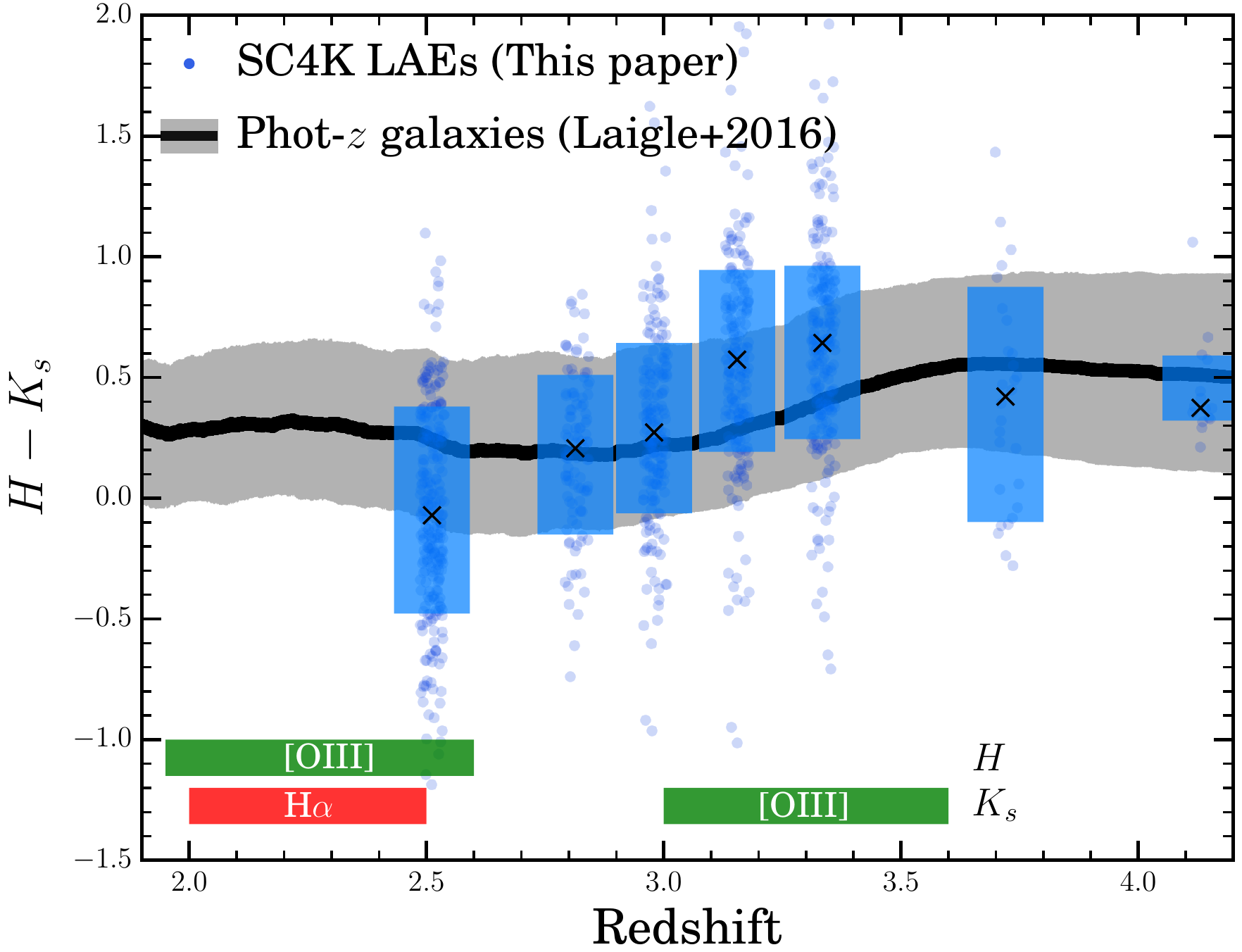} 
\caption{$H-K_s$ colours as a function of redshift for our sample of LAEs at $z\sim2.5-4$. The grey region shows the 16-84 percentile range of the colours of the general galaxy population \citep{Laigle2016}, while the blue boxes show the percentiles for the SC4K LAEs. We use photometric-redshifts for the general galaxy sample, but assign the redshift where Ly$\alpha$ falls in the MB for the SC4K emitters (points are randomly shifted for visualisation purposes). The green and red boxes indicate the redshifts where the strong H$\alpha$ and [O{\sc iii}] lines fall in the $H$ and $K_s$ filters and can affect the colours. The LAEs at $z\approx3.1-3.4$ have systematically redder $H-K_s$ colours compared to the general galaxy population, indicative of strong [O{\sc iii}] emission in the $K_s$ filter.} \label{fig:HKredshift}
\end{figure} 

A diagnostic that provides information on the validity of the sample of LAEs, and simultaneously provides insight into their nature is the evolution of the $H-K_s$ colours with redshift. The flux in these filters may be boosted by strong H$\alpha$ and [O{\sc iii}]+H$\beta$ emission lines \citep[e.g.][]{Faisst2016b}, depending on the redshift, affecting the $H-K_s$ colours (see also \citealt{Forrest2017}). If our sample had significant number of redshift interlopers, such effects on the colours would not be seen, as interlopers will not show them. 

Figure $\ref{fig:HKredshift}$ shows the median $H-K_s$ colours of the general galaxy population in the COSMOS field \citep{Laigle2016} and of the SC4K sample of LAEs from $z\sim2$ to $z\sim4$. Several interesting trends can be seen. The sample of LAEs at $z\approx2.5$ has systematically bluer $H-K_s$ colours than the general galaxy sample, which indicates that the $H$ band is significantly boosted by strong [O{\sc iii}]+H$\beta$ emission, while the majority of the sample does not have H$\alpha$ falling in the $K_s$ filter. The LAEs at $z=3.1-3.4$ have systematically redder $H-K_s$ colours than typical galaxies. This indicates LAEs have relatively strong [O{\sc iii}]+H$\beta$ emission, which is similar to the spectroscopic results from \cite{Nakajima2016}. As no strong lines affect the $H-K_s$ colours at $z\approx2.7-3.0$ and $z>3.6$, the colours of LAEs at these redshifts are similar to the colours of the general population.

\section{Ly$\alpha$ Luminosity functions} \label{tab:ap_LFvalues}

\begin{figure}
\centering
\includegraphics[width=8cm]{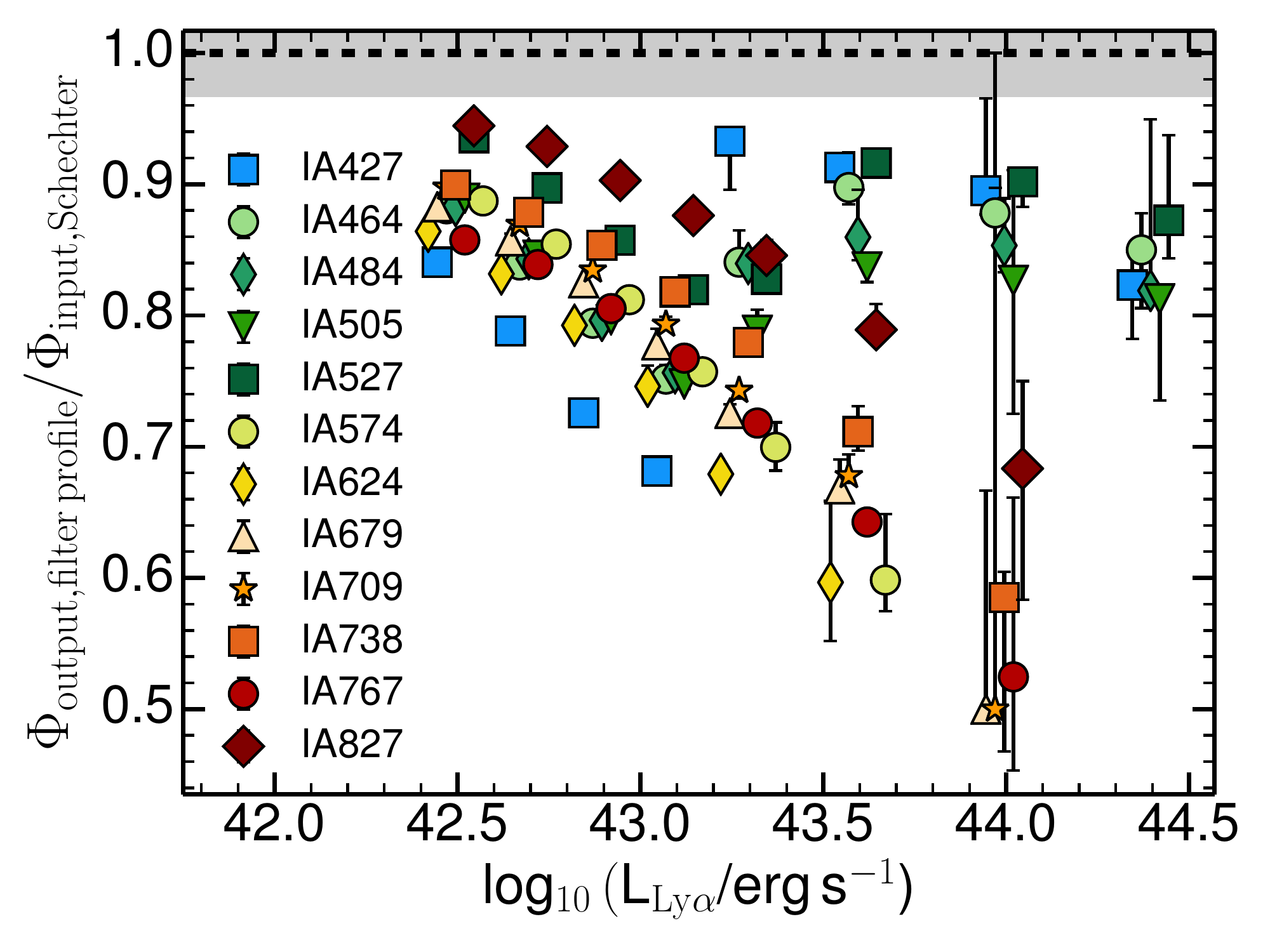}
\caption{The ratio between the observed Ly$\alpha$ luminosity function through the real filter profiles and input simulated sample of LAEs. We shift the bins by $\pm0.08$ for visibility. For each filter, we distribute simulated sources over a large redshift range (wider than what the filter can detect), with a number density distribution given by the first pass/observed Ly$\alpha$ luminosity function. This includes both a Schechter component and a power-law component at the highest luminosities, for $2.2<z<3.5$, and a Schechter component only for $3.5<z<6$. Points are offset in luminosity for visibility. We find that the bright end of the Schechter component of the LF leads to a significant observed underestimation of the LF, while the power-law component is more easily recovered (see Section \ref{sec:apdx_filtercor} for full details.}
\label{fig:filter_cor_all}
\end{figure}

\subsection{Filter profile corrections} \label{sec:apdx_filtercor}

In order to evaluate the necessary potential corrections to the Ly$\alpha$ LF due to the use of different real filters in comparison to idealized top-hat versions of them, we follow the procedure fully described in Section \ref{sec:filtercor} \citep[see also][]{Sobral2012}. Here we provide the full results of our simulations, presented in Figure \ref{fig:filter_cor_all} (see also Figure \ref{fig:filter_cor_816_827}) and discuss them. We find that the number density of sources recovered by folding through a population of LAEs with a luminosity function described by a Schechter function is always underestimated, and strongly underestimated for the highest luminosities. This is a relatively easy effect to understand, and becomes particularly important for the state-of-the-art large volume surveys that can now probe significantly above L$^{\star}$, where the number counts may drop exponentially (contrarily to the behavior of the sub-L$^{\star}$ component of the luminosity function, which behaves like a power-law).

Our results are a consequence of observed fluxes being the convolution between real input fluxes and the filter profile transmission. On average, this always results in a drop of flux except at the very peak of filter profile transmission. For medium-band filters there is still considerable volume under these conditions, while for narrow-bands such fraction is lower. For the evaluation of the luminosity function, this means that the observed number densities of sources at some luminosity L are always lower than reality, as most sources of that luminosity actually contribute to bins of fainter luminosities (they are observed to be fainter). The effect is not always extreme because it is partially compensated by sources at even higher luminosities that count towards a bin at luminosity L; this is why the intrinsic shape is crucial. The global result is that such corrections depend on the shape of the intrinsic luminosity function and how steep number density counts drop as a function of luminosity. This means that while for some shallow faint-end slopes the number of sources making it from higher luminosities and those making it away from a given bin is close to 1 (meaning the recovered number density is close to the input one), for steeper functions (and even more so for exponential declines), the effect starts to become very strong, as sources move to lower luminosity bins and almost no other sources come from brighter bins to compensate (because they are simply too rare). The effect is therefore the strongest beyond L$^{\star}$ for an intrinsic Schechter LF distribution of sources being observed through any filter profile that is not a perfect top-hat.

Overall, our results show that for any reasonable Schechter function prior (the observed LF is a good prior which does not require any assumptions), the bright end of the Ly$\alpha$ LF must be corrected more than the faint end, and that such corrections are much larger with narrow-band filters than with medium band filters (see Figure \ref{fig:filter_cor_816_827}). Moreover, once corrections are applied, narrow- and medium-band independent estimates agree. Our results also show that the exponential decline part of the LF becomes more and more underestimated the more Gaussian and the narrower a filter is (compared to assuming a top-hat transmission for the flux and volume). We note that another way to correct for the filter profile effects is to shift the luminosity function by some constant (to correct for the fact that a fraction of sources are not measured at full transmission). This is a relatively good way to do this in the case of a self-similar LF and/or when one is only measuring the power-law component of the Schechter function and when such component is not tremendously steep. However, for large volumes that can trace the exponential decline, the corrections are simply not the same: for the power-law (faint) component there are sources coming in from higher luminosities and going away to fainter luminosities, but in the exponential part there is much more migration away from the bin to lower L than there is migration into the bin from brighter sources. 

We also find that if the decline of the number densities at high L is described by a relatively shallow power-law (such as the cases found at $z\sim2-3$), then the corrections (Figure \ref{fig:filter_cor_all}) are close to unity (again, due to the same effect: there are brighter sources which are still numerous enough to make it into the bin and roughly compensate for sources that are observed to be fainter). Overall, our results show the importance of correcting for this effect specifically for Schechter-like functions, and less so for the case of a shallow power-law decline with increasing luminosity.

\subsection{Luminosity functions: this study and the S-SC4K compilation/comparison}

We provide the derived Ly$\alpha$ LFs (one example realisation; Table \ref{tab:lum_bins}), including the observed number of sources and number densities obtained after completeness and filter profile corrections (see Table \ref{tab:lum_bins}) and the full error propagation steps (see Section \ref{corrections_errs_LF}). Table \ref{SSC4K_refs_IDs} presents the full S-SC4K compilation which we use to compare our results and to derive our synergy LF (S-SC4K). We also present the results from the 10,000 fits to each perturbed LF in Table \ref{S-SC4K_full_results}.

%
%
\begin{table*}
\caption{The global Ly$\alpha$ LF and for each of the medium-band filters in SC4K/this study (full LFs provided as a {\sc fits} catalogue with the refereed version of the paper). Here we present the first LF (global) with the first 13 entries in the table. We show the sample/filter name, followed by the Ly$\alpha$ luminosity bin. We also present the number of observed sources in each bin and the volume densities with the chain following the steps described in the paper: observed, observed + perturbing selection, completeness corrected and filter profile corrected (final). In addition, we also show the full sequential error calculation/propagation (see full details in Section \ref{corrections_errs_LF}). We note that we set the error to 1.0 whenever it is not defined in log space (for the odd bins which are just populated by one source); for these bins the error propagation is not conducted.} \label{tab:lum_bins}
\begin{center}
\begin{tabular}{@{}cccccccc@{}}
\hline
Sample &  $\rm \log_{10}\,L_{Ly\alpha}$ & Sources & $\rm \Phi_{observed}$ & $\rm \Delta\Phi_{obs+pert}$   & $\rm \Phi_{comp. corr}$ &  $\rm \Phi_{final}$ & $\rm \Delta\Phi_{final}$    \\
&  (erg\,s$^{-1}$)  &   (\#)   &  (Mpc$^{-3}$) &  (Mpc$^{-3}$)  &  (Mpc$^{-3}$)  &  (Mpc$^{-3}$) &  (Mpc$^{-3}$)  \\
\hline
 SC4K All MBs &  $42.60\pm0.05$ &   $156\pm12$   &   $-3.41^{+0.03}_{-0.04}$ & $^{+0.07}_{-0.07}$   &   $-3.02^{+0.08}_{-0.10}$ &  $-2.93^{+0.08}_{-0.11}$ & $^{+0.09}_{-0.11}$  \\
 SC4K All MBs &  $42.70\pm0.05$ &   $134\pm11$   &   $-3.47^{+0.04}_{-0.04}$ & $^{+0.06}_{-0.06}$   &   $-3.23^{+0.06}_{-0.08}$ &  $-3.12^{+0.07}_{-0.09}$ & $^{+0.07}_{-0.09}$  \\
 SC4K All MBs &  $42.80\pm0.05$ &   $607\pm24$   &   $-3.45^{+0.02}_{-0.02}$ & $^{+0.03}_{-0.03}$   &   $-3.12^{+0.04}_{-0.04}$ &  $-3.05^{+0.04}_{-0.04}$ & $^{+0.05}_{-0.05}$  \\
 SC4K All MBs &  $42.90\pm0.05$ &   $463\pm21$   &   $-3.68^{+0.02}_{-0.02}$ & $^{+0.03}_{-0.03}$   &   $-3.47^{+0.04}_{-0.04}$ &  $-3.37^{+0.04}_{-0.04}$ & $^{+0.05}_{-0.05}$  \\
 SC4K All MBs &  $43.00\pm0.05$ &   $405\pm20$   &   $-3.89^{+0.02}_{-0.02}$ & $^{+0.05}_{-0.05}$   &   $-3.70^{+0.05}_{-0.06}$ &  $-3.60^{+0.05}_{-0.06}$ & $^{+0.06}_{-0.07}$  \\
 SC4K All MBs &  $43.10\pm0.05$ &   $220\pm14$   &   $-4.22^{+0.03}_{-0.03}$ & $^{+0.05}_{-0.05}$   &   $-4.12^{+0.06}_{-0.06}$ &  $-4.00^{+0.06}_{-0.07}$ & $^{+0.07}_{-0.07}$  \\
 SC4K All MBs &  $43.20\pm0.05$ &   $188\pm13$   &   $-4.35^{+0.03}_{-0.03}$ & $^{+0.09}_{-0.09}$   &   $-4.22^{+0.10}_{-0.11}$ &  $-4.11^{+0.10}_{-0.12}$ & $^{+0.11}_{-0.12}$  \\
 SC4K All MBs &  $43.30\pm0.05$ &   $113\pm10$   &   $-4.57^{+0.04}_{-0.04}$ & $^{+0.11}_{-0.11}$   &   $-4.48^{+0.11}_{-0.13}$ &  $-4.37^{+0.12}_{-0.14}$ & $^{+0.12}_{-0.15}$  \\
 SC4K All MBs &  $43.40\pm0.05$ &   $57\pm7$   &   $-4.92^{+0.05}_{-0.06}$ & $^{+0.10}_{-0.10}$   &   $-4.80^{+0.10}_{-0.13}$ &  $-4.68^{+0.11}_{-0.14}$ & $^{+0.11}_{-0.14}$  \\
 SC4K All MBs &  $43.50\pm0.05$ &   $50\pm7$   &   $-5.06^{+0.06}_{-0.07}$ & $^{+0.13}_{-0.13}$   &   $-4.93^{+0.14}_{-0.20}$ &  $-4.82^{+0.15}_{-0.21}$ & $^{+0.15}_{-0.22}$  \\
 SC4K All MBs &  $43.60\pm0.05$ &   $35\pm5$   &   $-5.21^{+0.07}_{-0.08}$ & $^{+0.16}_{-0.16}$   &   $-5.14^{+0.16}_{-0.25}$ &  $-5.02^{+0.17}_{-0.27}$ & $^{+0.18}_{-0.27}$  \\
 SC4K All MBs &  $43.75\pm0.05$ &   $14\pm3$   &   $-5.61^{+0.10}_{-0.14}$ & $^{+0.12}_{-0.12}$   &   $-5.58^{+0.12}_{-0.17}$ &  $-5.47^{+0.13}_{-0.19}$ & $^{+0.13}_{-0.19}$  \\
 SC4K All MBs &  $44.00\pm0.15$ &   $24\pm4$   &   $-5.86^{+0.08}_{-0.10}$ & $^{+0.11}_{-0.11}$   &   $-5.85^{+0.11}_{-0.14}$ &  $-5.79^{+0.12}_{-0.15}$ & $^{+0.12}_{-0.15}$  \\
 SC4K All MBs &  $44.30\pm0.15$ &   $5\pm2$   &   $-6.54^{+0.16}_{-0.26}$ & $^{+0.16}_{-0.16}$   &   $-6.54^{+0.16}_{-0.27}$ &  $-6.46^{+0.17}_{-0.28}$ & $^{+0.17}_{-0.29}$  \\
 SC4K All MBs &  $44.60\pm0.15$ &   $\bf 1^{+1}_{-0.8}$   &   $-7.24^{+0.30}_{-1.00}$ & $^{+0.30}_{-0.30}$   &   $-7.24^{+0.30}_{-1.00}$ &  $-7.24^{+0.30}_{-1.00}$ & $^{+0.30}_{-1.00}$  \\
\hline
\end{tabular}
\end{center}
\end{table*}

%
%
\begin{table*}
\caption{A compilation of Ly$\alpha$ LFs used or compared with in this study, by alphabetical order. We provide references to the original papers and also references for LFs generated to make them more comparable with those we present in this paper when appropriate (e.g. by correcting for potential contamination or by applying consistent filter profile corrections for comparison). We provide all these LFs as a {\sc fits} format catalogue with the full refereed version of the paper. The redshifts are the average when studies have used redshift bins with the $\pm$ representing the maximum and minimum redshifts in the studies, and not the standard deviation. Note that for NB surveys this is given/rounded to 0.1, but typically the redshift range is lower than that. The minimum and maximum luminosity bins probed by each study are given in $\rm \log_{10}(L_{Ly\alpha}$/erg\,s$^{-1}$). CC: correction for potential contamination by lower redshift emitters \citep[see][]{Sobral2017}; FPC: correction for filter profile effects \citep[see this study and][]{Matthee2015,Santos2016}.}\label{SSC4K_refs_IDs}
\begin{center}
\begin{tabular}{cccccc}
\hline
 Study \# & Reference(s) & Technique/ & Redshift & $\rm L_{Ly\alpha,min}$& $\rm L_{Ly\alpha,max}$ \\
(This compilation) &  (Original or w/ correction) & Instrument & ($z$)  &  ($\rm log_{10}$)  & ($\rm log_{10}$)  \\
\hline
1  & \cite{Bina2016}   &    IFU MUSE-All  & $z=4.8\pm1.8$ & $41.3$ & $42.2$ \\
2.1  & \cite{Cassata2011}   &    Slit VIMOS-bin  & $z=2.5\pm0.5$ & $41.3$ & $42.9$ \\
2.2  & \cite{Cassata2011}   &    Slit VIMOS-bin  & $z=3.8\pm0.8$ & $41.8$ & $42.8$ \\
2.3  & \cite{Cassata2011}   &    Slit VIMOS-bin  & $z=5.5\pm1.0$ & $42.1$ & $43.3$ \\
3  & \cite{Dawson2007}   &    NB Mosaic-CCD MT  & $z=4.5\pm0.1$ & $42.2$ & $43.4$ \\
4  & \cite{Drake2017a}   &    IFU MUSE-All  & $z=4.8\pm1.8$ & $41.9$ & $42.9$ \\
5  & \cite{Drake2017b}   &    IFU MUSE-All  & $z=4.7\pm1.9$ & $41.2$ & $42.8$ \\
5.1  & \cite{Drake2017b}   &    IFU MUSE-Bin  & $z=3.5\pm0.5$ & $41.6$ & $42.8$ \\
5.2  & \cite{Drake2017b}   &    IFU MUSE-Bin  & $z=4.5\pm0.5$ & $41.6$ & $43.3$ \\
5.3  & \cite{Drake2017b}   &    IFU MUSE-Bin  & $z=5.8\pm0.8$ & $41.6$ & $43.2$ \\
6  & \cite{Konno2016}   &    NB S-cam Subaru  & $z=2.2\pm0.1$ & $41.7$ & $44.4$ \\
6.1  & \cite{Konno2016,Sobral2017}   &    NB S-cam Subaru CC  & $z=2.2\pm0.1$ & $41.7$ & $44.4$ \\
7  & \cite{Konno2017}   &    NB HSC Subaru  & $z=5.7\pm0.1$ & $43.0$ & $43.8$ \\
7.1  & \cite{Konno2017,Santos2016}   &    NB HSC Subaru FPC  & $z=5.7\pm0.1$ & $43.0$ & $43.8$ \\
8.1  & \cite{MattheeBOOTES2017}   &    NB WFC INT  & $z=2.2\pm0.1$ & $42.8$ & $43.5$ \\
8.2  & \cite{MattheeBOOTES2017}   &    NB WFC INT  & $z=2.4\pm0.1$ & $43.4$ & $44.7$ \\
8.3  & \cite{MattheeBOOTES2017}   &    NB WFC INT  & $z=3.1\pm0.1$ & $43.0$ & $43.6$ \\
9.1  & \cite{Ouchi2008}   &    NB S-cam Subaru  & $z=3.1\pm0.1$ & $42.2$ & $43.6$ \\
9.2  & \cite{Ouchi2008}   &    NB S-cam Subaru  & $z=3.7\pm0.1$ & $42.7$ & $43.5$ \\
9.3  & \cite{Ouchi2008}   &    NB S-cam Subaru  & $z=5.7\pm0.1$ & $42.5$ & $43.5$ \\
9.4  & \cite{Ouchi2008,Santos2016}   &    NB S-cam Subaru FPC  & $z=5.7\pm0.1$ & $42.5$ & $43.5$ \\
10  & (Perez et al. in prep.)   &    NB S-cam Subaru  & $z=4.8\pm0.1$ & $43.1$ & $43.5$ \\
11  & \cite{Santos2016}   &    NB S-cam Subaru  & $z=5.7\pm0.1$ & $42.5$ & $43.7$ \\
12  & \cite{Sobral2017}   &    NB WFC INT  & $z=2.2\pm0.1$ & $42.3$ & $43.5$ \\
\hline
\end{tabular}
\end{center}
\end{table*}

%
%
\begin{table*}
\caption{The results of fitting different Ly$\alpha$ LFs 10,000 times with a Schechter function (and a single power-law, for comparison) at the appropriate luminosity range ($^*$ fitting only up to 10$^{43.3}$\,erg\,s$^{-1}$), when using SC4K only and when combining SC4K with deeper surveys (S-SC4K). As part of each fit we also integrate our Ly$\alpha$ LFs to obtain $\rho_{\rm Ly\alpha}$, derived for different redshift bins, down to $1.75\times10^{41}$\,erg\,s$^{-1}$, corresponding to 0.04\,$L^\star_{z=3}$ from \citet{Gronwall2007}; see Section \ref{sec:Lya_rho}. All errors are the 16th and 84th percentiles for all 10,000 realisations per LF estimation which, due to degeneracies in the parameters, can sometimes exaggerate the errors on individual parameters, so these can be seen as conservative. We also provide a comparison (ratio) between reduced $\chi^2$ for Schechter and power-law fits ($\rm \chi^2_{Sch}/\chi^2_{PL}$); values below 1 indicate that a Schechter fit performs better, while a large value indicates that a simple power-law fit provides a lower reduced $\chi^2$.}\label{S-SC4K_full_results}
\begin{tabular}{@{}lccccccc@{}}
\hline
Redshift slice &  $\alpha$ &  $\rm \log_{10}\,L^*_{Ly\alpha}$  & $\log_{10}\,\Phi^*_{\rm Ly\alpha}$ &  $\rho_{\rm Ly\alpha}/10^{40}$ Sch & Power-law (PL)   & $\rm \chi^2_{Sch}/$  & Reference(s)  \\
 (S-)SC4K  &    &  (erg\,s$^{-1}$)  & (Mpc$^{-3}$)  &   (erg\,s$^{-1}$\,Mpc$^{-3}$) &  (A\,$\log_{10}\,$L+B)  &  $\rm \chi^2_{PL}$ & (Table \ref{SSC4K_refs_IDs})    \\
\hline 
 $z=2.2\pm0.1^*$ & $-1.8\pm0.2$ (fix) &  $42.69^{+0.13}_{-0.11}$  & $-3.33^{+0.21}_{-0.26}$  &  $0.48^{+0.04}_{-0.04}$ & $-1.24^{+0.08}_{-0.09}$, $49.3^{+3.6}_{-3.6}$  & 0.3   &  2.1  \\
 $z=2.5\pm0.1^*$ & $-1.8\pm0.2$ (fix) &  $42.76^{+0.07}_{-0.07}$  & $-3.23^{+0.14}_{-0.15}$  &  $0.73^{+0.18}_{-0.13}$ & $-2.34^{+0.19}_{-0.20}$, $96.9^{+8.5}_{-8.1}$  & 2.3   &  SC4K only  \\
 $z=2.8\pm0.1^*$ & $-1.8\pm0.2$ (fix) &  $42.83^{+0.36}_{-0.19}$  & $-3.27^{+0.58}_{-0.75}$  &  $0.84^{+1.12}_{-0.41}$ & $-2.66^{+1.03}_{-1.01}$, $110.7^{+43.4}_{-44.2}$  & 1.0   &  SC4K only  \\
 $z=3.0\pm0.1^*$ & $-1.8\pm0.2$ (fix) &  $42.64^{+0.06}_{-0.05}$  & $-2.54^{+0.16}_{-0.16}$  &  $2.54^{+0.87}_{-0.62}$ & $-3.17^{+0.28}_{-0.29}$, $132.9^{+12.3}_{-12.2}$  & 1.1   &  SC4K only  \\
 $z=3.2\pm0.1^*$ & $-1.8\pm0.2$ (fix) &  $42.80^{+0.09}_{-0.07}$  & $-3.01^{+0.16}_{-0.19}$  &  $1.35^{+0.41}_{-0.29}$ & $-2.41^{+0.25}_{-0.27}$, $100.0^{+11.4}_{-10.6}$  & 0.6   &  SC4K only  \\
 $z=3.3\pm0.1^*$ & $-1.8\pm0.2$ (fix) &  $42.68^{+0.07}_{-0.06}$  & $-2.70^{+0.16}_{-0.16}$  &  $1.95^{+0.64}_{-0.44}$ & $-2.98^{+0.25}_{-0.26}$, $124.7^{+11.3}_{-10.6}$  & 1.5   &  SC4K only  \\
 $z=3.7\pm0.1$ & $-1.8\pm0.2$ (fix) &  $43.03^{+0.18}_{-0.15}$  & $-4.09^{+0.41}_{-0.40}$  &  $0.21^{+0.17}_{-0.09}$ & $-3.18^{+0.68}_{-0.85}$, $133.0^{+36.7}_{-29.3}$  & 2.4   &  SC4K only  \\
 $z=4.1\pm0.1$ & $-1.8\pm0.2$ (fix) &  $42.83^{+0.17}_{-0.15}$  & $-3.49^{+0.46}_{-0.43}$  &  $0.49^{+0.49}_{-0.19}$ & $-3.11^{+0.71}_{-0.85}$, $129.8^{+36.8}_{-30.7}$  & 0.8   &  SC4K only  \\
 $z=4.6\pm0.1$ & $-1.8\pm0.2$ (fix) &  $43.15^{+0.16}_{-0.15}$  & $-3.92^{+0.37}_{-0.38}$  &  $0.42^{+0.32}_{-0.16}$ & $-2.98^{+0.62}_{-0.68}$, $124.6^{+29.6}_{-27.0}$  & 1.5   &  SC4K only  \\
 $z=4.8\pm0.1$ & $-1.8\pm0.2$ (fix) &  $42.98^{+0.17}_{-0.14}$  & $-3.62^{+0.48}_{-0.46}$  &  $0.56^{+0.65}_{-0.28}$ & $-3.99^{+0.89}_{-1.01}$, $168.1^{+43.9}_{-38.5}$  & 1.4   &  SC4K only  \\
 $z=5.1\pm0.1$ & $-1.8\pm0.2$ (fix) &  $43.30^{+0.23}_{-0.19}$  & $-4.36^{+0.59}_{-0.54}$  &  $0.24^{+0.33}_{-0.13}$ & $-3.88^{+1.09}_{-1.51}$, $163.8^{+65.7}_{-47.4}$  & 2.0   &  SC4K only  \\
 $z=5.3\pm0.1$ & $-1.8\pm0.2$ (fix) &  $43.30^{+0.28}_{-0.20}$  & $-4.22^{+0.71}_{-0.73}$  &  $0.33^{+0.75}_{-0.21}$ & $-3.88^{+1.47}_{-1.68}$, $164.2^{+73.2}_{-64.0}$  & 0.8   &  SC4K only  \\
 $z=5.8\pm0.1$ & $-1.8\pm0.2$ (fix) &  $43.35^{+0.24}_{-0.19}$  & $-4.19^{+0.66}_{-0.67}$  &  $0.39^{+0.83}_{-0.25}$ & $-3.55^{+1.15}_{-1.49}$, $149.7^{+65.3}_{-50.2}$  & 0.9   &  SC4K only  \\
\hline
 $z=2.2\pm0.1^*$ & $-2.00^{+0.15}_{-0.15}$  &  $42.82^{+0.13}_{-0.11}$  & $-3.59^{+0.22}_{-0.28}$  &  $0.52^{+0.05}_{-0.05}$ & $-1.54^{+0.07}_{-0.07}$, $62.1^{+3.1}_{-3.0}$  & 0.6   &  2.1, 6.1, 12  \\
 $z=2.5\pm0.1^*$ & $-1.72^{+0.15}_{-0.15}$  &  $42.70^{+0.09}_{-0.08}$  & $-3.10^{+0.17}_{-0.20}$  &  $0.74^{+0.08}_{-0.07}$ & $-1.33^{+0.07}_{-0.07}$, $53.6^{+2.9}_{-3.0}$  & 0.6   &  2.1, 5.1  \\
 $z=2.8\pm0.1^*$ & $-1.73^{+0.20}_{-0.21}$  &  $42.78^{+0.16}_{-0.12}$  & $-3.18^{+0.27}_{-0.35}$  &  $0.77^{+0.10}_{-0.09}$ & $-1.28^{+0.08}_{-0.08}$, $51.3^{+3.5}_{-3.5}$  & 0.8   &  2.1, 5.1  \\
 $z=3.0\pm0.1^*$ & $-1.58^{+0.17}_{-0.17}$  &  $42.75^{+0.12}_{-0.09}$  & $-3.00^{+0.21}_{-0.25}$  &  $0.88^{+0.10}_{-0.09}$ & $-1.15^{+0.07}_{-0.07}$, $46.0^{+3.0}_{-2.8}$  & 0.6   &  2.1, 5.1  \\
 $z=3.2\pm0.1^*$ & $-1.70^{+0.17}_{-0.17}$  &  $42.85^{+0.15}_{-0.11}$  & $-3.20^{+0.24}_{-0.31}$  &  $0.84^{+0.09}_{-0.09}$ & $-1.15^{+0.07}_{-0.07}$, $45.9^{+2.9}_{-3.0}$  & 0.7   &  2.1, 5.1  \\
 $z=3.3\pm0.1^*$ & $-1.62^{+0.17}_{-0.17}$  &  $42.76^{+0.12}_{-0.10}$  & $-3.05^{+0.22}_{-0.26}$  &  $0.85^{+0.10}_{-0.09}$ & $-1.17^{+0.07}_{-0.07}$, $46.9^{+3.0}_{-2.9}$  & 0.6   &  2.1, 5.1  \\
 $z=3.7\pm0.1$ & $-2.57^{+0.23}_{-0.21}$  &  $43.23^{+0.37}_{-0.23}$  & $-4.54^{+0.61}_{-0.91}$  &  $1.01^{+0.20}_{-0.16}$ & $-2.01^{+0.12}_{-0.14}$, $82.2^{+5.7}_{-5.2}$  & 0.8   &  2.2, 5.1  \\
 $z=4.1\pm0.1$ & $-2.23^{+0.30}_{-0.24}$  &  $42.96^{+0.28}_{-0.22}$  & $-3.79^{+0.53}_{-0.66}$  &  $0.87^{+0.15}_{-0.11}$ & $-1.93^{+0.12}_{-0.14}$, $78.8^{+6.1}_{-5.2}$  & 0.9   &  2.2, 5.1  \\
 $z=4.6\pm0.1$ & $-2.38^{+0.20}_{-0.19}$  &  $43.32^{+0.24}_{-0.16}$  & $-4.34^{+0.39}_{-0.57}$  &  $1.19^{+0.40}_{-0.30}$ & $-1.80^{+0.10}_{-0.10}$, $73.6^{+4.1}_{-4.1}$  & 0.9   &  2.3, 3, 5.2, 10  \\
 $z=4.8\pm0.1$ & $-2.28^{+0.22}_{-0.22}$  &  $43.14^{+0.19}_{-0.15}$  & $-3.98^{+0.36}_{-0.46}$  &  $1.12^{+0.37}_{-0.27}$ & $-1.92^{+0.12}_{-0.12}$, $78.5^{+5.2}_{-4.9}$  & 0.8   &  2.3, 3, 5.2, 10  \\
 $z=5.1\pm0.1$ & $-2.46^{+0.22}_{-0.20}$  &  $43.41^{+0.28}_{-0.21}$  & $-4.58^{+0.54}_{-0.67}$  &  $1.27^{+0.48}_{-0.32}$ & $-2.00^{+0.13}_{-0.15}$, $82.1^{+6.3}_{-5.8}$  & 0.7   &  2.3, 3, 5.2, 10  \\
 $z=5.3\pm0.1$ & $-1.92^{+0.22}_{-0.19}$  &  $43.21^{+0.14}_{-0.13}$  & $-3.70^{+0.30}_{-0.32}$  &  $1.08^{+0.21}_{-0.16}$ & $-1.80^{+0.13}_{-0.14}$, $73.6^{+6.2}_{-5.3}$  & 0.2   &  5.3, 9.4, 11  \\
 $z=5.8\pm0.1$ & $-1.95^{+0.20}_{-0.18}$  &  $43.26^{+0.13}_{-0.13}$  & $-3.78^{+0.29}_{-0.30}$  &  $1.10^{+0.21}_{-0.16}$ & $-1.74^{+0.12}_{-0.13}$, $71.0^{+5.4}_{-5.0}$  & 0.2   &  5.3, 9.4, 11  \\
\hline

\end{tabular}
\end{table*}

\bsp	
\label{lastpage}
\end{document}